\documentclass[a4paper, 10pt, superscriptaddress, onecolumn, accepted=2023-01-11]{quantumarticle}

\pdfoutput=1

\usepackage{amsmath}
\usepackage{amssymb}
\usepackage{bbm}
\usepackage{framed}
\usepackage{epstopdf}
\usepackage{color}
\usepackage[colorlinks = true, urlcolor = blue]{hyperref}
\usepackage{epstopdf}
\usepackage{dsfont}
\usepackage{amsthm}
\usepackage{wrapfig}
\usepackage{relsize}
\usepackage{bm} 
\usepackage[latin1]{inputenc}
\usepackage{enumitem}
\usepackage[numbers]{natbib}
\usepackage{graphicx}
\usepackage{physics}
\usepackage{xcolor}
\usepackage{hyperref}
\usepackage{subcaption}
\usepackage[left=2.5cm,right=2.5cm]{geometry}

\newcommand{\kb}[2]{\ket{#1}\!\!\bra{#2}}	
\newtheorem{thm}{Theorem}

\begin{document}

	\title{Resonant Multilevel Amplitude Damping Channels}
	\author{Stefano Chessa}
	\orcid{0000-0003-2771-8330}
	\affiliation{NEST, Scuola Normale Superiore and Istituto Nanoscienze-CNR, I-56126 Pisa, Italy}
	\affiliation{Electrical and Computer Engineering, University of Illinois Urbana-Champaign, Urbana, Illinois, 61801, USA}
	\email{schessa@illinois.edu}
	\author{Vittorio Giovannetti}
	\orcid{0000-0002-7636-9002}
	\affiliation{NEST, Scuola Normale Superiore and Istituto Nanoscienze-CNR, I-56126 Pisa, Italy}
	\date{16 Jan 2023}

\begin{abstract}
We introduce a new set of quantum channels: resonant multilevel amplitude damping (ReMAD) channels. Among other instances, they can describe energy dissipation effects in multilevel atomic systems induced by the interaction with a zero-temperature bosonic environment. At variance with the already known class of multilevel amplitude damping (MAD) channels,
this new class of maps allows the presence of an environment unable to discriminate transitions with identical energy gaps. After characterizing the algebra of their composition rules, by analyzing the qutrit case, we show that this new set of channels can exhibit degradability and antidegradability in vast regions of the allowed parameter space. There we compute their quantum capacity and private classical capacity. We show that these capacities can be computed exactly also in regions of the parameter space where the channels aren't degradable nor antidegradable.
\end{abstract}

\maketitle

\section{Intro}

While two-level quantum systems (qubit) represent the fundamental building block for
any  Quantum Information processing, there are indications that 
working with qudits (i.e. quantum systems with an Hilbert space of dimension $d>2$) may bring in  some  advantages, 
both in terms of communication and cryptography (see e.g. \cite{QUDIT_COMM}  and references therein) and of computation (see e.g. \cite{QUDIT_COMP}  and references therein), with qutrits that recently made their first appearance on commercial quantum devices \cite{QUTRIT_RIGETTI}. Despite this fact the landscape of mathematical models describing the physical noises affecting these systems is still relatively unexplored especially in terms of their associated information capacities, see \cite{QUDIT_DEP, QUDIT_PAULI, DEGRADABLE, d_arrigo_2007, MAD, PCDS, PLATYPUS, STR_DEG_CH, DET_POS_CAP} for results on basic noise models in higher dimensions. 
 Capacities are figures of merit developed in the context of Quantum Shannon Theory \cite{HOLEVOBOOK, WILDEBOOK, WATROUSBOOK, HAYASHIBOOK, NC,  ADV_Q_COMM, HOLEGIOV, SURVEY}, which allow one to quantitatively  measure the level of deterioration that a given noise process induces on the quantum system it acts upon. A formal definition of such quantities is properly constructed in a specific communication scenario. There one describes the effect of the noise as an information loss during a signaling process that connects a sender (Alice) who is controlling the state of the system before the action of the noise, and a receiver (Bob), who instead has access to the deteriorated version of the qudit. In this context, depending on the type of messages one is considering (e.g. classical, private classical, or quantum) and on the type of side resources one allocates to the task (e.g. shared entanglement, two-way communication), the capacity of the noise channel is defined. Formally it corresponds to the optimal rate which gauges the maximum number of bits, secret bits, or qubtis that Alice can reliably transfer to Bob per use of the channel in the communication setting.  
 Unfortunately for the vast majority of models such optimal rates are not computable neither analytically nor algorithmically \cite{Huang_2014, Cubitt2015, Elkouss2018} due to superadditivity and superactivation effects~\cite{SUPERADD, SUPERACT, SUPERADD_PRIV, SUPERADD_TRAD, SUPERADD_ENT}. In this sense the literature has evolved to find capacities bounds and to find channel properties to be leveraged in order to overcome the hurdles of the direct computation or at least provide meaningful upper bounds. Among others, concerning the unassisted quantum and private classical capacities, we find: degradability \cite{DEGRADABLE}, antidegradability \citep{ANTIDEGRADABLE}, weak degradability \cite{ANTIDEGRADABLE}, additive extensions \cite{ADD_EXT}, conjugate degradability \cite{CONJ_DEG}, less noisy or more capable channels \cite{LESS_N_MORE_C}, partial degradability \cite{PARTIAL_DEG}, approximate degradability \cite{APPROX_DEG}, teleportation-covariant channels \cite{PLOB}, unital channels \cite{UNITAL}, low noise approximations \cite{LOW_NOISE}. In this sense a corpus of literature is being built with the aim to produce efficiently computable bounds and approximations of these capacities, see e.g. \cite{7586115, 7807212, Christandl2017, 8482492, 2011.05949, Fang2021, 2202.11688, 2203.02127}. From this perspective, this paper approaches the evaluation of the quantum capacity $Q$ and the private classical capacity $C_{\text{p}}$ of a new class of qudit channels, providing their exact expression for a wide range of noise parameters.

Specifically, we introduce a  family of noisy transformations that mimic energy loss in a multilevel quantum system $\text{S}$ (e.g. an atom or a molecule coupled with a low temperature e.m. field). 
 Our construction builds up from the Multilevel Amplitude Damping (MAD) proposal of \cite{MAD} where a $d$-dimensional generalization of the qubit Amplitude Damping Channel (ADC)~\cite{NC} is modeled as a collection of independent two-level processes that induce transitions from higher energy levels of the system to the lower ones. These transitions are mediated by an exchange interaction of the type $\sigma_{\text{S}}^-\otimes \sigma_{\text{E}}^+ + \sigma_{\text{S}}^+\otimes \sigma_{\text{E}}^-$ between $\text{S}$ and its environment $\text{E}$ ($\sigma_{\text{X}}^{\pm}$ representing raising and lowering operators for the system  $\text{X}$).
 The major difference between the transformations  we discuss here and the ones analyzed in  \cite{MAD}, is that we now allow for the possibility that  the transitions of S involving identical energy gaps will couple with  the same type of excitations in the environment. 
We dub this  class of the processes Resonant Multilevel Amplitude Damping (ReMAD) channels:
they behave as usual MAD channels on the populations of each energy level of $\text{S}$, but exhibit different effects on the coherence terms of the system. 
 Specifically due to the peculiar nature of the selected $\text{S}-\text{E}$ coupling, 
 under the action of a ReMAD channel the system $\text{S}$ will be typically slightly 
 less prone to dephasing than under the action of a MAD channel characterized by the same
 transition probabilities. At the mathematical level this corresponds to a net reduction of the minimal number of Kraus operators~\cite{HOLEVOBOOK} required to describe the effect of the associated noise. This leads to some simplification in the characterization of their quantum capacities.
 What makes this noise model interesting is its simplicity and the fact that it can emerge in a variety of scenarios. To enumerate some of the ones relevant in quantum information processing and quantum communications we can mention: atomic systems in quantum memories and quantum repeaters \cite{Q_MEMORY_ATOMS,Q_REPEATERS_ATOMS,Q_NETWORK_ATOMS}, optical qudits transmitted through lines such as optical fibers with polarization dependent losses \cite{POLARIZATION_LOSS}, optical qudits interacting with beamsplitters and qudits encoded in harmonic oscillators \cite{QUBIT_OSC,ENC_SPIN_OSC}, see \cite{BOS_CODES, BOS_CODES1} and references therein for applications and implementations of bosonic codes, quantum computation and simulation with molecular spins \cite{MOLECULAR_SPINS} and references therein.\\

The article is structured as follows: in Sec.~\ref{sec: Definition}  we introduce ReMAD channels, their complementary channels and their composition rules; in Sec.~\ref{sec: Deg, Qcap&Prcap} we briefly review the issues of degradability, antidegradability and the computation of quantum and private classical capacities $Q$ and $C_{\text{p}}$; in Sec.~\ref{sec: Degradable chan} we provide an analysis of $Q$ and $C_{\text{p}}$ for qutrit ReMAD channels; conclusions are drawn in Sec.~\ref{sec: Conclusions}.

\section{Definitions}\label{sec: Definition}

\begin{figure}[t!]
     \centering
     \includegraphics[width=0.42\textwidth]{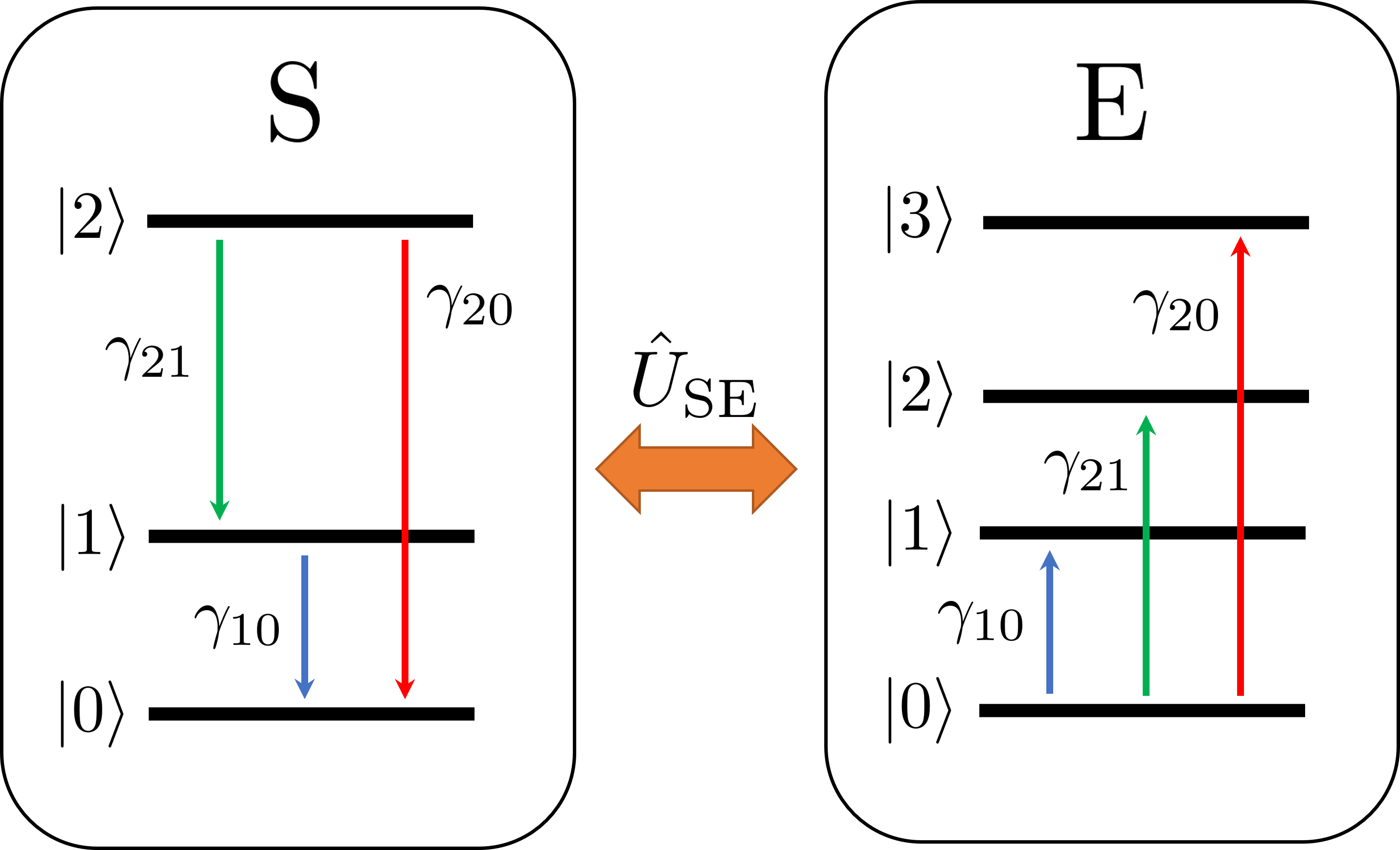} 
     \hspace*{1.5cm}
     \includegraphics[width=0.42\textwidth]{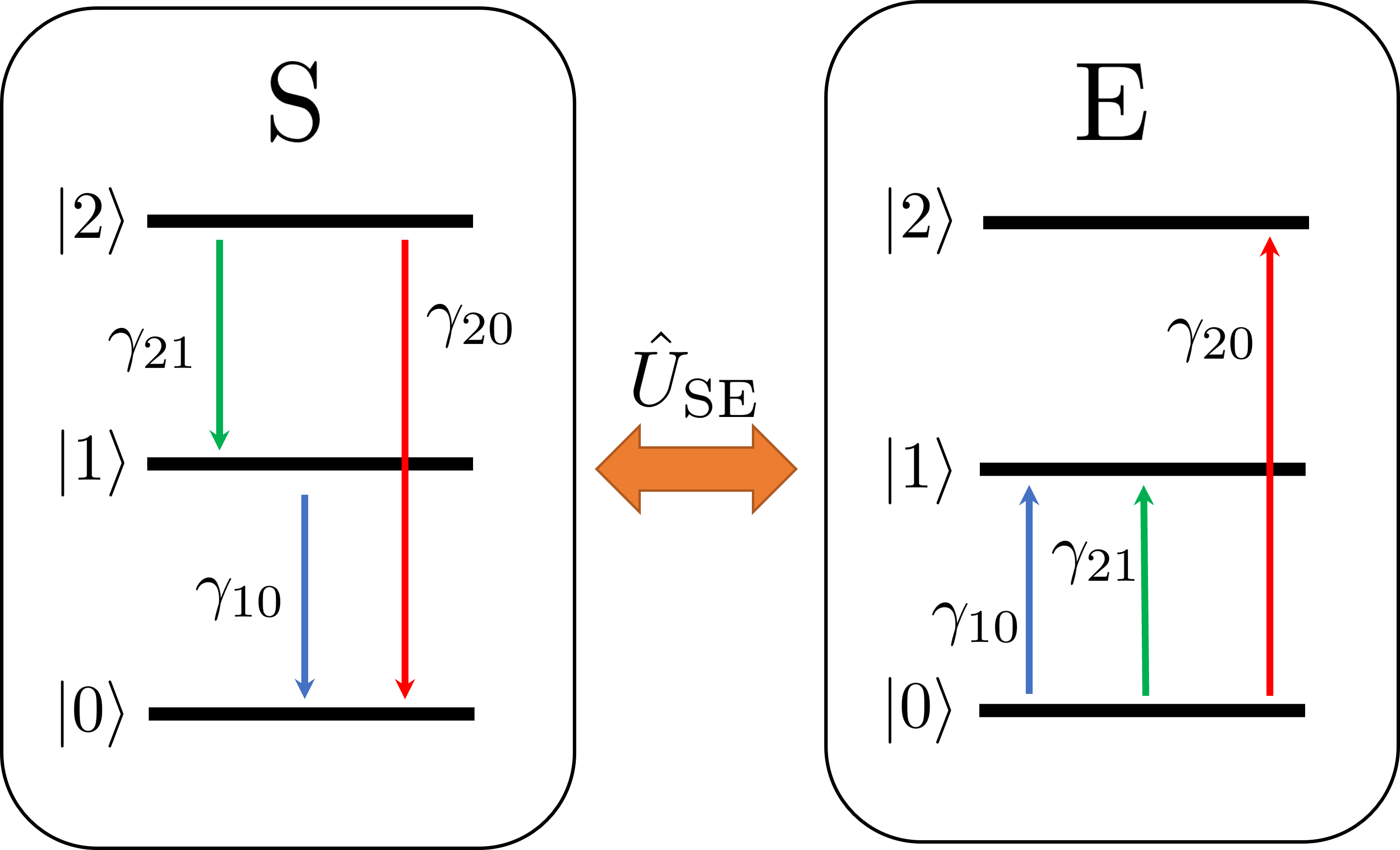}        
     \caption{\textbf{Left:} depiction of the MAD excitations exchange between the system S and its environment E. \textbf{Right:} depiction of the ReMAD excitations exchange between the system S and the environment E, notice the environment size. In both examples we set the total number of energy level equal to $d=3$.}
        \label{fig: qutritMADvsReMAD}
\end{figure}

Let $\mathcal{H}_{\text{S}}$ be the Hilbert space associated with a $d$-dimensional quantum system ${\text{S}}$ characterized by an energetically ordered canonical basis. This basis is represented by a collection of orthonormal levels 
$\{ \ket{0}_\text{S}, \ket{1}_\text{S}, \cdots, \ket{d-1}_\text{S}\}$ with $E_j < E_{j+1}$, being $E_j$ the energy associated with the $j$-th level. 
In this context we can formally describe energy damping processes by specifying a lower-triangular $d\times d$ transition matrix 
\begin{equation}\label{transM} 
\Gamma:=\begin{pmatrix}
\gamma_{0,0} & 0 & 0 &0 &  \cdots & 0\\
\gamma_{1,0}  & \gamma_{1,1} & 0 &0 &\cdots & 0\\
\gamma_{2,0}  & \gamma_{2,1} & \gamma_{2,2} &0 & \cdots & 0 \\
\gamma_{3,0}  & \gamma_{3,1}  & \gamma_{3,2} & \gamma_{3,3}& \cdots & 0\\
\vdots & \vdots & \vdots & \vdots & \vdots  & \vdots\\
\gamma_{d-1,0}  & \gamma_{d-1,1}  & \gamma_{d,2} &  \gamma_{d-1,3}  & \cdots & \gamma_{d-1,d-1}
\end{pmatrix} ,\end{equation}
 whose elements 
$\gamma_{j,k}$ for $j\leq k\in\{0,\cdots, d-1\}$ are positive semidefinite quantities that define the transition probabilities, i.e. the probability for the energy level $|j\rangle$ to be
 mapped into the lower energy level $|k\rangle$. Consistency requirements are such that 
 \begin{align} 
\sum_{k=0}^{j}\gamma_{j,k}= 1 \; , \qquad \forall j\in\{ 0, 1,\cdots, d-1\} \;, \label{probcons} 
\end{align}
that for $j\geq 1$ identify  $\gamma_{j,j} = 1- \sum_{k=0}^{j-1}\gamma_{j,k}$ as the survival probability  of the level $j$ (notice that by construction $\gamma_{0,0}=1$, being $|0\rangle_{\text{S}}$ a fixed point of the channels model).  
The MAD channels introduced in Ref.~\cite{MAD} assign to each  matrix $\Gamma$ 
a special 
Linear, Completely Positive, Trace Preserving (LCPTP)
transformation $\Phi^{({\text{\tiny MAD}})}_{\Gamma}$ that, 
at the level of Stinespring representation~\cite{STINE},  can be seen as a coupling with a zero-temperature external bath 
${\text{B}}$ absorbing each individual energy jump 
$E_{j} \rightarrow  E_{k}$ into a distinct (orthogonal) degree of freedom (see left panel of Fig.~\ref{fig: qutritMADvsReMAD}). 

\begin{figure}[t!]
     \centering
     \begin{subfigure}[b]{0.4\textwidth}
         \centering
         \includegraphics[width=\textwidth]{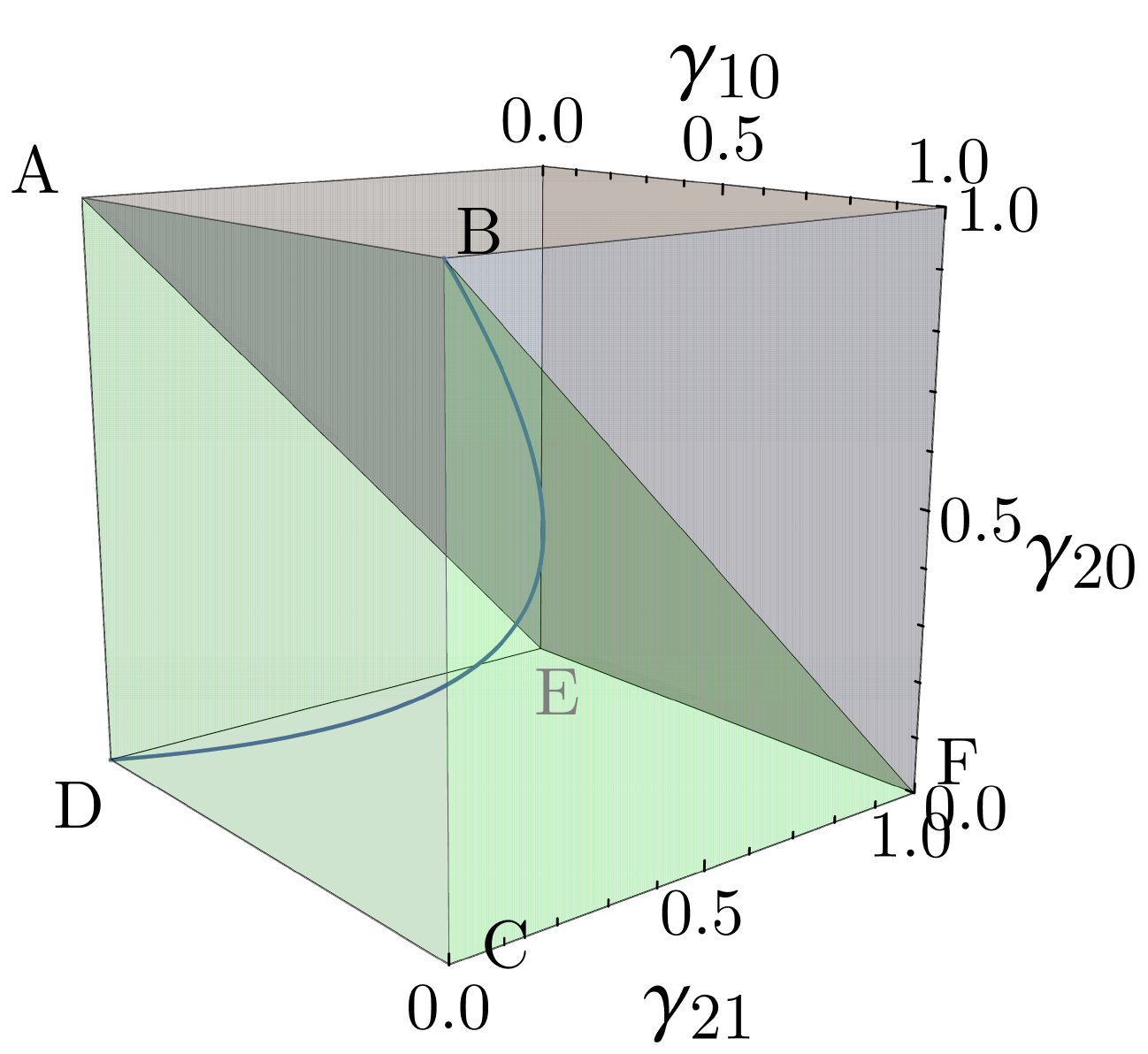}         
     \end{subfigure}
     \hspace*{5mm}
     \begin{subfigure}[b]{0.4\textwidth}
         \centering
         \includegraphics[width=\textwidth]{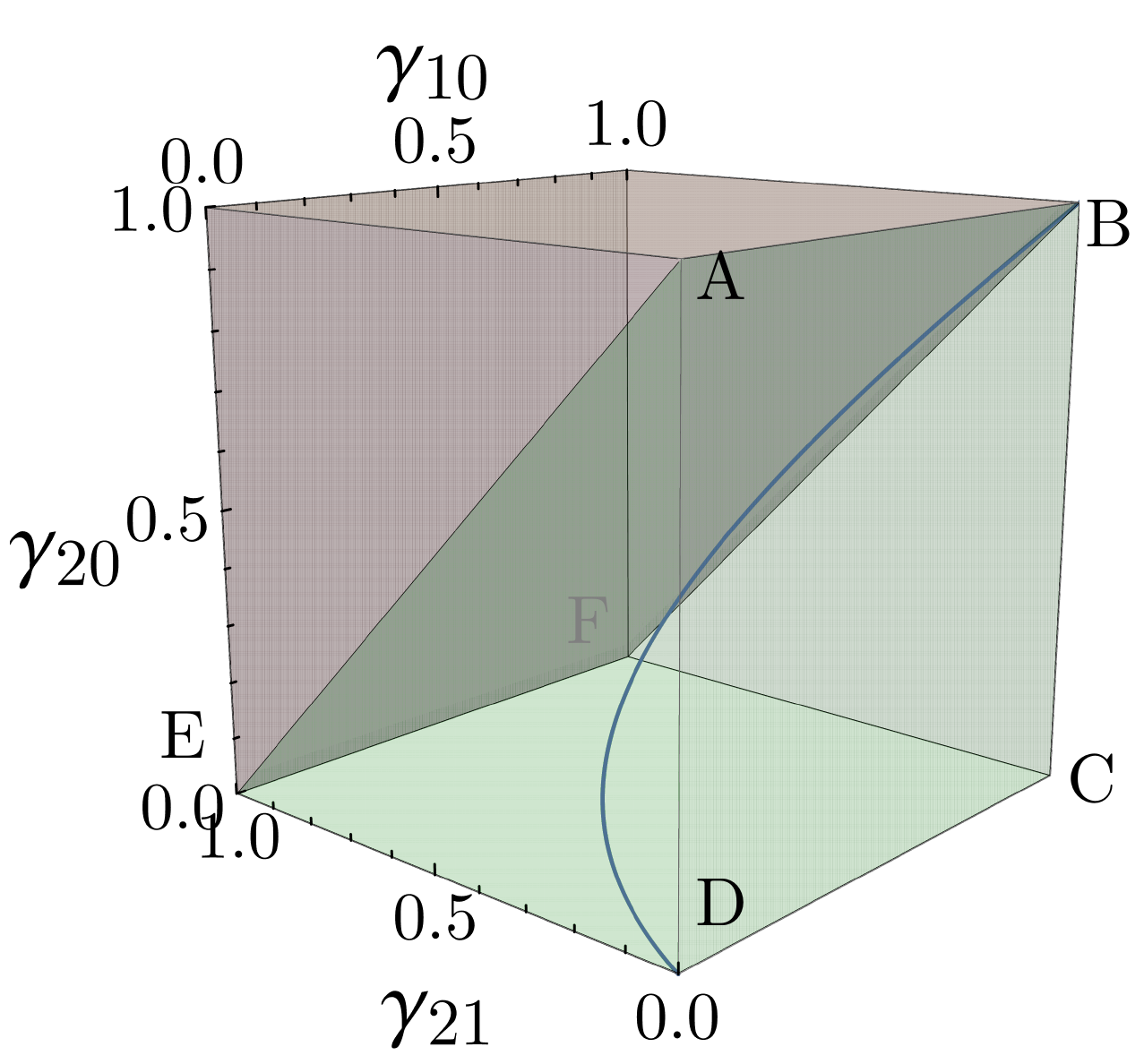}         
     \end{subfigure}
        \caption{Plot of the domain ${\mathbb D}_3$ of Eq.~(\ref{DOMAIN3}) which identifies the set of ReMAD channels for $d=3$ (green region, the grey volume corresponds to the inaccessible parameter region); the beamsplitter type ADC of Ref.~\cite{BS_ADC} corresponds to the  blue line that connects the vertexes ${\bf D}$ and ${\bf B}$. } 
        \label{fig: BS_ADC}
\end{figure}

Specifically, given $\hat{\rho}\in\mathfrak{S}({\cal H}_{\text{S}})$ a generic density operator of ${\text{S}}$, we express its evolution under the action of $\Phi^{({\text{\tiny MAD}})}_{\Gamma}$ as,
\begin{equation} \label{STINE1} 
\Phi^{({\text{\tiny MAD}})}_{\Gamma}(\hat{\rho}) = \mbox{Tr}_{\text{B}}[ \hat{V}_{\Gamma} (
\hat{\rho} \otimes \ket{0}\!\!\bra{0}_\text{B} ) \hat{V}^\dag_{\Gamma}] \;, \end{equation}
with $\ket{0}_\text{B}$ being the ground state of the bath, $\mbox{Tr}_{\text{B}}[\cdots]$ the partial trace over the environment, and $\hat{V}_{\Gamma}$ the unitary transformation that induces the mappings
\begin{eqnarray}\nonumber 
\hat{V}_{\Gamma} \ket{0}_\text{S}\ket{0}_\text{B} &:=&  \ket{0}_\text{S}\ket{0}_\text{B}\; , \\
 \hat{V}_{\Gamma} \ket{j}_\text{S}\ket{0}_\text{B} &:=& 
 \sqrt{\gamma_{j,j}}\ket{j}_\text{S}\ket{0}_\text{B} +  \sum_{k=1}^j  \sqrt{\gamma_{j,j-k}}\ket{j-k}_\text{S}\ket{j,k}_\text{B}\; , \qquad \forall j\in\{ 1,\cdots, d-1\} \label{defV} \;.
\end{eqnarray}
In the above expression  for $j\in \{1,\cdots, d-1\}$ and $k\in \{ 1,\cdots, j\}$, kets  $\ket{j,k}_\text{B}$ describe a collection of vectors that are mutually orthogonal and orthogonal with $\ket{0}_\text{B}$ as well. Since they represent the states where the environment stores the energy   $E_j-E_k$ lost by $\text{S}$ when moving from level $\ket{j}_\text{S}$ to level $\ket{k}_\text{S}$, the construction in Eq.~(\ref{defV}) implicitly assumes that the environment of the model is capable to discriminate among the different energy jumps. This condition is physically realized e.g. when the energy spectrum of $\text{S}$ is composed by incommensurable levels.
  Notice also that this construction fixes the dimension of the Hilbert space 
 $\mathcal{H}_{\text{B}}$ of 
$\text{B}$, as well as the minimum number of Kraus operators  needed to represent $\Phi^{({\text{\tiny MAD}})}_{\Gamma}$ in the operator-sum representation~\cite{KRAUS} 
equal to $d(d-1)/2+1$, i.e.
\begin{equation} \label{operatorsum} 
\Phi^{({\text{\tiny MAD}})}_{\Gamma}(\hat{\rho}) = \hat{M}_{\Gamma}^{(0)}\; \hat{\rho}\;  \hat{M}_{\Gamma}^{(0)\dagger} + 
 \sum_{j=1}^{d-1}\sum_{k=1}^j  \hat{M}_{\Gamma}^{(j,k)}  \; \hat{\rho}  \; \hat{M}_{\Gamma}^{(j,k)\dagger}  \; ,
\end{equation}
with
\begin{eqnarray}\label{eq: Kraus general MAD}
\hat{M}_{\Gamma}^{(0)}&:=&{_{\text{B}}\!\bra{0}} \hat{V}_{\Gamma} \ket{0}_{\text{B}} =
\sum_{l=0}^{d-1}\sqrt{\gamma_{l,l} }  \ket{l}\!\!\bra{l}_{\text{S}}\; ,\nonumber \\
 \hat{M}_{\Gamma}^{(j,k)} 
&:=&{_{\text{B}}\!\bra{j,k}} \hat{V}_{\Gamma} \ket{0}_{\text{B}} =
\sqrt{\gamma_{j,j-k}}\;  \ket{j-k}\!\!\bra{j}_{\text{S}} \; , \qquad  \forall j\in\{ 1,\cdots, d-1\}, \forall k\in \{ 1, \cdots,j\} \;.
\end{eqnarray}
By introducing ReMAD channels $\Phi_{\Gamma}$ we now allow for the possibility that some of the transition events of $\text{S}$ will excite the same internal degrees of freedom of 
the bath, a condition which can be achieved e.g. when the energy levels of the system are equally spaced, i.e. $E_j =j  \Delta E$ for all $j\in \{0,\cdots, d-1\}$. In this case we can replace 
the unitary coupling in Eq.~(\ref{defV}) with the new interaction 
\begin{eqnarray}\label{eq: transitions general}
 \hat{U}_{\Gamma} \ket{j}_\text{S}\ket{0}_\text{E} &:=& \sum_{k=0}^j  \sqrt{\gamma_{j,j-k}}\ket{j-k}_\text{S}\ket{k}_\text{E}\; , \qquad \forall j\in\{ 0,\cdots, d-1\}
\end{eqnarray}
where $\{ \ket{0}_\text{E}, \ket{1}_\text{E}, \cdots, \ket{d-1}_\text{E}\}$ are a (possibly energetically ordered) orthonormal basis of the environment ${\text{E}}$. The resulting LCPTP transformation is hence obtained by replacing Eqs.~(\ref{STINE1}) and (\ref{operatorsum}) with 
\begin{equation} \label{REMAD} 
\Phi_{\Gamma}(\hat{\rho}) = \mbox{Tr}_{\text{E}}[ \hat{U}_{\Gamma} (
\hat{\rho} \otimes \ket{0}\!\!\bra{0}_\text{E}  \hat{U}^\dag_{\Gamma}] 
=
\sum_{i=0}^{d-1}\hat{K}_{\Gamma}^{(i)} \hat{\rho} \hat{K}_{\Gamma}^{(i)^\dagger} \; ,
\end{equation}
where 
\begin{eqnarray}\label{eq: Kraus general}
 \hat{K}_{\Gamma}^{(i)}
&:=&{_{\text{E}}\!\bra{i}} \hat{U}_{\Gamma} \ket{0}_{\text{E}} =
\sum_{l=0}^{d-i-1}\sqrt{\gamma_{i+l,l}}\;  \ket{l}\!\!\bra{i+l}_{\text{S}} \; , \qquad \forall i\in\{ 0,\cdots, d-1\}\;,
\end{eqnarray}
is a Kraus set characterized by at most $d$ non-zero elements.
It is easy to check that wile for $d=2$, the ReMAD channel $\Phi_{\Gamma}$ coincides with its corresponding 
MAD counterpart $\Phi^{({\text{\tiny MAD}})}_{\Gamma}$ (indeed both reduce to the conventional qubit ADC).
Also one can easily verify that for arbitrary $d$, both $\Phi_{\Gamma}$ and $\Phi^{({\text{\tiny MAD}})}_{\Gamma}$
produce the same diagonal output states per diagonal input states $\hat{\rho}^{(\text{diag})} := \sum_{j=0}^{d-1} \rho_{j,j} \ket{j}\!\!\bra{j}_{\text{S}}$, indeed
  \begin{equation}
\Phi_{\Gamma} \left(\hat{\rho}^{(\text{diag})} \right)=\Phi^{({\text{\tiny MAD}})}_{\Gamma} \left(\hat{\rho}^{(\text{diag})} \right)
= 
\sum_{j=0}^{d-1} \sum_{k=0}^{j} \rho_{j,j}\gamma_{j,k}  \ket{k}\!\!\bra{k}_{\text{S}} \;. \label{DIACG} 
\end{equation} 
However for  $d\geq 3$ the two sets of transformations have rather different impact on  the off-diagonal term of the input state.
Consider for instance   the qutrit ($d=3$) scenario where, thanks to the constraint~(\ref{probcons}) 
  the transition matrix $\Gamma$   can be parametrized by 3 non-negative terms, e.g. $\gamma_{10}$, $\gamma_{21}$, and $\gamma_{20}$, forming a 3-dimensional vector $(\gamma_{10},\gamma_{21},\gamma_{20})$
  spanning  the domain 
\begin{eqnarray}{\mathbb D}_3: = \{ (\gamma_{10},\gamma_{21},\gamma_{20})\in \mathbb{R}^{3}: \label{DOMAIN3} \gamma_{10},\gamma_{20}, \gamma_{21} \in[0,1] \; \mbox{and}\;  \gamma_{20}+ \gamma_{21}  \leq 1\} \;, \end{eqnarray} 
represented in Fig.~\ref{fig: BS_ADC} by the green rectangular right wedge  delimited by the vertexes ${\bf A}$, ${\bf B}$, ${\bf C}$, ${\bf D}$, ${\bf E}$, and ${\bf F}$.  In this case 
Eq.~(\ref{eq: transitions general}) reduces to
\begin{align}
\hat{U}_{\Gamma} \ket{0}_\text{S}\ket{0}_\text{E}&= \ket{0}_\text{S}\ket{0}_\text{E}\;,  \nonumber \\
\hat{U}_{\Gamma} \ket{1}_\text{S}\ket{0}_\text{E}&= \sqrt{1-\gamma_{10}}\ket{1}_\text{S}\ket{0}_\text{E} +  \sqrt{\gamma_{10}}\ket{0}_\text{S}\ket{1}_\text{E} \;, \nonumber \\ 
\hat{U}_{\Gamma} \ket{2}_\text{S}\ket{0}_\text{E}&=  \sqrt{1-\gamma_{21} - \gamma_{20}}\ket{2}_\text{S}\ket{0}_\text{E}+ \sqrt{\gamma_{20}}\ket{0}_\text{S}\ket{2}_\text{E}  + \sqrt{\gamma_{21}}\ket{1}_\text{S}\ket{1}_\text{E}\;,
\end{align}
while, identifying the canonical basis states $|0\rangle_{\text{S}}$, $|1\rangle_{\text{S}}$, and 
$|2\rangle_{\text{S}}$ with the column vectors $(1,0,0)^T$, $(0,1,0)^T$, and $(0,0,1)^T$, we can express the associated Kraus operators in Eq.~(\ref{eq: Kraus general}) as
\begin{equation}
\begin{split}
&\hat{K}_{\Gamma}^{(0)}
=\begin{pmatrix}
1 & 0 & 0\\
0 & \sqrt{1-\gamma_{10}} & 0\\
0 & 0 & \sqrt{1-\gamma_{21}-\gamma_{20}}
\end{pmatrix} ,\quad \hat{K}_{\Gamma}^{(1)}
=\begin{pmatrix}
0 & \sqrt{\gamma_{10}} & 0\\
0 & 0 & \sqrt{\gamma_{21}}\\
0 & 0 & 0
\end{pmatrix},\quad
\hat{K}_{\Gamma}^{(2)}
=\begin{pmatrix}
0 & 0 & \sqrt{\gamma_{20}}\\
0 & 0 & 0\\
0 & 0 & 0
\end{pmatrix}. 
\end{split}
\end{equation}
Accordingly the action of $\Phi_{{\Gamma}}$ on a generic density matrix $\hat{\rho}$ of $\text{S}$ produces the output state of the form 
\begin{widetext}
\begin{equation}\label{eq: channel action}
\Phi_{{\Gamma}}(\hat{\rho})=\left(
\begin{array}{ccc}
\rho_{00} + \gamma_{10} \rho_{11}+\gamma_{20} \rho_{22} &  \sqrt{1-\gamma_{10}} \rho_{01} + \sqrt{\gamma_{10} \gamma_{21}}\rho_{12} &
    \sqrt{1-\gamma_{21}-\gamma_{20}} \rho_{02} \\
\sqrt{1-\gamma_{10}} \rho_{01}^* + \sqrt{\gamma_{10} \gamma_{21}} \rho_{12}^* & (1-\gamma_{10}) \rho_{11}+\gamma_{21} \rho_{22} &
    \sqrt{(1-\gamma_{10}) (1-\gamma_{21}-\gamma_{20})} \rho_{12} \\
 \sqrt{1-\gamma_{21}-\gamma_{20}} \rho_{02}^* &  \sqrt{(1-\gamma_{10}) (1-\gamma_{21}-\gamma_{20})}\rho_{12}^* &  (1-\gamma_{21}-\gamma_{20}) \rho_{22} \\
\end{array}
\right) \; ,
\end{equation}
\end{widetext}
to be compared with the associated transformation induced by the MAD counterpart
\begin{widetext}
\begin{equation}\label{eq: channel actionMAD}
\hspace*{-0.3cm}
\Phi^{({\text{\tiny MAD}})}_{\Gamma}(\hat{\rho})=\left(
\begin{array}{ccc}
\rho_{00} + \gamma_{10} \rho_{11}+\gamma_{20} \rho_{22} &  \sqrt{1-\gamma_{10}} \rho_{01}  &
    \sqrt{1-\gamma_{21}-\gamma_{20}} \rho_{02} \\
\sqrt{1-\gamma_{10}} \rho_{01}^*  & (1-\gamma_{10}) \rho_{11}+\gamma_{21} \rho_{22} &
    \sqrt{(1-\gamma_{10}) (1-\gamma_{21}-\gamma_{20})} \rho_{12} \\
 \sqrt{1-\gamma_{21}-\gamma_{20}} \rho_{02}^* &  \sqrt{(1-\gamma_{10}) (1-\gamma_{21}-\gamma_{20})}\rho_{12}^* &  (1-\gamma_{21}-\gamma_{20}) \rho_{22} \\
\end{array}
\right) \; ,
\end{equation}
\end{widetext}
with $\rho_{ij} := {_{\text{S}}\langle} i| \hat{\rho} |j\rangle_{\text{S}}$.
As one can observe, while on the diagonal elements in both cases we have the usual decay of populations predicted by
Eq.~(\ref{DIACG}),  the ReMAD output state in Eq.~(\ref{eq: channel action}) exhibits a transfer of coherence that mixes the terms $\rho_{12}$ and $\rho_{10}$ which is not contemplated in the MAD output of Eq.~(\ref{eq: channel actionMAD}). It is also worth noticing that
in case either $\gamma_{21}=0$ or $\gamma_{10}=0$ the qutrit map
$\Phi^{({\text{\tiny MAD}})}_{\Gamma}$ and the qutrit map $\Phi_{\Gamma}$ describe the same physical process.

As a final remark we notice  that a particular example of ReMAD channel is provided by the so called \textit{beamsplitter type} ADC ${\Psi}_{\eta}$
introduced  in Ref.~\cite{BS_ADC}. This subclass of ReMAD channels describes the evolution of a qudit encoded in the first $d$ states of the Fock basis of an harmonic oscillator passing through a beamsplitter of transmittance $\eta$. It's straightforward to verify that such mappings are a special instance of the ReMAD class characterized by a transition matrix $\Gamma[\eta]$ whose elements can be parametrized by the formula  
\begin{eqnarray} \gamma_{j,k}[\eta]:=\binom{j}{k}\eta^{j-k}(1-\eta)^k\;, \qquad \label{fdsf} 
\forall j\in \{ 0,\cdots, d-1\}\;,  \forall k\in \{0,\cdots, j\} \;,\end{eqnarray}  
so that ${\Psi}_{\eta} = \Phi_{\Gamma[\eta]}$. 
For $d=3$ this corresponds to having $\gamma_{10}[\eta]=\eta$, $\gamma_{21}[\eta]= 2 \eta(1-\eta)$, and $\gamma_{20}[\eta]= \eta^2$: a plot of the parameter region spanned by beamsplitter type ADC for $d=3$ is reported in Fig.~\ref{fig: BS_ADC}\;.

\subsection{Complementary maps} 
At variance with what happens with MAD channels, the complementary map~\cite{HOLEVOBOOK}  $\tilde{\Phi}_{\Gamma}$ of a generic ReMAD transformation ${\Phi}_{\Gamma}$
is also a ReMAD channel  (up to an isometry). This can be seen by recalling that $\tilde{\Phi}_{\Gamma}$ can be obtained from Eq.~(\ref{REMAD}) by 
replacing the partial trace over $\text{E}$ with a partial trace over $\text{S}$, i.e. 
\begin{equation} \label{REMADcomp} 
\tilde{\Phi}_{\Gamma}(\hat{\rho}) = \mbox{Tr}_{\text{S}}[ \hat{U}_{\Gamma} (
\hat{\rho} \otimes \ket{0}\!\!\bra{0}_\text{E} \hat{U}^\dag_{\Gamma}] 
=
\sum_{i=0}^{d-1}\hat{Q}_{\Gamma}^{(i)} \hat{\rho} \hat{Q}_{\Gamma}^{(i)\dagger} \; ,
\end{equation}
where now $\hat{Q}_{\Gamma}^{(i)}:{\cal H}_{\text{S}} \rightarrow {\cal H}_{\text{E}}$ are the operators  
\begin{eqnarray}\label{eq: Kraus general complementary}
 \hat{Q}_{\Gamma}^{(i)}
&:=&{_{\text{S}}\!\bra{i}} \hat{U}_{\Gamma} \ket{0}_{\text{E}} =
\sum_{l=0}^{d-i-1}\sqrt{\gamma_{i+l,i}}\;  \ket{l}_{\text{E\;S}}\!\bra{i+l} \; , \qquad \forall i\in\{ 0,\cdots, d-1\}\;.
\end{eqnarray}
Notice next that up to an isometry $\hat{V}_{SE}$ mapping the energy levels of E into the corresponding levels of S (i.e.
$\hat{V}_{SE} |l\rangle_E \rightarrow |l\rangle_S$, $\forall l$), $\hat{Q}_{\Gamma}^{(i)}$ has exactly the same form of the operators in Eq.~(\ref{eq: Kraus general}) computed with a new lower-triangular transition matrix $\tilde{\Gamma}$.
The non-zero elements of $\tilde{\Gamma}$, row by row, are obtained by a simple reordering of those of $\Gamma$, i.e. 
\begin{eqnarray} \label{Defgammadelta} 
\tilde{\gamma}_{j,k} := \gamma_{j,j-k} \;,  \qquad  \forall j\in\{ 1,\cdots, d-1\}, \forall k\in \{ 1, \cdots,j\}\;.
\end{eqnarray} 
Specifically we can write  
\begin{eqnarray}
\hat{V}_{SE} \hat{Q}_{\Gamma}^{(i)} = \hat{K}_{\tilde{\Gamma}}^{(i)} \;, \qquad \forall i\in\{ 0,\cdots, d-1\}\;.
\end{eqnarray}
and hence  \begin{eqnarray}\label{impoIDE} 
\hat{V}_{SE} \tilde{\Phi}_{\Gamma}(\cdots) \hat{V}^\dag_{SE}= {\Phi}_{\tilde{\Gamma}}(\cdots)\;.
\end{eqnarray} 
Again, as an example we provide the  explicit form of the complementary channel $\tilde{\Phi}_{{\Gamma}}$ of the qutrit ReMAD map $\Phi_{\Gamma}$ of Eq.~(\ref{eq: channel action}), i.e. 
\begin{equation}\label{eq: compl channel action}
\tilde{\Phi}_{{\Gamma}}(\hat{\rho})=\left(
\begin{array}{ccc}
 \rho_{00}+(1-\gamma_{10}) \rho_{11}+ (1-\gamma_{21}-\gamma_{20})\rho_{22} & \sqrt{\gamma_{10}} \rho_{01} +  \sqrt{(1-\gamma_{10}) \gamma_{21}}\rho_{12} & \sqrt{\gamma_{20}} \rho_{02} \\
  \sqrt{\gamma_{10}} \rho_{01}^* +\sqrt{(1-\gamma_{10}) \gamma_{21}} \rho_{12}^*& \gamma_{10} \rho_{11}+\gamma_{21} \rho_{22} &
    \sqrt{\gamma_{10} \gamma_{20}} \rho_{12}\\
 \sqrt{\gamma_{20}} \rho_{02}^* & \sqrt{\gamma_{10} \gamma_{20}} \rho_{12}^* & \gamma_{20} \rho_{22} \\
\end{array}
\right) \; .
\end{equation}
Notice finally that, as a consequence of Eq.~(\ref{Defgammadelta}), it follows that in the case of a beamsplitter type ADC ${\Psi}_{\eta}$  defined by Eq.~(\ref{fdsf}), one gets $\tilde{\gamma}_{j,k}[\eta]={\gamma}_{j,k}[1-\eta]$. This, thanks to Eq.~(\ref{impoIDE}), allows us to recover the well known fact that the complementary map  $\tilde{\Psi}_{\eta}$ of ${\Psi}_{\eta}$ is isometrically equivalent to ${\Psi}_{1-\eta}$.

\subsection{Composition rules}\label{sec: Composition rules}

In Ref.~\cite{MAD} it was shown that MAD channels are closed under composition rules, a property that proves useful in studying their information capacities. 
In this section we investigate whether ReMAD channels behave similarly.  Maybe surprisingly, 
it turns out that it's not always the case. To begin with, let's observe that from the composition rules of beamsplitters one can easily verify that the following identity holds true
\begin{eqnarray}\label{dfdsfa} 
\Phi_{\Gamma[\eta']} \circ \Phi_{\Gamma[\eta]} = \Phi_{\Gamma[\eta'\eta]} \;, \qquad 
\forall \eta',\eta\in[0,1]\;,
\end{eqnarray} 
(the symbol  ``$\circ$"  represents channel composition).
Apart from this special case, the analysis is slightly convoluted. We therefore focus on the simplest, nontrivial case of $d=3$ qutrit systems defined by the input-output relations of Eq.~(\ref{eq: channel action}). 
By explicit computation it follows that being $\Gamma$ and $\Gamma'$ two $3\times 3$ transition matrices as in Eq.~(\ref{transM}) we have 
\begin{align}
&[\Phi_{\Gamma'}\circ \Phi_{\Gamma}(\hat{\rho})]_{00}=\rho_{00} + (\gamma'_{10}\gamma_{10}+1-\gamma_{10}) \rho_{11}+ (\gamma'_{10}\gamma_{21}+\gamma'_{20}\gamma_{20}+1-\gamma_{21}-\gamma_{20}) \rho_{22} \; , \nonumber \\
&[\Phi_{{\Gamma'}}\circ \Phi_{\Gamma}(\hat{\rho})]_{01}=\sqrt{1-\gamma'_{10}}\sqrt{1-\gamma_{10}} \rho_{01}  + (\sqrt{\gamma'_{10} \gamma'_{21}(1-\gamma_{10}) (1-\gamma_{21}-\gamma_{20})}+ \sqrt{\gamma_{10} \gamma_{21} (1-\gamma'_{10})}) \rho_{12} \; , \nonumber \\
&[\Phi_{{\Gamma'}}\circ \Phi_{\Gamma}(\hat{\rho})]_{02}=\sqrt{1-\gamma'_{21}-\gamma'_{20}}\sqrt{1-\gamma_{21}-\gamma_{20}} \rho_{02} \; , \nonumber \\
&[\Phi_{{\Gamma'}}\circ \Phi_{\Gamma}(\hat{\rho})]_{11}=(1-\gamma'_{10})(1-\gamma_{10}) \rho_{11}+[\gamma'_{21}(1-\gamma_{21}-\gamma_{20})+(1-\gamma'_{10})\gamma_{21} ] \rho_{22} \; , \nonumber \\
&[\Phi_{{\Gamma'}}\circ \Phi_{\Gamma}(\hat{\rho})]_{12}=\sqrt{(1-\gamma'_{10}) (1-\gamma'_{21}-\gamma'_{20})}\sqrt{(1-\gamma_{10}) (1-\gamma_{21}-\gamma_{20})} \rho_{12}  \; , \nonumber \\
&[\Phi_{{\Gamma'}}\circ \Phi_{\Gamma}(\hat{\rho})]_{22}=(1-\gamma'_{21}-\gamma'_{20})(1-\gamma_{21}-\gamma_{20}) \rho_{22} \; ,
\end{align}
plus the Hermitian conjugate of off-diagonal elements. 
 We are interested in understanding under which conditions a new transition matrix ${\Gamma''}$ of elements $\gamma''_{j,k}$ exists such that 
 \begin{eqnarray} \label{dfdsfxxx} 
\Phi_{\Gamma''}=\Phi_{{\Gamma'}}\circ \Phi_{\Gamma}\;.
\end{eqnarray}
Among the equations above, those referring to elements 00, 02, 11, 12, 22 are all consistent with setting 
\begin{align}\label{eq: delta params}
&\gamma''_{10}=\gamma_{10}+\gamma'_{10}(1-\gamma_{10}) \; , \nonumber \\
&\gamma''_{20}=\gamma_{20}+\gamma'_{10}\gamma_{21} + \gamma'_{20}(1-\gamma_{21}-\gamma_{20})  \; , \nonumber \\
&\gamma''_{21}=(1-\gamma'_{10})\gamma_{21}+\gamma'_{21}(1-\gamma_{21}-\gamma_{20}) \; . 
\end{align}
(observe that such definitions are compatible with the request that $\gamma''_{j,k}\in [0,1]$, as well as with the normalization constraint $\sum_{k=0}^{j}\gamma''_{j,k}= 1$ for all $j$). 
Element 01 instead forces an additional constraint, i.e. 
\begin{equation}\label{eq: parameters constraint}
\sqrt{\gamma''_{21}\gamma''_{10}}=\sqrt{\gamma'_{10} \gamma'_{21}(1-\gamma_{10}) (1-\gamma_{21}-\gamma_{20})}+ \sqrt{\gamma_{10} \gamma_{21} (1-\gamma'_{10})} \; ,
\end{equation}
which is not necessarily granted. 
As a matter of fact by substituting in Eq.~(\ref{eq: parameters constraint})  the values for $\gamma''_{21}$ and $\gamma''_{10}$ obtained in Eq.~(\ref{eq: delta params}) we get that in order to satisfy it we need
\begin{equation}\label{eq: comp constraint}
\gamma_{10}\gamma'_{21}(1-\gamma_{21}-\gamma_{20})=\gamma_{21}\gamma'_{10}(1-\gamma_{10})(1-\gamma'_{10}) \; .
\end{equation}
This identifies a specific region for the parameters $\Gamma$ and ${\Gamma'}$ where the composition $\Phi_{\Gamma'}\circ \Phi_{\Gamma}$ corresponds to a ReMAD channel. Notice in particular that, in agreement with Eq.~(\ref{dfdsfa}), the identity in Eq.~(\ref{eq: comp constraint}) is always fulfilled for beamsplitter type ADC, i.e. for 
$\Gamma=\Gamma[\eta]$ and $\Gamma'=\Gamma[\eta']$. Furthermore we observe that 
if $\Gamma'$ is such that either $\gamma_{21}'=\gamma'_{10} =0$ or $\gamma_{21}'=1-\gamma'_{10} =0$, then
Eq.~(\ref{eq: comp constraint}) is always satisfied for all the choices of $\gamma'_{20}$. 
Of particular interest for our analysis is the first of these two configurations: here $\Phi_{{\Gamma'}}$ describes a ReMAD channel where the first two levels of S are untouched by the noise, while the third level gets mapped into the ground state with probability $\gamma'_{20}$. 
Looking at Eq.~(\ref{eq: delta params}) we notice that the transition probabilities  $\gamma''_{10}$ and $\gamma''_{21}$  of the ReMAD channel $\Phi_{\Gamma''}$, which emerge by the concatenation, coincide with those of $\Phi_{\Gamma}$ (i.e. $\gamma''_{10}=\gamma_{10}$ and $\gamma''_{21}=\gamma_{21}$), while the transition probability $\gamma''_{20}$ increases with respect to $\gamma_{20}$. Specifically $\gamma''_{20} = \gamma_{20}+ \gamma'_{20}(1-\gamma_{21}-\gamma_{20})$. Observing this, by varying $\gamma_{20}''\in [0,1]$, the last quantity can span over the entire interval $[1-\gamma_{21},1]$. 
 We can use this observation to claim that being $\Phi_{\Gamma}$ and $\Phi_{\Gamma''}$ two qutrits ReMAD channels having the same transition probabilities connecting level 2 to level 1 and level 1 to level 0, but with $\gamma''_{20}$ larger than or equal to $\gamma_{20}$, then they can be connected via a third ReMAD channel as in Eq.~(\ref{dfdsfxxx}).

\subsection{Covariance}\label{sec: Appendix averaged covariance}

The ReMAD channels exhibit a covariance property under suitable unitary transformations. Specifically consider
the set of unitary gates 
\begin{eqnarray} 
\hat{U}_S(\theta) = \sum_{j=0}^{d-1} e^{-i  j \theta}  \ket{j}\!\!\bra{j}_{\text{S}} \;, \\
\hat{U}_E(\theta) = \sum_{j=0}^{d-1} e^{-i  j \theta}  \ket{j}\!\!\bra{j}_{\text{E}} \;, 
\end{eqnarray} 
with $\theta$  real. It then follows that 
 \begin{equation} \label{COVAR} 
 {\Phi}_{\Gamma} (\hat{U}_{\text{S}}(\theta)  \hat{\rho} \hat{U}_{\text{S}}^{\dagger} (\theta))=
\sum_{k=0}^{d-1}  \sum_{j,j'=k}^{d-1}\;  \rho_{j,j'} \; e^{-i (j-j') \theta} \sqrt{\gamma_{j,k} \gamma_{j',k} } 
\;  \ket{j-k}\!\!\bra{j'-k}_{\text{S}}= 
 \hat{U}_{\text{S}}(\theta) {\Phi}_{\Gamma}( \hat{\rho}) \hat{U}_{\text{S}}^{\dagger} (\theta)\;,
 \end{equation} 
 and similarly for the complementary channel 
$\tilde{\Phi}$ we get 
\begin{equation}\label{eq:compl covariance} 
 \tilde{\Phi}_{\Gamma}(\hat{U}_{\text{S}}(\theta)  \hat{\rho} \hat{U}_{\text{S}}^{\dagger} (\theta))=
\sum_{k=0}^{d-1}  \sum_{j,j'=k}^{d-1}\;  \rho_{j,j'} \; e^{-i (j-j') \theta} \sqrt{\tilde{\gamma}_{j,k} \tilde{\gamma}_{j',k} } 
\;  \ket{j-k}\!\!\bra{j'-k}_{\text{E}}= 
 \hat{U}_{\text{E}}(\theta) \tilde{\Phi}_{\Gamma}( \hat{\rho}) \hat{U}_{\text{E}}^{\dagger} (\theta)\;.
%
\end{equation}

As we'll see in the following the identities above turn out to be extremely useful in the computation of the capacities of our channels.

\section{Review of quantum and private classical capacities}\label{sec: Deg, Qcap&Prcap}

This section is dedicated to reviewing some basic notions about channel capacities for those readers who may be not familiar with these quantities. While the plethora of capacities for quantum channels includes a large collection of different functionals, our analysis will be focused on the quantum capacity $Q$ and the private classical capacity $C_{\text{p}}$.

\subsection{Quantum Capacity and Private Classical Capacity}

The quantum capacity $Q(\Phi)$ of a quantum channel $\Phi$  defines the maximum rate of transmitted quantum information achievable per channel use, assuming $\Phi$ to act in the regime of i.i.d. noise \cite{HOLEVOBOOK,WILDEBOOK,WATROUSBOOK,HAYASHIBOOK,HOLEGIOV,NC,ADV_Q_COMM,SURVEY}. Intuitively it tells you how faithfully a quantum state, possibly correlated with an external system, can be sent and received by two communication parties if the communication line is noisy. Recently $Q$ has also been showed to provide a lower bound to the space overhead necessary for fault tolerant quantum computation in presence of noise \cite{Q_CAP_OVER}.
The formal definition of $Q$ relies on the coherent information $I_{\text{coh}}$ \cite{COH_INFO}.  Assuming $n$ uses of the channel $\Phi$ and a generic state $\hat{\rho}^{(n)}\in \mathfrak{S}({\cal H}_{\text{S}}^{\otimes n})$ we have 
\begin{equation}\label{eq:coherent info}
I_{\text{coh}} \left(\Phi^{\otimes n},\hat{\rho}^{(n)} \right):= S \left( \Phi^{\otimes n}(\hat{\rho}^{(n)}) \right)-S \left( \tilde{\Phi}^{\otimes n}(\hat{\rho}^{(n)}) \right).
\end{equation}
with $S(\hat{\rho}):= -\mbox{Tr}[ \hat{\rho} \log_2 \hat{\rho}]$ the von Neumann entropy of the state $\hat{\rho}$, and $\tilde{\Phi}$ the complementary channel of $\Phi$.
Maximizing over $\mathfrak{S}({\cal H}_{\text{S}}^{\otimes n})$ we get the maximized coherent information $Q^{(n)}$ for $n$ instances of the channel
\begin{equation}
Q^{(n)}(\Phi):= \max_{\hat{\rho}^{(n)}\in {\mathfrak{S}({\cal H}_{\text{S}}^{\otimes n})}}
I_{\text{coh}} \left( \Phi^{\otimes n},\hat{\rho}^{(n)} \right) \; ,
\end{equation}
and by regularization of this expression we get the maximized coherent information per channel use, that is the quantum capacity $Q(\Phi)$ \cite{QCAP1,QCAP2,QCAP3}
\begin{equation}\label{eq:QCapacity}
Q(\Phi)=\lim_{n\rightarrow \infty} \frac{Q^{(n)}(\Phi)}{n} \; . 
\end{equation}

 The private classical capacity $C_{\text{p}}(\Phi)$ quantifies 
 the maximum rate of classical information achievable per channel use assuming also the privacy of communication (where privacy is intended as limiting to arbitrarily small the amount of information that an eavesdropper can extract from the environment during the communication). The fundamental tool needed  to 
 compute $C_{\text{p}}(\Phi)$ is the Holevo information functional $\chi$ \cite{HOL_INFO}. Being ${\cal E}_n := \{ p_i , \hat{\rho}^{(n)}_i\}$ an ensemble of quantum states $\hat{\rho}^{(n)}_i \in \mathfrak{S}({\cal H}_{\text{S}}^{\otimes n})$, we have
\begin{equation}
 \chi(\Phi^{\otimes n}, {\cal E}_n) := S\left( \Phi^{\otimes n} \left(\sum_i p_i \hat{\rho}_i^{(n)} \right) \right)-\sum_i p_i S\left( \Phi^{\otimes n} \left( \hat{\rho}_i^{(n)} \right) \right) \; .
\end{equation} 
Through the Holevo information we define next the private information for $n$ uses $C_{\text{p}}^{(n)}(\Phi)$, that involves a maximization over all ensembles ${\cal E}_n$
\begin{equation}
C_{\text{p}}^{(n)}(\Phi) :=   \max_{{\cal E}_n} \left( \chi \left( \Phi^{\otimes n}, {\cal E}_n \right) - \chi \left( \tilde{\Phi}^{\otimes n}, {\cal E}_n \right) \right) \; ,
\end{equation}
from which finally $C_{\text{p}}(\Phi)$ can be computed via 
regularization over $n$, i.e. 
\cite{QCAP3, PRIV2}:
\begin{equation}\label{eq:private class}
C_{\text{p}}(\Phi)=\lim_{n\to \infty}  \frac{C_{\text{p}}^{(n)}(\Phi)}{n} \; . 
\end{equation}
We conclude our brief review by recalling that $Q$ and $C_{\text{p}}$ 
obey data processing inequalities~\cite{WILDEBOOK}. This means that, for ({\it any}) channel $\Phi$ that can be expressed
as a composition of other two LCPTP maps $\Phi_1$ and $\Phi_2$ (i.e. 
$\Phi= \Phi_1\circ \Phi_2$) it follows
\begin{eqnarray}
&Q(\Phi) \leq \min\{ Q(\Phi_1), Q(\Phi_2)\}\label{data} \;, \\
&C_{\text{p}}(\Phi) \leq \min\{ C_{\text{p}}(\Phi_1), C_{\text{p}}(\Phi_2)\} \; .
\end{eqnarray}

\subsection{Degradability and antidegradability} \label{sec: Appendix degradable}

The need of regularization in the evaluation of  $Q(\Phi)$ and $C_{\text{p}}(\Phi)$ poses a  well known problem which ultimately is the underlying reason of our efforts here. An exception to this predicament is given by degradable \cite{DEGRADABLE} and antidegradable \cite{ANTIDEGRADABLE} channels, whose definitions and properties are reviewed here.
A quantum channel $\Phi:\mathcal{L}(\mathcal{H}_{\text{S}})\rightarrow \mathcal{L}(\mathcal{H}_{\text{S}'})$ is said degradable if a LCPTP map $\mathcal{D}:\mathcal{L}(\mathcal{H}_{\text{S}'})\rightarrow \mathcal{L}(\mathcal{H}_{\text{E}})$ exists s.t.
\begin{equation}\label{eq:degradable}
\tilde{\Phi}=\mathcal{D}\circ \Phi \; ,
\end{equation}
while it's said antidegradable if it exists a LCPTP map $\mathcal{A}:\mathcal{L}(\mathcal{H}_{\text{E}})\rightarrow \mathcal{L}(\mathcal{H}_{\text{S}'})$ s.t.
\begin{equation}\label{eq:antidegradable}
\Phi=\mathcal{A}\circ\tilde{\Phi} \; .
\end{equation}
In case $\Phi$ is mathematically invertible, a simple direct way to determine whether it is degradable or not is to formally 
invert the composition in Eq.~(\ref{eq:degradable}). This is done (if possible) by constructing the super-operator $\mathcal{D}=\tilde{\Phi}\circ \Phi^{-1}$ and by checking whether such object is LCPTP \cite{INVERSE1,INVERSE}. The check can be performed by studying the positivity of its associated Choi matrix $\text{C}_{\mathcal{D}}$, i.e.
\begin{equation} \label{NECANDSUFF} 
\mbox{$\Phi$ invertible} \Longrightarrow \mbox{$\Phi$ degradable iff $\text{C}_{\mathcal{D}}\geq 0$} \;,
\end{equation} 
where given $\ket{\Gamma}_{\text{RS}}=\sum_{i=0}^{d-1}\ket{i}_\text{R}\ket{i}_\text{S}$ we set 
$\text{C}_{\mathcal{D}}:=(I_\text{R}\otimes \mathcal{D}_{\text{S}})\kb{\Gamma}{\Gamma}_{\text{RS}}$ (see App.~\ref{sec:channelinversion} for details).
 A completely similar argument can be made for antidegradability: in this case we can claim
\begin{equation} \label{NECANDSUFFanti} 
\mbox{$\tilde{\Phi}$ invertible} \Longrightarrow \mbox{$\Phi$ antidegradable iff $\text{C}_{\mathcal{A}}\geq 0$} \;,
\end{equation} 
where now $\text{C}_{\mathcal{A}}$ is the Choi operator of the map
 $\mathcal{A}={\Phi}\circ \tilde{\Phi}^{-1}$.

As already mentioned the evaluation of the quantum a private classical capacity simplifies for degradable channels.
To begin with, in this case  $Q$ and $C_{\text{p}}$ result to be additive, so the regularization over $n$ in Eq.~(\ref{eq:QCapacity})  isn't needed, leading to the single-letter formula~\cite{PRIV4}
\begin{equation}\label{eq:singLett}
C_{\text{p}}(\Phi)=Q(\Phi)= Q^{(1)}(\Phi) :=\max_{\hat{\rho} \in {\mathfrak{S}({\cal H}_{\text{S}})}}
I_{\text{coh}} \left( \Phi,\hat{\rho} \right) \;.
\end{equation}
Another important simplification arises from the fact that under degradability conditions 
the coherent information functional is concave w.r.t. the input state \cite{CONCAVITY}.
Accordingly if $\Phi$ and $\tilde{\Phi}$ turn out to be covariant under the action of some unitary group, the maximization in Eq.~(\ref{eq:singLett}) can be further simplified. In the case of ReMAD channels where the properties (\ref{COVAR}) and (\ref{eq:compl covariance}) hold true, under degradability conditions we can write 
\begin{eqnarray}\label{eq: coh info concave}
I_{\text{coh}}\left(  \Phi_{\Gamma} , \int \frac{d \theta }{2\pi} \hat{U}_S(\theta)  \hat{\rho}\hat{U}^\dag_S(\theta)    \right)&\geq& \int \frac{d \theta }{2\pi}  
I_{\text{coh}}\left(  \Phi_{\Gamma} ,\hat{U}_S(\theta)  \hat{\rho}\hat{U}^\dag_S(\theta)   \right) \nonumber\\ 
&=&\int \frac{d \theta }{2\pi}  
S(  \Phi_{\Gamma} (\hat{U}_S(\theta)  \hat{\rho}\hat{U}^\dag_S(\theta)  )-
S(  \tilde{\Phi}_{\Gamma} (\hat{U}_S(\theta)  \hat{\rho}\hat{U}^\dag_S(\theta)  ) \nonumber \\
&=& I_{\text{coh}}\left(  \Phi_{\Gamma} , \hat{\rho}  \right)\;,
\end{eqnarray}
where in the last equality we made use of the invariance of von Neumann entropy under unitary transformations.
Observing then that 
\begin{eqnarray}
\int \frac{d \theta }{2\pi} \hat{U}_S(\theta)  \hat{\rho}\hat{U}^\dag_S(\theta)  = 
\hat{\rho}^{(\text{diag})}  :=\sum_{j=0}^{d-1} \rho_{j,j} \;  \ket{j}_{\text{S}}\!\bra{j}\;, 
\end{eqnarray} 
we can conclude that for degradable ReMAD channels the maximization over of (\ref{eq:singLett}) can be restricted on the set of states
$\hat{\rho}^{(\text{diag})}$ which are diagonal in the canonical basis, i.e. 
\begin{equation}\label{eq:singLettsimp}
C_{\text{p}}(\Phi_{\Gamma})=Q(\Phi_{\Gamma})= Q^{(1)}(\Phi_{\Gamma}) =\max_{\hat{\rho}^{(\text{diag})}  \in {\mathfrak{S}({\cal H}_{\text{S}})}}
I_{\text{coh}} \left( \Phi_{\Gamma},\hat{\rho}^{(\text{diag})}  \right) \;.
\end{equation}
For antidegradable channels instead, due to a no-cloning argument \cite{ERAS_NO_CLO}, $Q(\Phi)=0$. Similarly, $C_{\text{p}}(\Phi)=0$: the environment can reconstruct the channel output simply by applying the antidegrading channel, so no private information can be transmitted. Therefore for channels exhibiting antidegradability no maximizations are needed.

\section{Degradability and antidegradability regions for $d=3$ ReMAD channels}\label{sec: Degradable chan}
In this section we study the degradability/antidegradability properties for qutrits  ReMAD channels. 
Mimicking the approach used for other amplitude damping channels, we tackle the problem working under the  
heuristic assumption that if a degrading (or antidegrading) channel of a ReMAD channel exists, it is itself a ReMAD channel. While this choice is potentially suboptimal, numerical tests based on the more rigorous (but analytically impractical)
matrix inversion method discussed in Sec.~\ref{sec: Appendix degradable}, reveals that this is not the case.

\begin{figure}[h]
     \centering
     \begin{subfigure}[b]{0.4\textwidth}
         \centering
         \includegraphics[width=\textwidth]{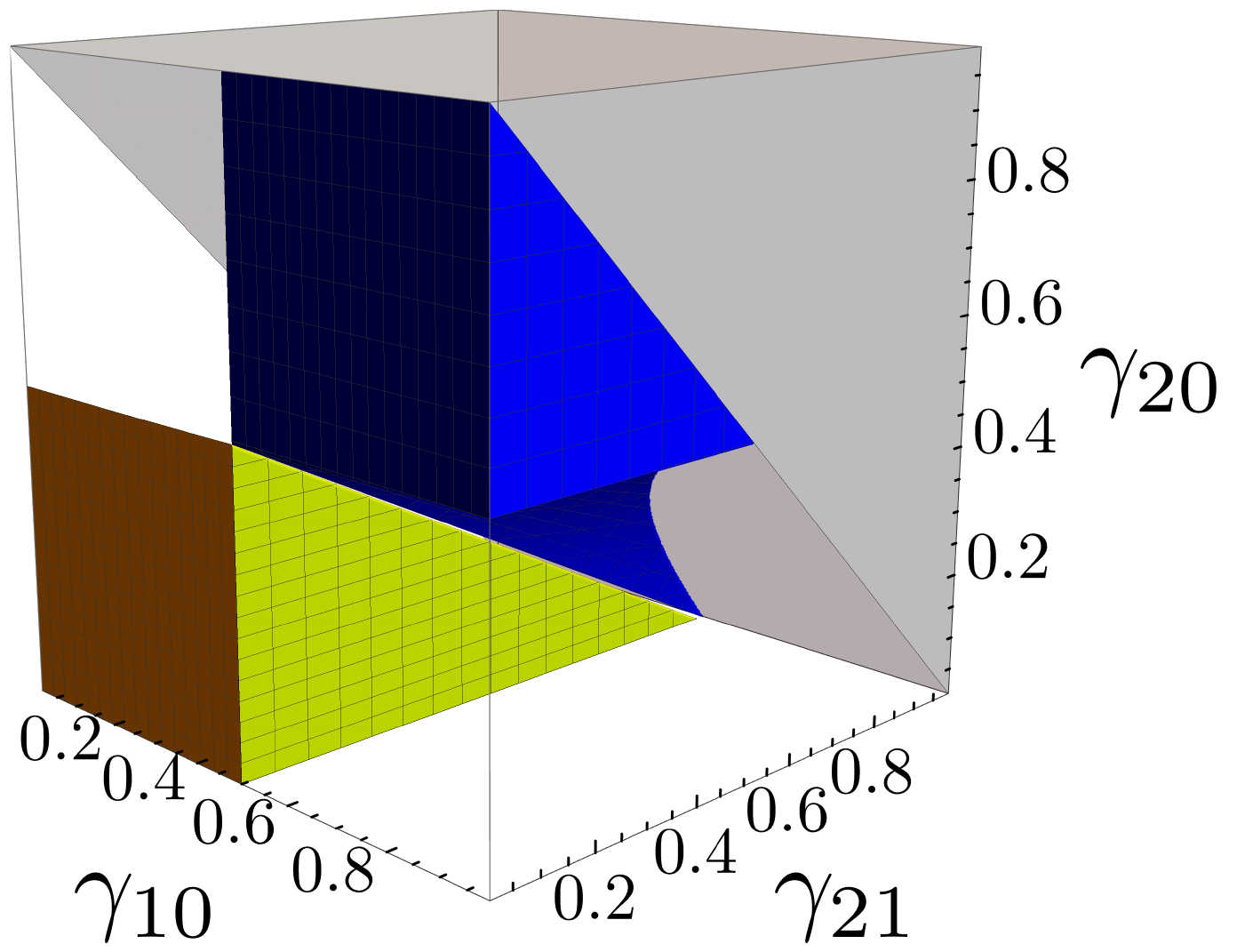}         
     \end{subfigure}
     \begin{subfigure}[b]{0.4\textwidth}
         \centering
         \includegraphics[width=\textwidth]{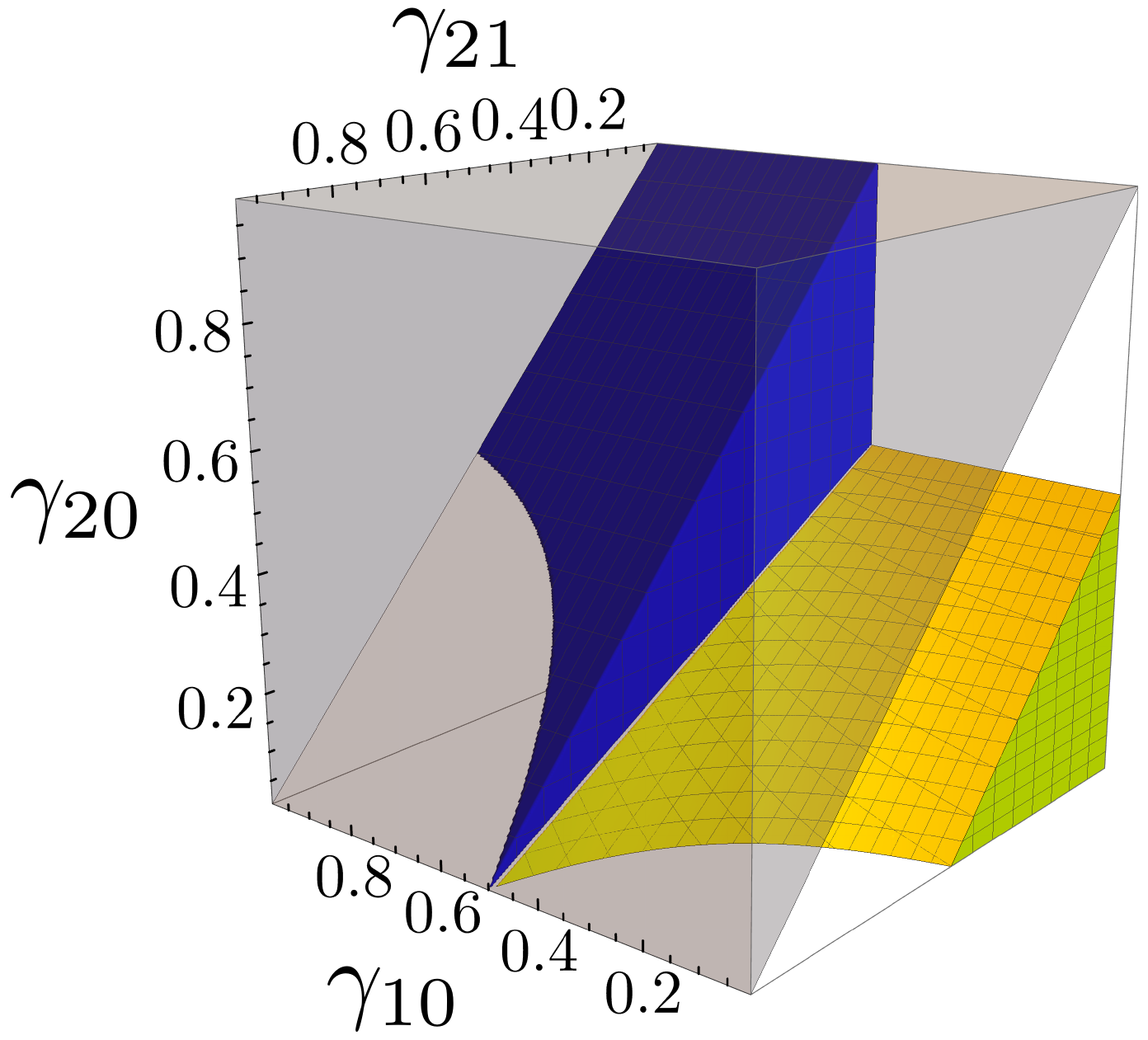}         
     \end{subfigure}
        \caption{Degradability region (yellow) and antidegradability region (blue) for a qutrit ReMAD channel, the grey volume representing the non-physical region of the parameter space (the two panels show different perspectives). Notice that the beamsplitter type ADC $\Psi_\eta$ (blue line of Fig.~\ref{fig: BS_ADC}) is fully contained in the above regions (it lies in the degradability region for $\eta \geq 1/2$, $\Phi_{\eta}$ and lies in the antidegradability region for $\eta <1/2$).} 
        \label{fig: degr and antidegr}
\end{figure}

\begin{figure}[h]
     \centering
     \vspace*{-1.cm}
     \begin{subfigure}[t]{0.48\textwidth}
         \centering
         \includegraphics[width=\textwidth]{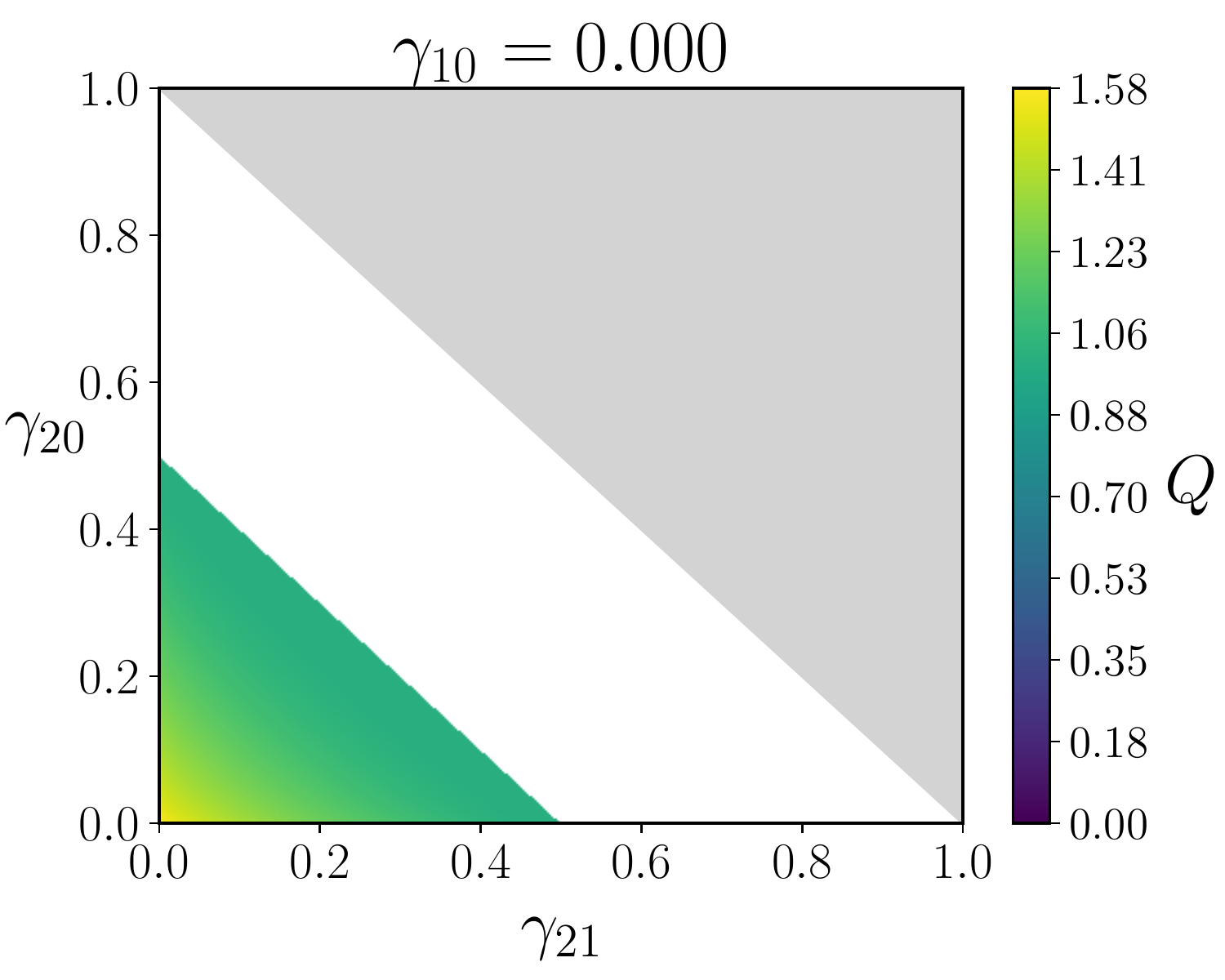}         
     \end{subfigure}     
     \begin{subfigure}[t]{0.48\textwidth}
         \centering
         \includegraphics[width=\textwidth]{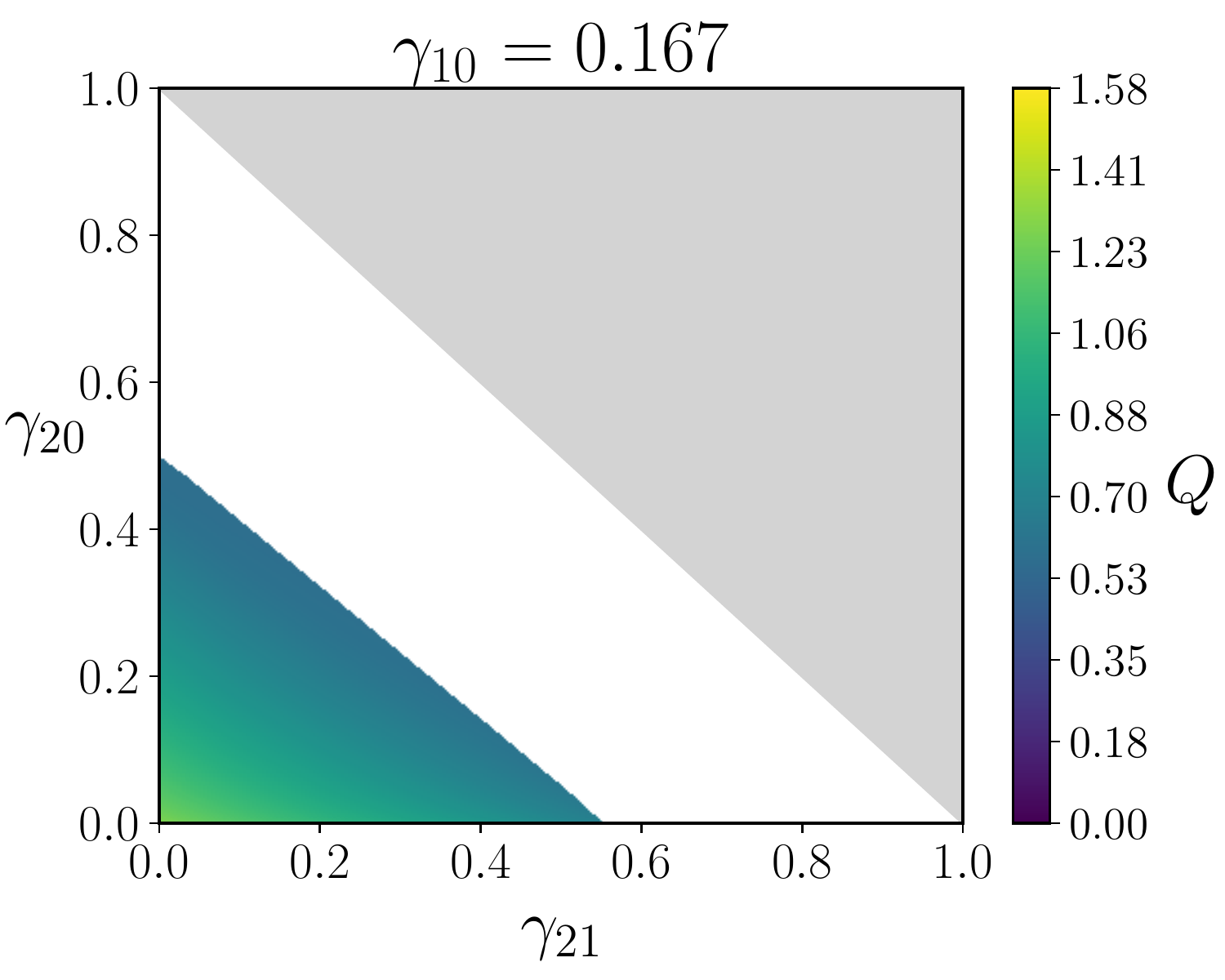}
     \end{subfigure}
     \begin{subfigure}[t]{0.48\textwidth}
         \centering
         \includegraphics[width=\textwidth]{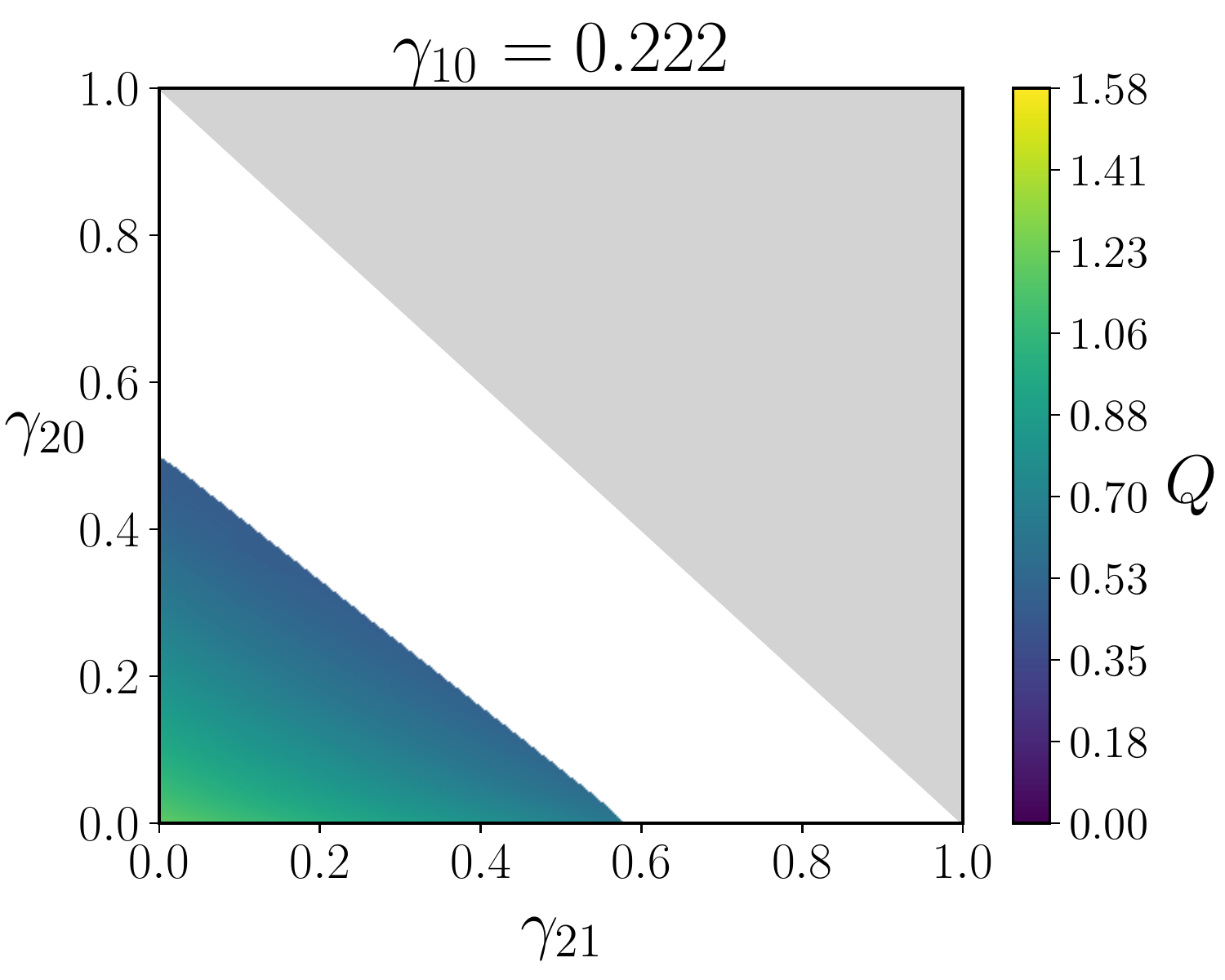}         
     \end{subfigure}    
     \begin{subfigure}[t]{0.48\textwidth}
         \centering
         \includegraphics[width=\textwidth]{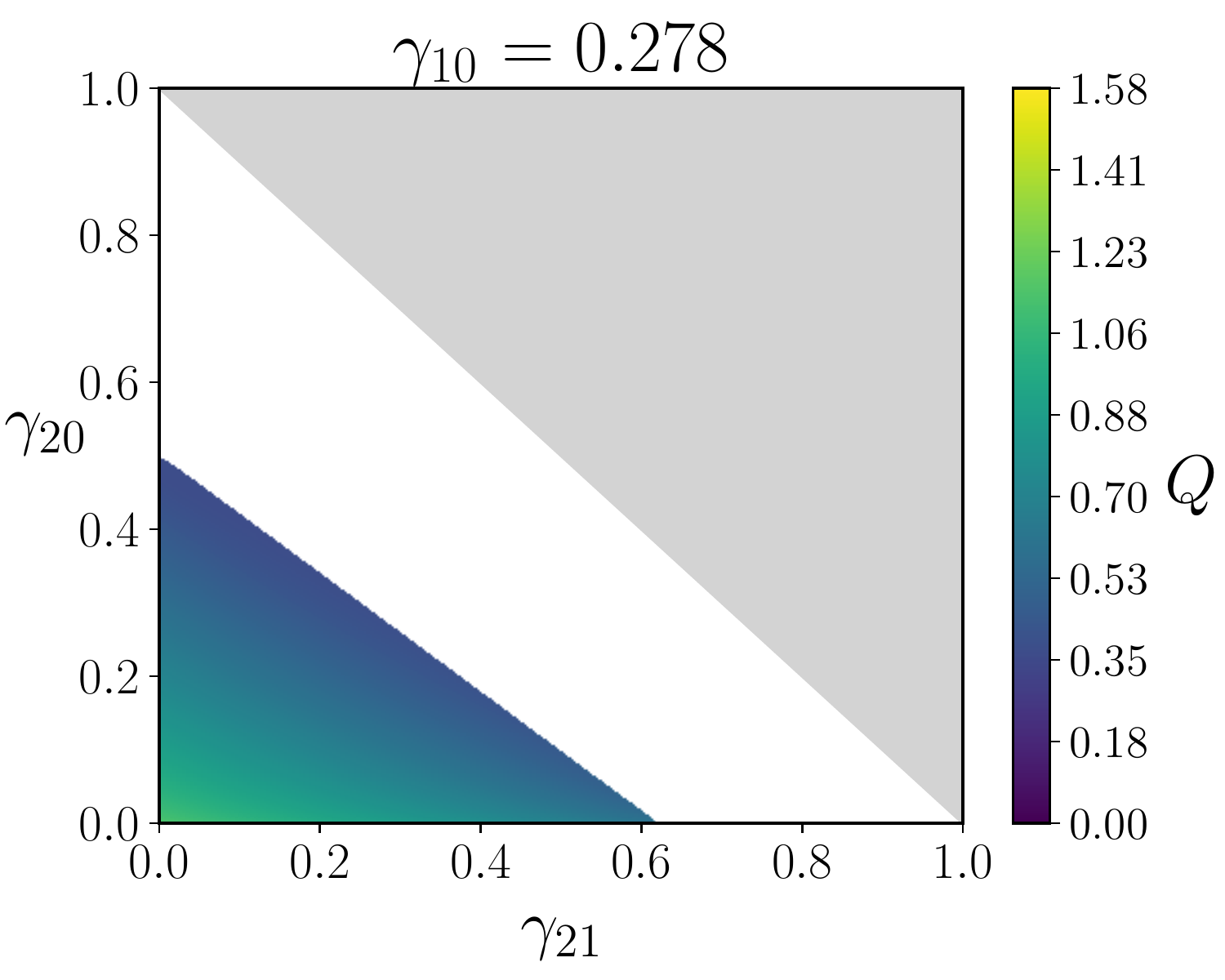}         
     \end{subfigure}     
     \begin{subfigure}[t]{0.48\textwidth}
         \centering
         \includegraphics[width=\textwidth]{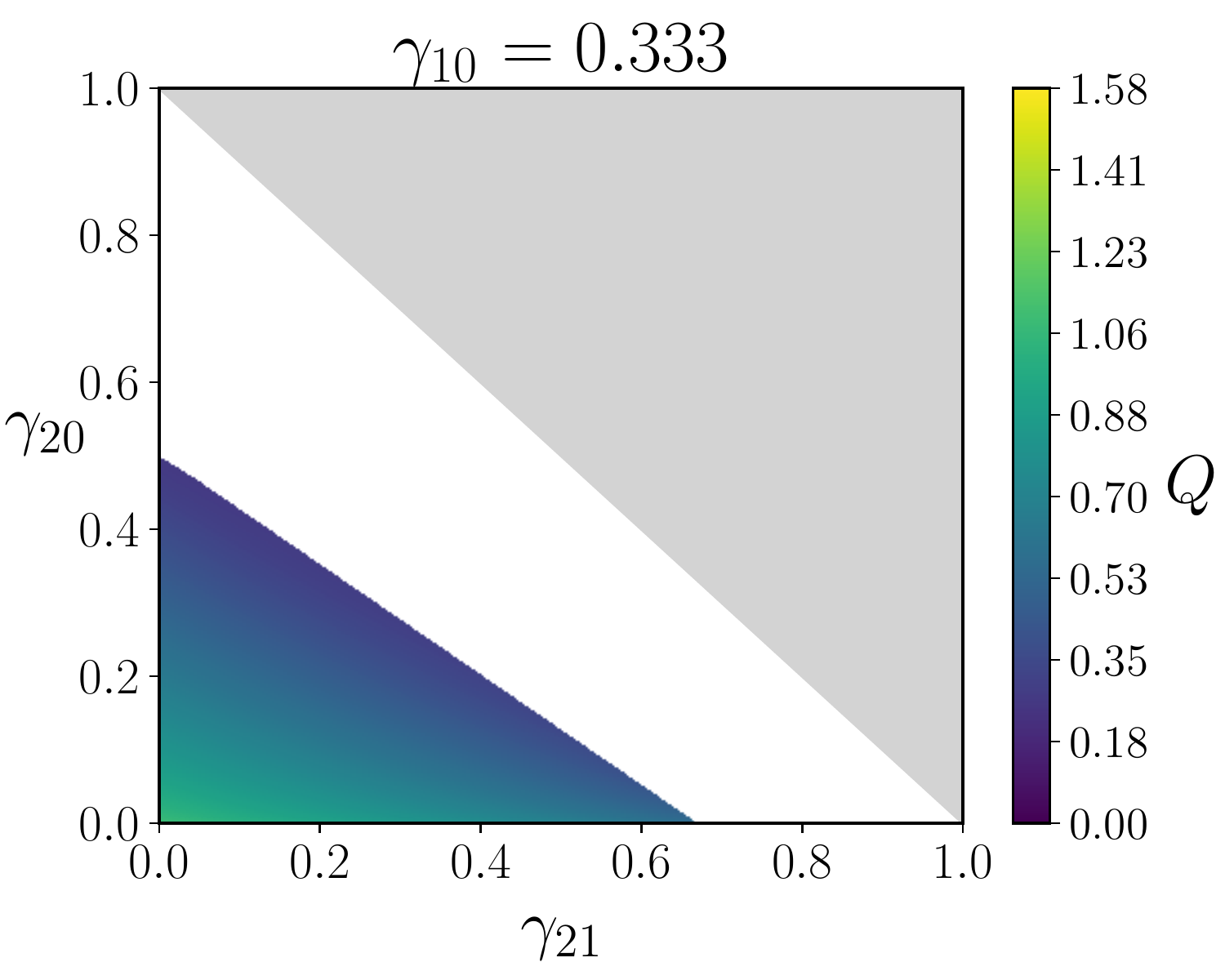}
     \end{subfigure}
     \begin{subfigure}[t]{0.48\textwidth}
         \centering
         \includegraphics[width=\textwidth]{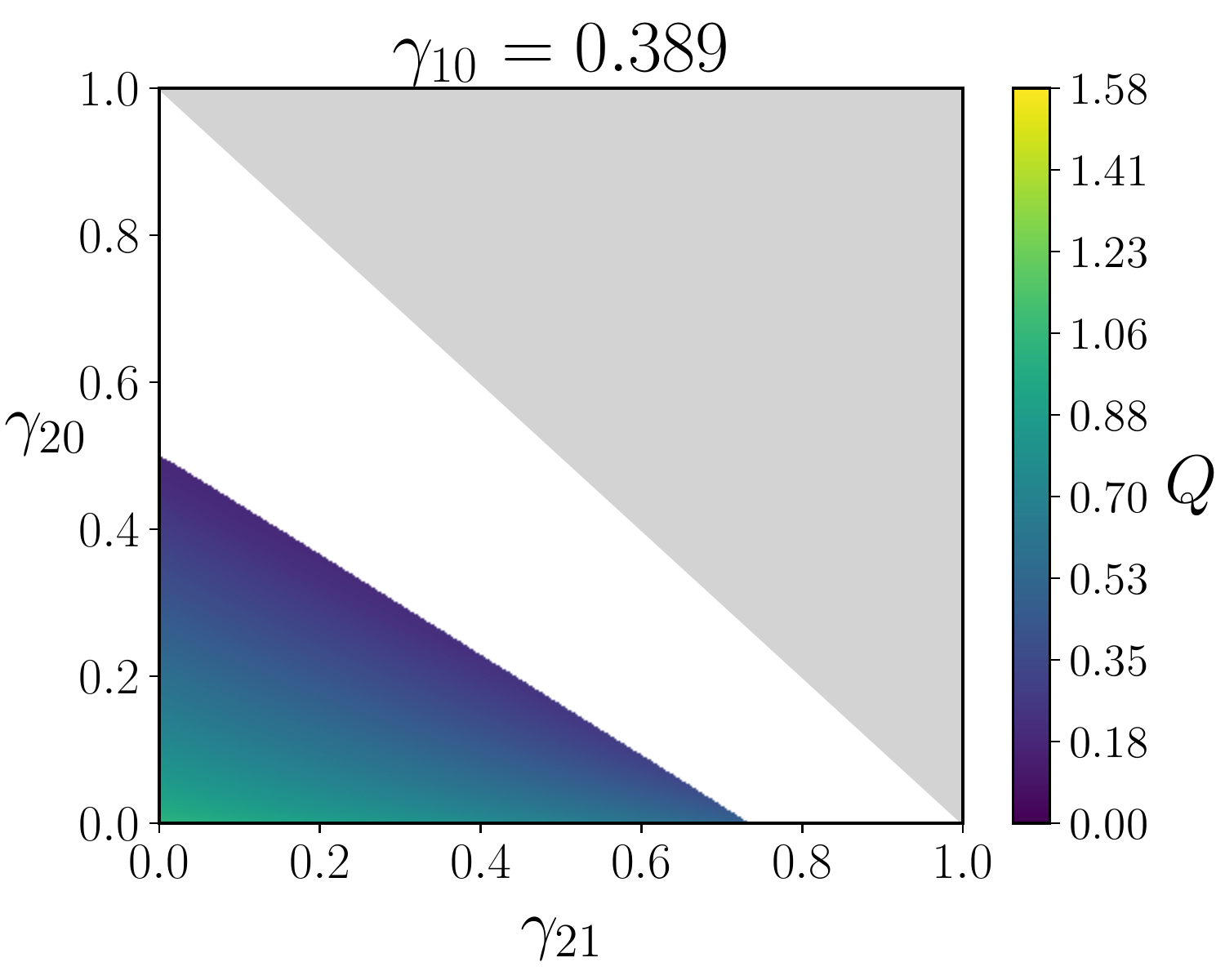}
     \end{subfigure}
     \begin{subfigure}[t]{0.48\textwidth}
         \centering
         \includegraphics[width=\textwidth]{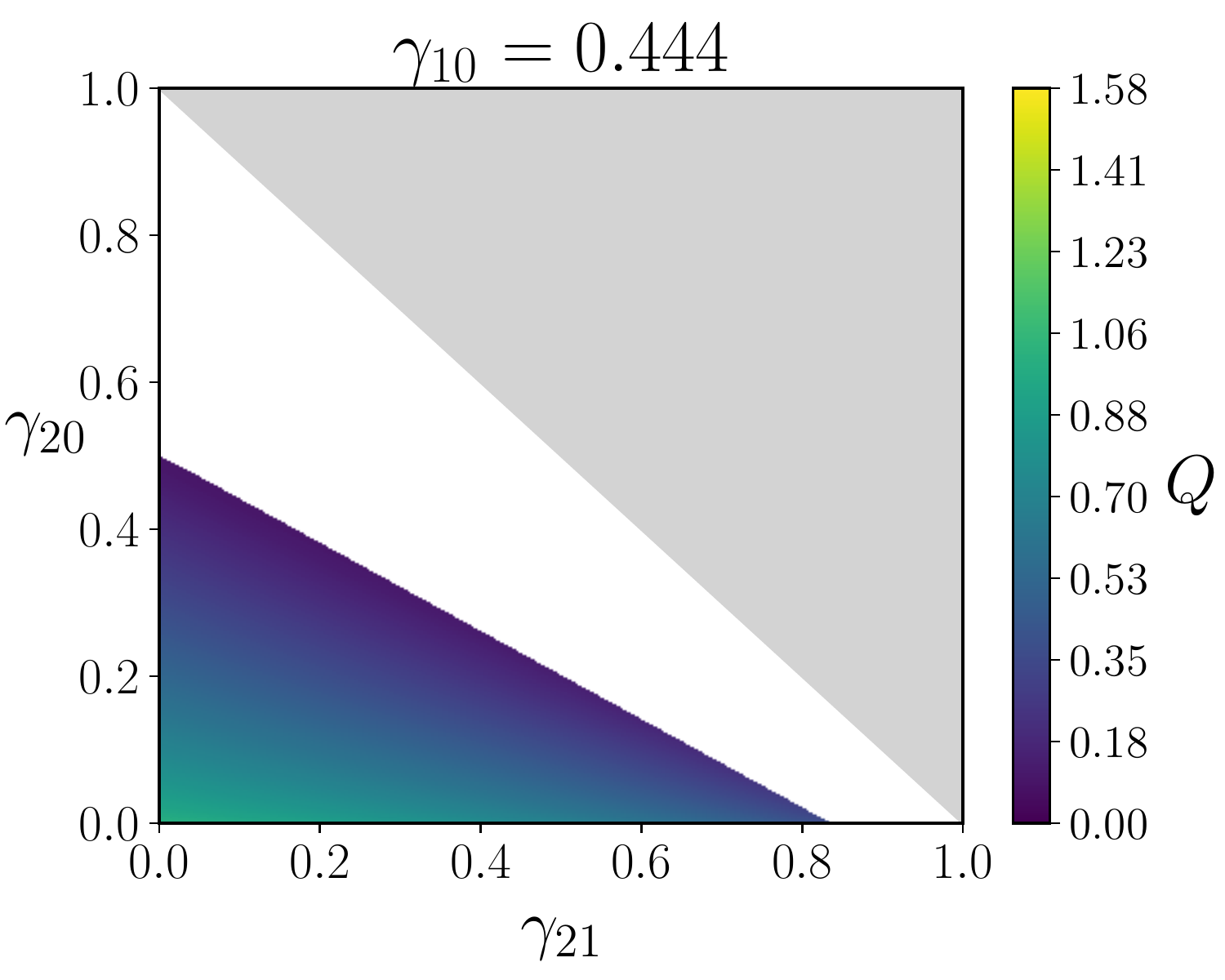}
     \end{subfigure}
     \begin{subfigure}[t]{0.48\textwidth}
         \centering
         \includegraphics[width=\textwidth]{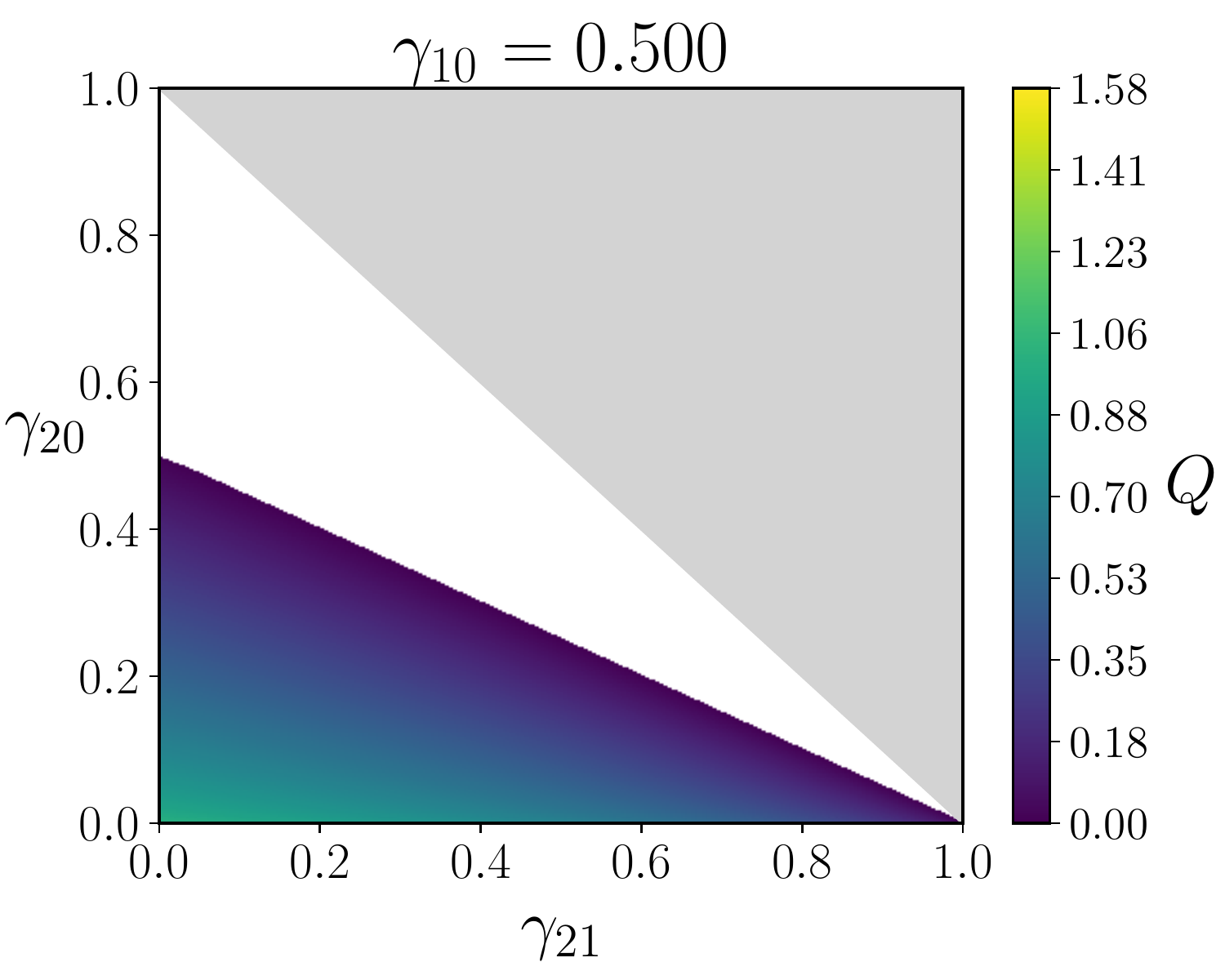}
     \end{subfigure}
        \caption{$Q$ (and $C_{\text{p}}$) for a qutrit ReMAD $\Phi_{\Gamma}$ at different $\gamma_{10},$ $\gamma_{21}$ and $\gamma_{20}$ in the degradable region. The white regions correspond to points associated with not degradable (nor antidegradable) channels,  grey regions are those outside the allowed parameter space ${\mathbb D}_3$. As we'll see in Sec. \ref{sec: cap non degradable}, at least for $\gamma_{10}=0.000$ (left-top panel) $Q$ and $C_{\text{p}}$ can also be extended in the white region.}
        \label{fig: Q capacity}
\end{figure}

Reminding the isometric connection given in Eq.~(\ref{impoIDE}), we hence translate the degradability condition
on $\Phi_{\Gamma}$ into the problem of 
identifying a  transition matrix ${\Gamma'}$ s.t. 
\begin{eqnarray}
{\Phi}_{\tilde{\Gamma}}=\Phi_{{\Gamma'}}\circ\Phi_{\Gamma}\;.
\end{eqnarray}  
With a procedure similar to the one employed in Sec.~\ref{sec: Composition rules} such equation can be mapped into the following constraints
\begin{align}\label{eq: delta params degradability}
&\tilde{\gamma}_{10}=\gamma_{10}+\gamma'_{10}(1-\gamma_{10}) \; , \nonumber \\
&\tilde{\gamma}_{20}=\gamma_{20}+\gamma'_{10}\gamma_{21} + \gamma'_{20}(1-\gamma_{21}-\gamma_{20})  \; , \nonumber \\
&\tilde{\gamma}_{21}=(1-\gamma'_{10})\gamma_{21}+\gamma'_{21}(1-\gamma_{21}-\gamma_{20}) \; ,
\end{align}

which using the connection as in Eq.~(\ref{Defgammadelta}) to replace $\tilde{\gamma}_{10}=\gamma_{11}= 1-\gamma_{10}$,
$\tilde{\gamma}_{20}=\gamma_{22}= 1-\gamma_{20}-\gamma_{21}$, and 
$\tilde{\gamma}_{21}=\gamma_{21}$, leads to 
\begin{align}
&\gamma'_{10} =\frac{1-2\gamma_{10}}{1-\gamma_{10}} \; , \nonumber \\
&\gamma'_{21}=\frac{1-2\gamma_{10}}{1-\gamma_{10}}\frac{\gamma_{21}}{1-\gamma_{21}-\gamma_{20}}  \; , \nonumber \\
&\gamma'_{20}=\frac{1-\gamma_{21}-2\gamma_{20}}{1-\gamma_{21}-\gamma_{20}}-\frac{\gamma_{21}}{1-\gamma_{21}-\gamma_{20}}\frac{1-2\gamma_{10}}{1-\gamma_{10}} \; . 
\end{align}
Imposing now the vector $(\gamma'_{10},\gamma'_{21},\gamma'_{20})$
 to belong to the domain ${\mathbb D}_3$ of Eq.~(\ref{DOMAIN3}) we get the degradability region which we depict as the yellow region of Fig.~\ref{fig: degr and antidegr}.
  In such region we can compute exactly the value of $Q(\Phi_{\Gamma})$ and $C_{\text{p}}(\Phi_{\Gamma})$
  using the single-letter formula in Eq.~(\ref{eq:singLettsimp}) only involving diagonal input matrices. 
Results of such optimization are reported in Fig.~\ref{fig: Q capacity} for different values of the parameters $\gamma_{10},$ $\gamma_{21}$ and $\gamma_{20}$. In the cases of $\gamma_{21}=0$ and $\gamma_{10}=0$ the obtained values were already known in the whole resulting parameter space (including some non-degradable regions), since there ReMAD channels reduce to double decay MAD channels, already studied in \citep{MAD}. 

To check antidegradability of a ReMAD channel we follow the same path, looking for a ${\Gamma'}$ that fulfills the identity 
\begin{eqnarray}
{\Phi}_{{\Gamma}}=\Phi_{{\Gamma'}}\circ\Phi_{\tilde{\Gamma}}\;,
\end{eqnarray}  
leading to a constraint that can be obtained from the one in Eq.~(\ref{eq: delta params degradability}) by exchanging
$\gamma_{j,k}$ with $\tilde{\gamma}_{j,k}$. This leads finally to  

\begin{align}
&\gamma'_{10}=\frac{2\gamma_{10}-1}{\gamma_{10}} \; , \nonumber \\
&\gamma'_{21}=\frac{\gamma_{21}}{\gamma_{20}}\frac{2\gamma_{10}-1}{\gamma_{10}}   \; , \nonumber \\
&\gamma'_{21}=2+\frac{\gamma_{21}(1-\gamma_{10})-\gamma_{10}}{\gamma_{10}\gamma_{20}} \; .
\end{align}
Imposing to have $(\gamma'_{10},\gamma'_{21},\gamma'_{20})$
in ${\mathbb D}_3$ brings us to the antidegradability region depicted in blue in Fig.~\ref{fig: degr and antidegr}.

\clearpage

\subsection{Capacities in non-degradable and non-antidegradable regions}\label{sec: cap non degradable}
In those regions of parameters for which degradability or antidegradability are not achieved a closed accessible expression for $Q$ and $C_{\text{p}}$ is lacking. 
A notable exception is provided by  the two planar regions identified respectively by $\gamma_{10}=0$ (triangular surface of Fig.~\ref{fig: BS_ADC}, identified by the vertexes ${\bf ADE}$) and $\gamma_{21}=0$ (rectangular region ${\bf ABCD}$). As one can observe from Fig.~\ref{fig: degr and antidegr} they overlap only partially with the degradable and antidegradable regions: yet, thanks to the fact that there $\Phi_{\Gamma}$ and $\Phi^{({\text{\tiny MAD}})}_{\Gamma}$ coincide (see comments in Sec.~\ref{sec: Definition}), we can use the same techniques of~\cite{MAD}, see Appendix \ref{app: Qcaps nondeg}, to compute the quantum capacity of the associated ReMAD channels. Even for those points that aren't explicitly degradable (or antidegradable), specifically: 

\begin{itemize}

\item Planar region $\gamma_{10}=0$: here the channel  $\Phi_{\Gamma}$ is neither degradable nor antidegradable for all points with $\gamma_{12}+\gamma_{20}\geq 1/2$, see Appendix \ref{app: uniqueness}. With a similar approach as the one in~\cite{MAD}, see Appendix \ref{app: Q gamma_10}, we can conclude that in this region  the capacities are constant and equal to 1, i.e.
\begin{eqnarray}
Q(\Phi_{\Gamma})=C_{\text{p}}(\Phi_{\Gamma}) = 1\;,  \qquad \forall \; \gamma_{12}+\gamma_{20}\geq 1/2\;, \label{planar1}
\end{eqnarray} 
see left panel of Fig. \ref{fig: Qcap nondeg}.

\item Planar region $\gamma_{21}=0$: here the channel  $\Phi_{\Gamma}$ is neither degradable nor antidegradable for all points which verify the conditions $\gamma_{01}\geq1/2\geq \gamma_{20}$ (right-lower quadrant in Fig. \ref{fig: Qcap nondeg}, right panel) 
or $\gamma_{20}\geq1/2\geq \gamma_{01}$  (left-upper quadrant in Fig. \ref{fig: Qcap nondeg}, right panel), see Appendix \ref{app: uniqueness}. 
Applying techniques of~\cite{MAD}, see Appendix \ref{app: Q gamma_21}, we can conclude that in these regions the capacities assume symmetric values, i.e. give $0\leq x\leq 1/2 \leq y\leq 1$, we have 
\begin{eqnarray}
&& Q(\Phi_{\Gamma})\Big|_{\begin{subarray}{l} \gamma_{01}=x \\
        \gamma_{21}=0\\\gamma_{20}=y
      \end{subarray}}=C_{\text{p}}(\Phi_{\Gamma})\Big|_{\begin{subarray}{l} \gamma_{01}=x \\
        \gamma_{21}=0\\\gamma_{20}=y
      \end{subarray}} =Q(\Phi_{\Gamma})\Big|_{\begin{subarray}{l} \gamma_{01}=y \\
        \gamma_{21}=0 \\\gamma_{20}=x 
      \end{subarray}} = C_{\text{p}}(\Phi_{\Gamma})\Big|_{\begin{subarray}{l} \gamma_{01}=y \\
        \gamma_{21}=0 \\\gamma_{20}=x 
      \end{subarray}}={\cal Q}(x) \;, \label{frontpanel} 
\end{eqnarray} 
with
\begin{eqnarray} \nonumber 
    {\cal Q}(x)  &:=&  \max_{ p_1,p_2 } \{ - [1-(1-x) p_1)]\log_2 [1-(1-x) p_1)]   - [(1-x) p_1]\log_2 [(1-x) p_1] \\ 
&&\qquad\quad +(1-x p_1-p_2)\log_2 (1-xp_1-p_2) +   x p_1 \log_2 (xp_1 )+p_2 \log_2 (p_2)\}\nonumber  \\ 
&=& \max_{p_1 \in [0,1]} \{ H_2((1-x) p_1)- H_2(x p_1 )\} \;, \label{eq: opt qubitnew}
\end{eqnarray} 
 see right panel of Fig. \ref{fig: Qcap nondeg}.
In the first expression the maximization is performed over the populations $p_1,p_2\in [0,1]$ of the input state with respect to levels $|1\rangle_{\text{S}}$ and $|2\rangle_{\text{S}}$ under the consistency constraint
 $p_1+p_2\leq 1$; the second identity instead follows from the observation that for $p_0$ fixed, the maximum of the r.h.s. term is always achieved by $p_2=0$ (no population assigned to the second excited level). Observe that the final expression in Eq.~(\ref{eq: opt qubit}) exactly matches the quantum capacity of a qubit ADC channel with damping probability equal to $1-x$  \cite{QUBIT_ADC}.
 
\hspace*{4cm}
\begin{figure}[t!]
     \centering
     \hspace*{-1.5cm}
     \begin{subfigure}[b]{0.49\textwidth}
         \centering
         \hspace*{0.5cm}
         \includegraphics[width=1.1\textwidth]{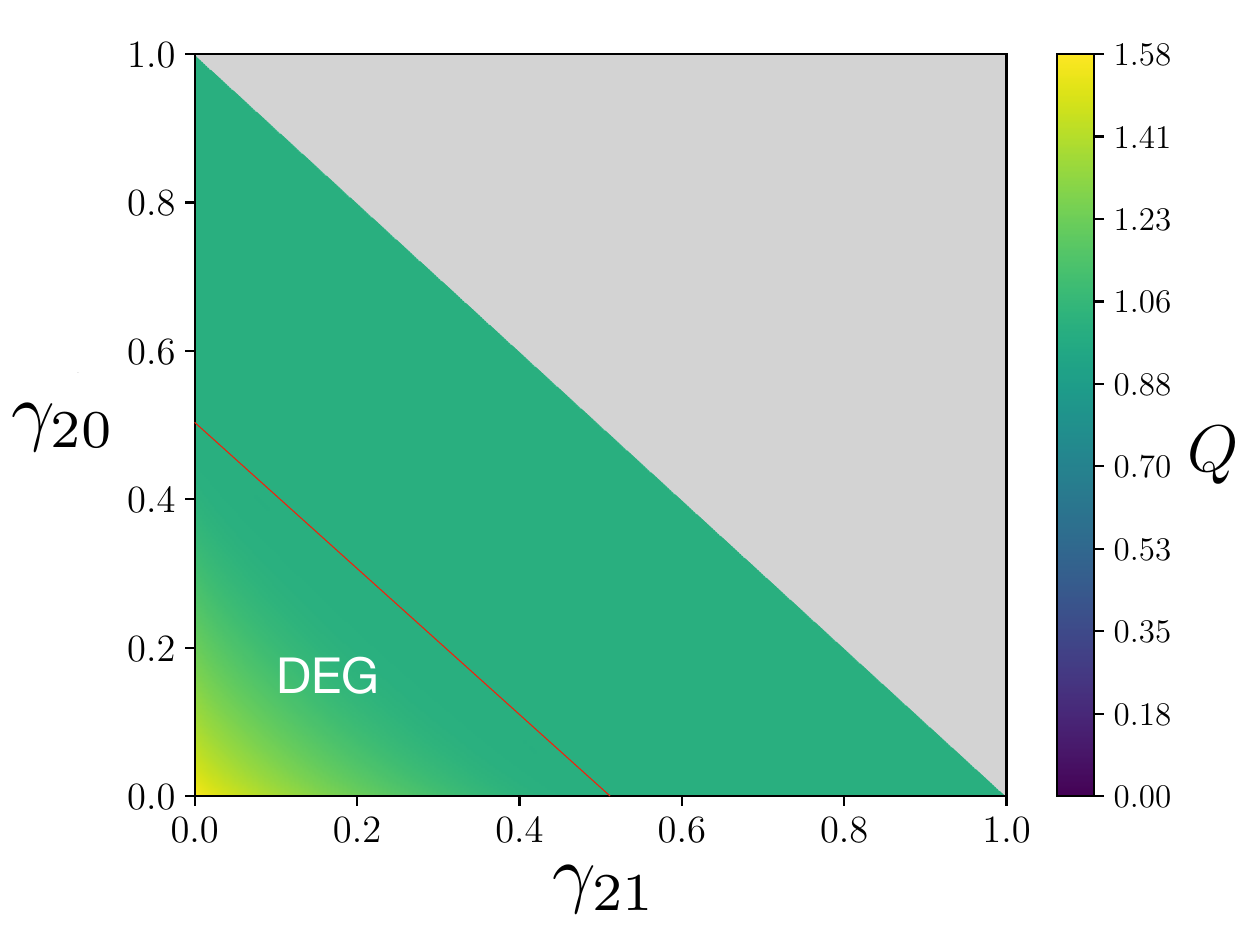}         
     \end{subfigure}
     \hspace*{1cm}
     \begin{subfigure}[b]{0.49\textwidth}
         \centering
         \hspace*{0.5cm}
         \includegraphics[width=1.1\textwidth]{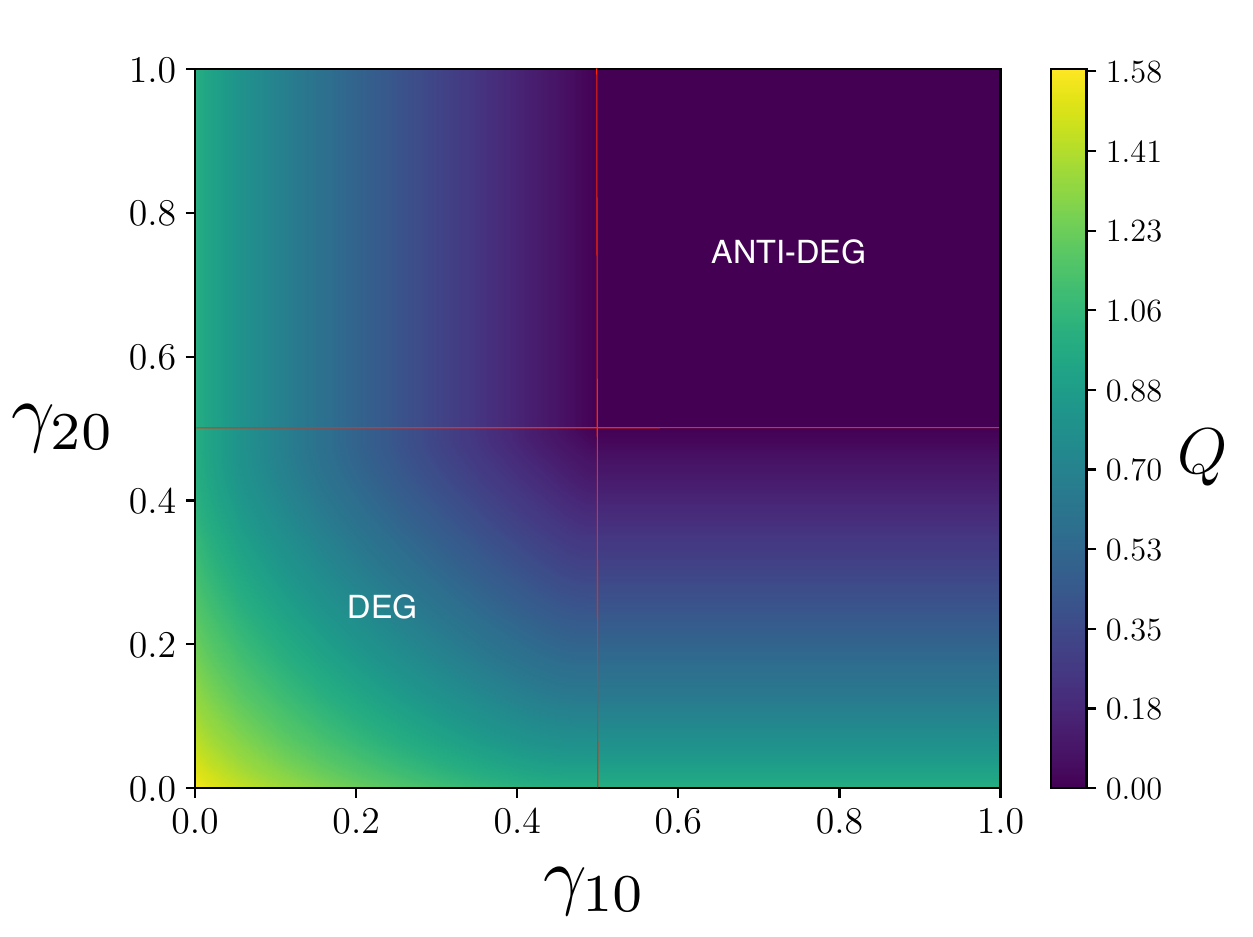}         
     \end{subfigure}
        \caption{\textbf{Left}: $Q(\Phi_\Gamma)$ (and $C_{\text{p}}(\Phi_\Gamma)$) for $\gamma_{10}=0$. The capacities are known also in the region $\gamma_{20}+\gamma_{21}\geq \frac{1}{2}$ that is not degradable nor antidegradable. \textbf{Right}: $Q(\Phi_\Gamma)$ (and $C_{\text{p}}(\Phi_\Gamma)$) for $\gamma_{21}=0$. The capacities are known also in two regions that are not degradable nor antidegradable, see left-top panel ($\gamma_{10}=0.000$) of Fig. \ref{fig: Q capacity} } 
        \label{fig: Qcap nondeg}
\end{figure}

\item With a little additional effort we can also determine the exact value of 
$Q(\Phi_{\Gamma})$ for the elements on the boundary of the ${\mathbb D}_3$ region identified  by the constraints
\begin{eqnarray}
\gamma_{10}=1\;,\qquad  \gamma_{21}+\gamma_{20}=1\;,
\end{eqnarray} 
(notice that while for $\gamma_{21} \leq 1/2$ the points belong to the antidegradable region, for $\gamma_{21}>1/2$ they are neither degradable nor antidegradable, see Appendix \ref{app: uniqueness}). 
For these values Eq.~(\ref{eq: channel action}) becomes 
\begin{equation}\label{eq: channel action EDGE}
\Phi_{{\Gamma}}(\hat{\rho})=\left(
\begin{array}{ccc}
\rho_{00} + \rho_{11}+(1-\gamma_{21}) \rho_{22} &  \sqrt{\gamma_{21}}\rho_{12} &
   0 \\
 \sqrt{ \gamma_{21}} \rho_{12}^* & \gamma_{21} \rho_{22} &
    0 \\
0 &0 & 0 \\
\end{array}
\right) \; ,
\end{equation}
which bears a close resemblance to the MAD channel $\Phi^{({\text{\tiny MAD}})}_{\Gamma'}$
with values $\gamma_{21}'=0$, $\gamma_{20}'=1$, i.e. 
\begin{equation}\label{eq: channel actionMADedge}
\Phi^{({\text{\tiny MAD}})}_{\Gamma'}(\hat{\rho})=\left(
\begin{array}{ccc}
\rho_{00} + \gamma'_{10} \rho_{11}+ \rho_{22} &  \sqrt{1-\gamma'_{10}} \rho_{01}  &
  0 \\
\sqrt{1-\gamma_{10}} \rho_{01}^*  & (1-\gamma'_{10}) \rho_{11}&
0 \\
0&  0 &  0\\
\end{array}
\right) \; ,
\end{equation}
for which capacities $Q$ and $C_{\text{p}}$ have been computed for all values of $\gamma_{10}'$ \cite{MAD}, see Appendix \ref{sec:first}. The explicit connection between Eq.~(\ref{eq: channel action EDGE})  and Eq.~(\ref{eq: channel actionMADedge}) follows by observing that, setting $\gamma_{10}'=1-\gamma_{21}$  we can map the former into the latter via a unitary transformation on the input of the channel via the identity 
\begin{eqnarray} 
\Phi_{\Gamma}(\hat{\rho}) = \Phi^{({\text{\tiny MAD}})}_{\Gamma'}(\hat{V}_{\text{S}} \hat{\rho}\hat{V}^\dag_{\text{S}})\;, 
\end{eqnarray} 
being $\hat{V}_{\text{S}}$ the unitary operator reordering the canonical basis 
$\ket{0}_{\text{S}}$, $\ket{1}_{\text{S}}$, $\ket{2}_{\text{S}}$, into $\ket{1}_{\text{S}}$, $\ket{2}_{\text{S}}$, $\ket{0}_{\text{S}}$. Accordingly using Eq.~(\ref{eq:cohID13}) of Appendix \ref{app: Q gamma_21}, in the region $\frac{1}{2}\leq \gamma_{21} \leq 1$ we can write 
\begin{align}\nonumber 
& &Q(\Phi_{\Gamma}) = C_{\text{p}}(\Phi_{\Gamma}) = \max_{ p_0,p_1 } \{ & - (1-\gamma_{21}p_0)\log_2 (1-\gamma_{21}p_0)
- \gamma_{21} p_0 \log_2 ( \gamma_{21}p_0) \nonumber \\   
& & &+(1-(1-\gamma_{21})p_0-p_1)\log_2 (1-(1-\gamma_{21})p_0-p_1) \nonumber \\
& & &+ (1-\gamma_{21}) p_0 \log_2 ((1-\gamma_{21})p_0) + p_1 \log_2 p_1\} \nonumber \\ 
& & ={\cal Q}(1-\gamma_{21})  \;, &\label{eq: opt qubit}
\end{align} 
in the region $0\leq \gamma_{21} \leq \frac{1}{2}$ instead, $Q=C_{\text{p}}=0$ as the channel is antidegradable. The overall profile of the capacities in this case reduces then to that of a qubit ADC with damping parameter $1-\gamma_{21}$ \cite{QUBIT_ADC}. We report its evaluation in Fig. \ref{fig: Q capacity qubit}, left panel.
Notice that in the first expression the maximization is performed over the populations $p_0,p_1\in [0,1]$ of the input state with respect to levels $|0\rangle_{\text{S}}$ and $|1\rangle_{\text{S}}$ under the consistency constraint $p_0+p_1\leq 1$. The second identity then follows by noticing that, similarly to what happens in Eq.~(\ref{eq: opt qubitnew}), for $p_0$ fixed the maximum of the r.h.s. term is always achieved by $p_1=0$ (no population assigned to the first excited level).
Comparing  Eq.~(\ref{eq: opt qubit})  with  Eq.~(\ref{frontpanel}) we observe that on the  edges ${\bf BC}$ and ${\bf BF}$, that bound $\mathbb{D}_3$ on the plane $\gamma_{10}=1$, the capacity of the channel takes the same values as those for fixed $\gamma_{20}$. Indeed we get 
\begin{eqnarray}
&&Q(\Phi_{\Gamma})\Big|_{\begin{subarray}{l} 
\gamma_{01}=1 \\
        \gamma_{21}=0\\
        \gamma_{20}
      \end{subarray}
      }
 =Q(\Phi_{\Gamma})\Big|_{\begin{subarray}{l} 
\gamma_{01}=1 \\
        \gamma_{21}=1- \gamma_{20}\\
        \gamma_{20}
      \end{subarray}
      }=
{\cal Q}(\gamma_{20})  \;, \label{eq: opt qubit-newedge}
\end{eqnarray} 
which again is the capacity profile of a qubit ADC with damping parameter $\gamma_{20}$. See \ref{fig: Q capacity qubit}, right panel.
  
\begin{figure}[t!]
     \centering
     \begin{subfigure}[t]{0.45\textwidth}
         \centering
         \hspace*{-0.3cm}
         \includegraphics[width=\textwidth]{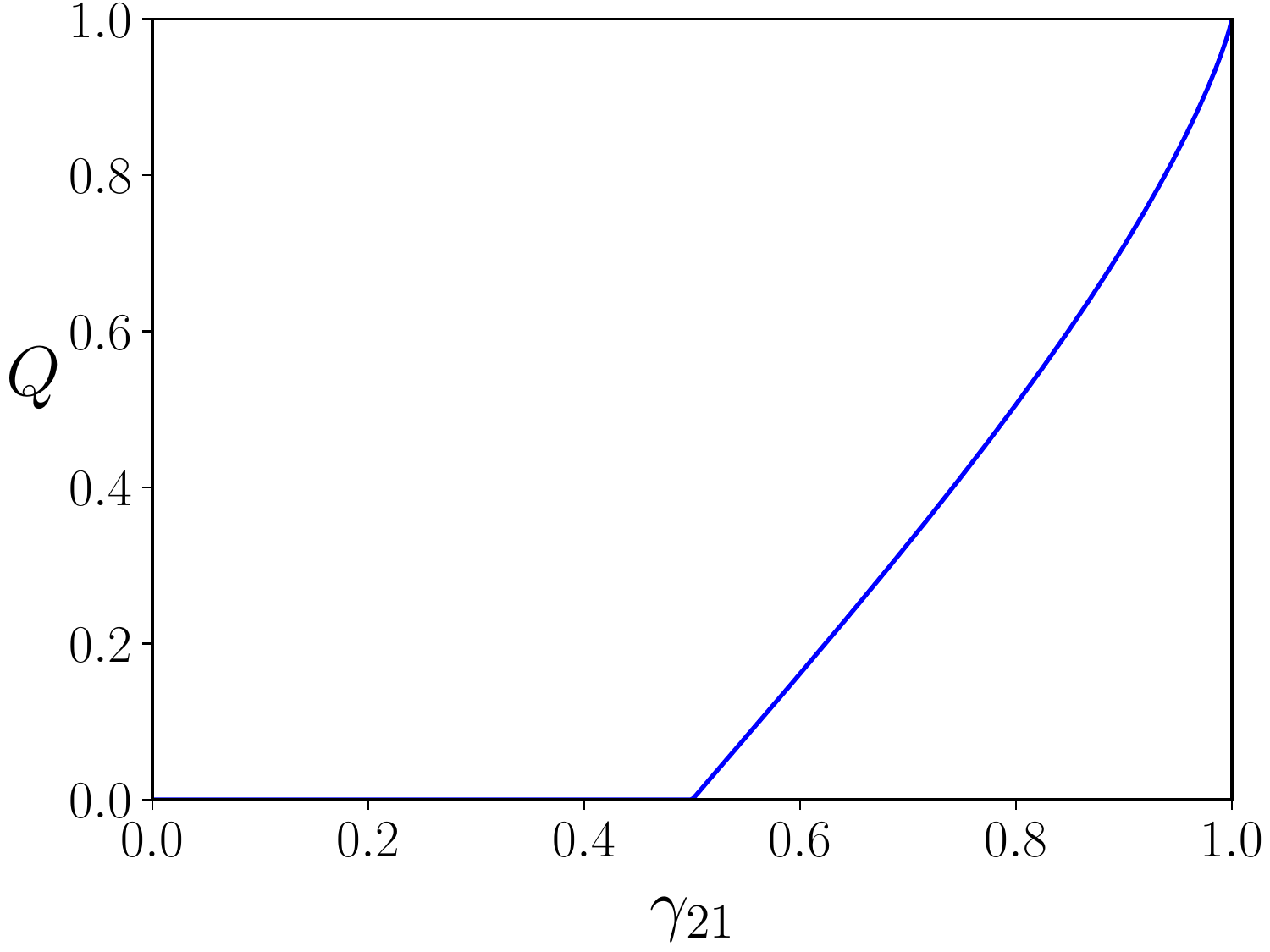}         
     \end{subfigure}     
     \begin{subfigure}[t]{0.45\textwidth}
         \centering
         \hspace*{0.3cm}
         \includegraphics[width=\textwidth]{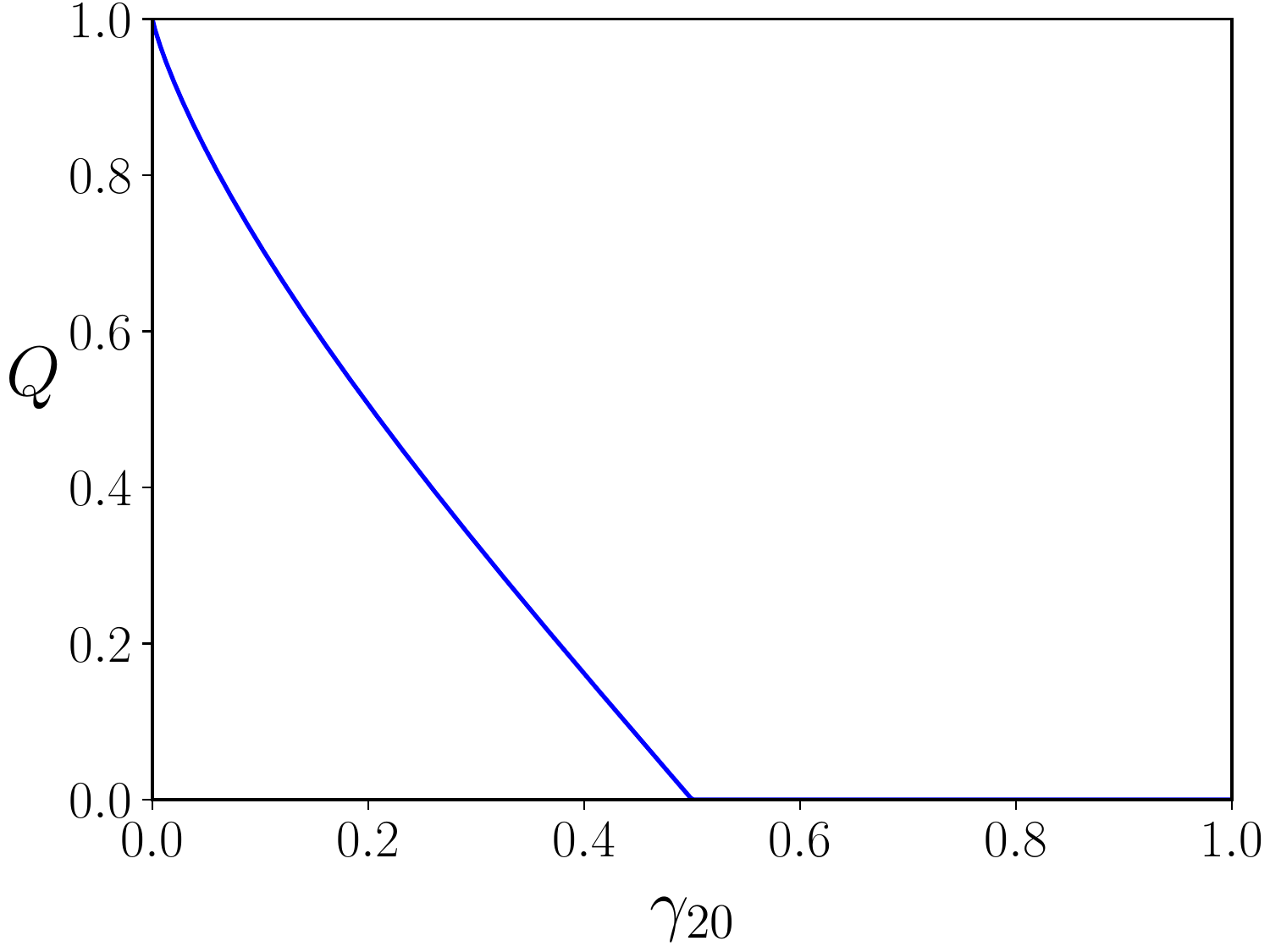}         
     \end{subfigure}
     \caption{ \textbf{Left}: $Q(1-\gamma_{21})$ as in Eq.~(\ref{eq: opt qubit}), in the region $\gamma_{21}\geq \frac{1}{2}$ the capacity is found for provably not degradable nor antidegradable channels. \textbf{Right}: $Q(\gamma_{20})$ as in Eq.~(\ref{eq: opt qubit-newedge}), in the region $\gamma_{20}\geq \frac{1}{2}$ the capacity is found for provably not degradable nor antidegradable channels. }
        \label{fig: Q capacity qubit}
\end{figure}

\item In the other sectors of the parameter space 
it is still possible to exploit some information theoretic properties  to obtain computationally efficient upper bounds. For instance exploiting the considerations underlined in the final paragraphs of Sec.~\ref{sec: Composition rules} and the data processing inequality (\ref{data}) we can claim that, for fixed values of $\gamma_{10}$ and $\gamma_{21}$, both $Q(\Phi_{\Gamma})$ and $C_{\text{p}}(\Phi_{\Gamma})$ are non-increasing functions of $\gamma_{20}$, e.g. 
\begin{eqnarray}
Q(\Phi_{\Gamma})\geq Q(\Phi_{\Gamma'})\;, \qquad \forall \gamma_{10}'=\gamma_{10}, 
\gamma_{21}'= \gamma_{21}, \gamma_{20}'\geq \gamma_{20}\;. 
\end{eqnarray} 

\end{itemize}

\section{Conclusions}\label{sec: Conclusions}

We introduced and characterized a new class of physically relevant noise models for high-dimensional quantum systems. ReMAD channels enlarge the limited class of quantum channels for which an exact analytical or numerical approach in terms of quantum and private classical capacities is feasible. This has been shown by focusing on the simplest, yet nontrivial, case in which the system of interest is a qutrit. In such case we have shown the existence of a substantial region of the noise parameter space where the channels $\Phi_{\Gamma}$ are degradable/antidegradable, allowing hence the direct expression of the associated quantum and private classical capacities. Also, by resorting to formal mapping to the case of MAD channels discussed in \cite{MAD}, we computed the value of these capacities in regions where $\Phi_{\Gamma}$ is provably neither degradable nor antidegradable. 
The full characterization in terms of information capacities, such as for instance classical capacity and two-way capacities, is  still missing for ReMAD channels and will require further investigations. A semi-analytical treatment is instead achievable for the characterization of entanglement assisted quantum and classical capacities $Q_E$ and $C_E$ of ReMAD channels, see Appendix \ref{sec: app ent ass cap}. 
\\

We acknowledge financial support by MIUR (Ministero dell' Istruzione, dell' Universit{\'a} e della Ricerca) by PRIN 2017 {\it Taming complexity via Quantum Strategies: a Hybrid Integrated Photonic approach} (QUSHIP) Id. 2017SRN- BRK, and via project PRO3 Quantum Pathfinder.

\clearpage

\appendix
\section{Channel inversion}\label{sec:channelinversion}  
To compute $C_{\cal D}$ we need to identify the inverse of the channel $\Phi$. 
Concretely the inversion of ${\Phi}$ can be done by exploiting the fact that quantum channels are linear maps connecting vector spaces of linear operators. They can in turn be represented as matrices acting on vector spaces. This through the following ``vectorization'' isomorphism, also called Liouville representation:
\begin{eqnarray}
\hat{\rho}_{\text{S}}=\sum_{ij} \rho_{ij}\ket{i}\!\!\bra{j}_{\text{S}}&\longrightarrow&\ket{\rho\rangle}=\sum_{ij} \rho_{ij}\ket{i}_{\text{S}}\otimes \ket{j}_{\text{S}}\in 
{\cal H}_{\text{S}}^{\otimes 2} \; , \nonumber \\
\label{eq:isomorphism}\\
\Phi(\hat{\rho}_{\text{S}})&\longrightarrow& \hat{\text{M}}_{\Phi} \ket{\rho\rangle} \; , \nonumber 
\end{eqnarray}
where now $\hat{\text{M}}_{\Phi}$ is a $d_{\text{S}'}^2\times d_{\text{S}}^2$ matrix connecting ${\cal H}_{\text{S}}^{\otimes 2}$ and 
 ${\cal H}_{\text{S}'}^{\otimes 2}$ ($d_{\text{S}}$ and $d_{\text{S}'}$ being respectively the dimensions of ${\cal H}_{\text{S}}$ and ${\cal H}_{\text{S}'}$).
 Given a Kraus set $\{\hat{K}_i\}_i$ for $\Phi$, $\hat{\text{M}}_{\Phi}$ can be explicitly expressed as 
\begin{equation}
\hat{\text{M}}_{\Phi}=\sum_i \hat{K}_i\otimes \hat{K}_i^* \; .
\end{equation}
Following Eq.~(\ref{eq:degradable}) we have hence that  for a degradable channel the following identity must apply 
\begin{equation}
\hat{\text{M}}_{\tilde{\Phi}}=\hat{\text{M}}_{\mathcal{D}}\hat{\text{M}}_{\Phi} \; ,
\end{equation}
with $\hat{\text{M}}_{\mathcal{D}}$ the matrix representation of the LCPTP degrading channel ${\cal D}$, 
implying that the super-operator $\tilde{\Phi}\circ \Phi^{-1}$ is now represented by 
\begin{equation}\label{eq: degr channel matrix}
\hat{\text{M}}_{\mathcal{D}} = \hat{\text{M}}_{\tilde{\Phi}}\hat{\text{M}}_{\Phi}^{-1} \; .
\end{equation}

In the case of the qutrits ReMAD channel of (\ref{eq: channel action}), the inverse channel can also be expressed in the form 
\begin{widetext}
\begin{equation}\label{eq: channel action INV}
\hspace*{-0.5cm}
\Phi^{-1}_{{\Gamma}}(\hat{\rho})=\left(
\begin{array}{ccc}
\rho_{00} - \tfrac{ \gamma_{10}}{1-\gamma_{10}} \rho_{11}+ \tfrac{\gamma_{10}\gamma_{21}-\gamma_{20} (1-\gamma_{10})}{(1-\gamma_{10})(1-\gamma_{21}-\gamma_{20})} \rho_{22} & \tfrac{\rho_{01}}{ \sqrt{1-\gamma_{10}} } - \tfrac{\sqrt{\gamma_{10} \gamma_{21}}\rho_{12}  }{(1-\gamma_{10}) \sqrt{(1-\gamma_{20}-\gamma_{21})}}
 &
   \tfrac{\rho_{02} }{\sqrt{1-\gamma_{21}-\gamma_{20}}}\\
\tfrac{\rho^*_{01}}{ \sqrt{1-\gamma_{10}} } - \tfrac{\sqrt{\gamma_{10} \gamma_{21}} \rho^*_{12}  }{(1-\gamma_{10}) \sqrt{(1-\gamma_{20}-\gamma_{21})}}
&\tfrac{ \rho_{11}}{1-\gamma_{10}} -\tfrac{ \gamma_{21} \rho_{22}}{(1-\gamma_{10}) (1-\gamma_{21}-\gamma_{20})} &
  \tfrac{ \rho_{12} }{  \sqrt{(1-\gamma_{10}) (1-\gamma_{21}-\gamma_{20})}}\\
 \tfrac{\rho_{02}^*}{\sqrt{1-\gamma_{21}-\gamma_{20}} } &  \tfrac{\rho_{12}^*}{\sqrt{(1-\gamma_{10}) (1-\gamma_{21}-\gamma_{20})}}&   \tfrac{\rho_{22}}{1-\gamma_{21}-\gamma_{20}} \\
\end{array}
\right) \; ,
\end{equation}
\end{widetext}
as one can check by direct computation. 

\subsection{Uniqueness of the degrading channel}\label{app: uniqueness}

In the most general case, even when the inverse of the channel considered exists, degrading maps may not be unique \cite{PITFALLS}. We show here how in the case of qutrit ReMAD channels, if these are degradable, the degrading channel must be unique and in the form of Eq.~(\ref{eq: degr channel matrix}). For our analysis this is also equivalent to say that if the degrading channel $\mathcal{D}$ can be found as in Eq.~(\ref{eq: degr channel matrix}) but it's not LCPTP, then the channel is not degradable.  
\noindent A proof of can be derived from \cite[Theorem 3]{PITFALLS}:
\setcounter{thm}{3}
\begin{thm}[\cite{PITFALLS}]
Let $\Phi:\mathfrak{S}({\cal H}_{\text{A}})\rightarrow \mathfrak{S}({\cal H}_{\text{B}})$ be a quantum channel and $\tilde{\Phi}:\mathfrak{S}({\cal H}_{\text{A}})\rightarrow \mathfrak{S}({\cal H}_{\text{E}})$ its complementary channel and let the corresponding super-operator $\hat{\text{M}}_{\Phi}$ of $\Phi$ be full rank: $\text{rank}\; \hat{\text{M}}_{\Phi} = \min [d_A^2, d_B^2]$. Then, if a degrading map $\mathcal{D}:\mathfrak{S}({\cal H}_{\text{B}})\rightarrow \mathfrak{S}({\cal H}_{\text{E}})$ exists,  it is unique iff $d_B \leq d_A$.
\end{thm}

 \noindent In our case $d_B = d_A$ and the `superoperator' $\hat{\text{M}}_{\Phi}$ is always full rank except when $ \gamma_{10} = 1$ or $\gamma_{20}+\gamma_{21}=1$, since 
\begin{equation*}
\hspace*{-1cm}\mathsf{M} = \hat{\text{M}}_{\Phi_\Gamma} = \scalebox{0.7}{$\begin{pmatrix}
1 & 0 & 0 & 0 & \gamma_{10} & 0 & 0 & 0 & \gamma_{20} \\
0 & \sqrt{1 - \gamma_{10}} & 0 & 0 & 0 & \sqrt{\gamma_{10}\gamma_{21}} & 0 & 0 & 0 \\ 
0 & 0 & \sqrt{1-\gamma_{21}-\gamma_{20}} & 0 & 0 & 0 & 0 & 0 & 0 \\ 
0 & 0 & 0 & \sqrt{1 - \gamma_{10}} & 0 & 0 & 0 & \sqrt{\gamma_{10}\gamma_{21}} & 0 \\ 
0 & 0 & 0 & 0 &  1 - \gamma_{10} & 0 & 0 & 0 & \gamma_{21} \\ 
0 & 0 & 0 & 0 & 0 & \sqrt{(1 - \gamma_{10})(1-\gamma_{21}-\gamma_{20})} & 0 & 0 & 0 \\ 
0 & 0 & 0 & 0 & 0 & 0 & \sqrt{1-\gamma_{21}-\gamma_{20}} & 0 & 0 \\ 
0 & 0 & 0 & 0 & 0 & 0 & 0 & \sqrt{(1 - \gamma_{10})(1-\gamma_{21}-\gamma_{20})} & 0 \\ 
0 & 0 & 0 & 0 & 0 & 0 & 0 & 0 & 1-\gamma_{21}-\gamma_{20}
\end{pmatrix}$}
\end{equation*}
is upper triangular and full rank, where we ordered the $\rho_{ij}$ `basis' as $\{ \rho_{00}, \rho_{01}, \rho_{02}, \rho_{10}, \rho_{11}, \rho_{12}, \rho_{20}, \rho_{21}, \rho_{22} \}$. So, whenever $\gamma_{10}\neq 1$ and $ \gamma_{21}+ \gamma_{20} \neq 1$ we have that $\hat{\text{M}}_{\Phi_\Gamma}$ is full rank (hence invertible) and if $\hat{\text{M}}_{\mathcal{D}} = \hat{\text{M}}_{\tilde{\Phi}_\Gamma}\hat{\text{M}}_{\Phi_\Gamma}^{-1}$ doesn't represent a LCPTP map then $\Phi_\Gamma$ is not degradable.\\

\noindent Now the cases $ \gamma_{10} = 1$ and $\gamma_{20}+\gamma_{21}=1$ remain, but it's straightforward to show that these channels can't degradable. Specifically we just need to show that $\text{ker}\; \Phi_\Gamma \not\subset \text{ker}\; \tilde{\Phi}_\Gamma$, see \cite[Sec. 2.1]{STR_DEG_CH}. Looking at the expressions of the channels and their complementaries
\begin{eqnarray}
&\Phi^{(\gamma_{10}=0)}_{{\Gamma}}(\hat{\rho})&=\left(
\begin{array}{ccc}
\rho_{00} +  \rho_{11}+\gamma_{20} \rho_{22} &   \sqrt{\gamma_{21}}\rho_{12} & \sqrt{1-\gamma_{21}-\gamma_{20}} \rho_{02} \\
 \sqrt{\gamma_{21}} \rho_{12}^* & \gamma_{21} \rho_{22} &  0 \\
 \sqrt{1-\gamma_{21}-\gamma_{20}} \rho_{02}^* & 0 &  (1-\gamma_{21}-\gamma_{20}) \rho_{22} \\
\end{array}
\right) \; \nonumber\\
&\tilde{\Phi}^{(\gamma_{10}=0)}_{{\Gamma}}(\hat{\rho})&= \left(
\begin{array}{ccc}
 \rho_{00} + (1-\gamma_{21}-\gamma_{20})\rho_{22} & \rho_{01} & \sqrt{\gamma_{20}} \rho_{02} \\
  \rho_{01}^* &  \rho_{11}+\gamma_{21} \rho_{22} & \sqrt{ \gamma_{20}}\rho_{12}\\
 \sqrt{\gamma_{20}} \rho_{02}^* & \sqrt{\gamma_{20}} \rho_{12}^* & \gamma_{20} \rho_{22} \\
\end{array}
\right) \; ,
\end{eqnarray}

\begin{align}
&\Phi^{(\gamma_{21}+\gamma_{20}=1)}_{{\Gamma}}(\hat{\rho})=\left(
\begin{array}{ccc}
\rho_{00} + \gamma_{10} \rho_{11}+\gamma_{20} \rho_{22} \sqrt{1-\gamma_{10}} \rho_{01} + \sqrt{\gamma_{10} \gamma_{21}}\rho_{12} & 0 \\
\sqrt{1-\gamma_{10}} \rho_{01}^* + \sqrt{\gamma_{10} \gamma_{21}} \rho_{12}^* & (1-\gamma_{10}) \rho_{11}+\gamma_{21} \rho_{22} &  0 \\
 0 &  0 &  0 \\
\end{array}
\right) \; \nonumber \\
& \tilde{\Phi}^{(\gamma_{21}+\gamma_{20}=1)}_{{\Gamma}}(\hat{\rho}) = \scalebox{1}{$ \left(
\begin{array}{ccc}
 \rho_{00}+(1-\gamma_{10}) \rho_{11} & \sqrt{\gamma_{10}} \rho_{01} +  \sqrt{(1-\gamma_{10}) \gamma_{21}}\rho_{12} & \sqrt{\gamma_{20}} \rho_{02} \\
  \sqrt{\gamma_{10}} \rho_{01}^* +\sqrt{(1-\gamma_{10}) \gamma_{21}} \rho_{12}^*& \gamma_{10} \rho_{11}+\gamma_{21} \rho_{22} &
    \sqrt{\gamma_{10} \gamma_{20}} \rho_{12}\\
 \sqrt{\gamma_{20}} \rho_{02}^* & \sqrt{\gamma_{10} \gamma_{20}} \rho_{12}^* & \gamma_{20} \rho_{22} \\
\end{array}
\right) $}\; ,
\end{align}

we see that the element $\ket{1}\!\!\bra{2} \in \text{ker}\; \Phi^{(\gamma_{10}=0)}_\Gamma \not\subset \text{ker}\; \tilde{\Phi}^{(\gamma_{10}=0)}_\Gamma$ and that the element $\ket{0}\!\!\bra{2} \in \text{ker}\; \Phi^{(\gamma_{21}+\gamma_{20}=1)}_\Gamma \not\subset \text{ker}\; \tilde{\Phi}^{(\gamma_{21}+\gamma_{20}=1)}_\Gamma$. Therefore both cannot be degradable.

\noindent In synthesis, for ReMAD channels the degrading map can be uniquely found via the inverse matrix of the channel. If it doesn't exist or the derived degrading map is not LCPTP then the channel is not degradable.\\

A similar analysis can be applied to antidegradability but typically we don't need it in our discussion: where we can't prove degradability, in those parameter regions we consider, ReMAD channels have positive capacity and hence cannot be antidegradable.\\

\section{Entanglement assisted quantum and classical capacities}\label{sec: app ent ass cap}

The discovery of protocols such as quantum teleportation \cite{TELEPORT} and superdense coding \cite{SUPERDENSE} showed how entanglement could be leveraged as an additional resource in order to boost the communication performance between two communicating parties. The formalization of these entanglement-assisted protocols in Shannon-theoretic terms was given in \cite{ENT_ASS, ENT_ASS1}, where the entanglement-assisted classical capacity $C_E$ and entanglement-assisted quantum capacity $Q_E$ were introduced. The peculiarity and the advantage with the definition of these capacities is that they are additive quantities and don't need a regularization. Specifically, recalling the definition of the quantum mutual information $I(\Phi, \hat{\rho})$ 
\begin{equation}\label{eq: mutual info}
I(\Phi, \hat{\rho})=S(\hat{\rho}) + I_{\text{coh}}(\Phi, \hat{\rho}) \; ,
\end{equation}
we have:
\begin{equation}
C_E(\Phi)=\max\limits_{\hat{\rho}\in \mathfrak{S}({\cal H}_{\text{S}})} I(\Phi, \hat{\rho}) \; , \qquad Q_E(\Phi)=\frac{1}{2} C_E(\Phi) \; ,
\end{equation}
where the definition of $Q_E(\Phi)$ is justified by the fact that in presence of entanglement a qudit quantum state can be teleported `spending' two classical $d$its (quantum teleportation) and viceversa two classical $d$its can be communicated by sending a single qudit (superdense coding). 

\begin{figure}[h!]
     \centering
     \hspace*{-1.5cm}
     \begin{subfigure}[t]{0.33\textwidth}
         \centering
         \includegraphics[width=1.15\textwidth]{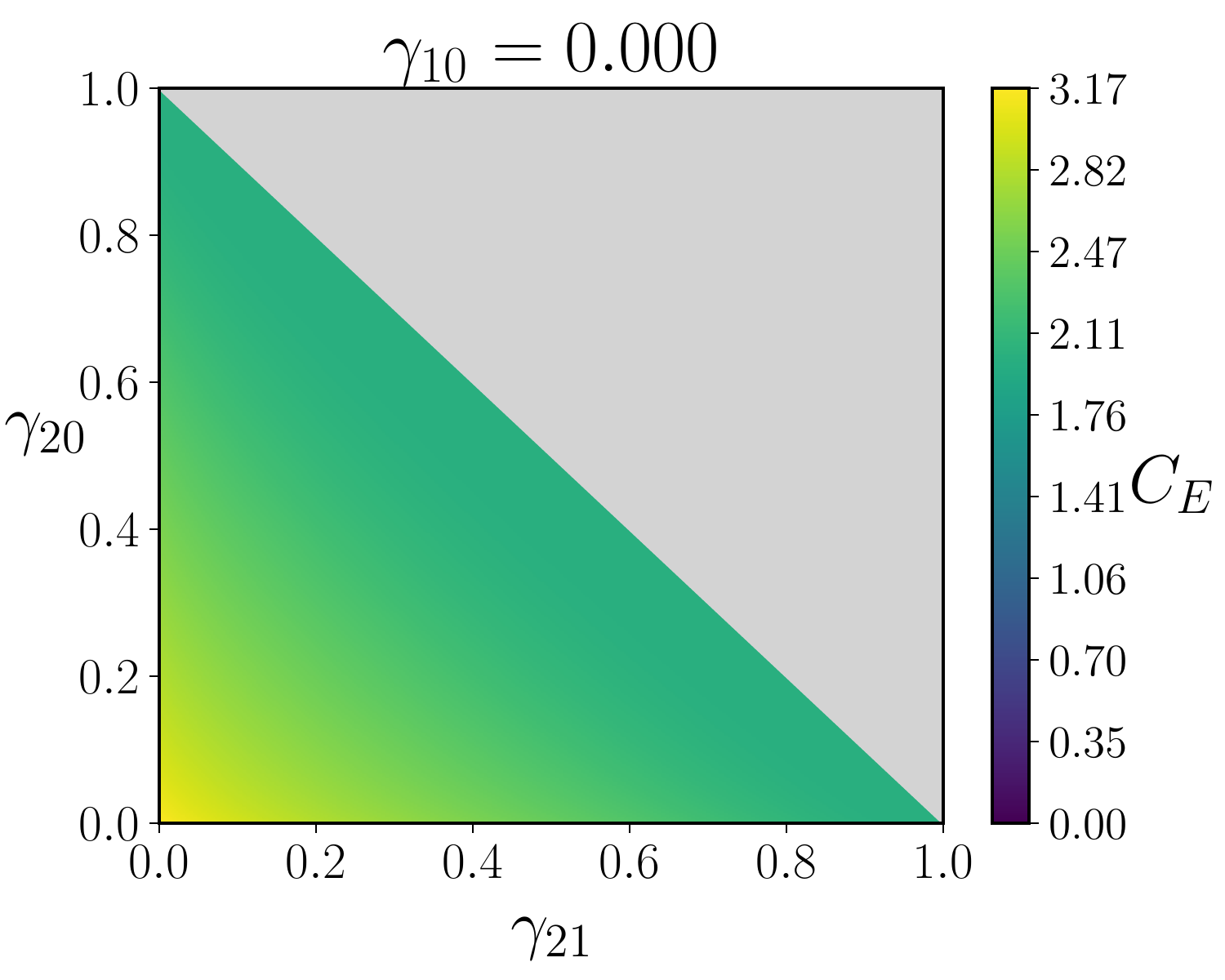}         
     \end{subfigure}     
     \hspace*{0.7cm}
     \begin{subfigure}[t]{0.33\textwidth}
         \centering
         \includegraphics[width=1.15\textwidth]{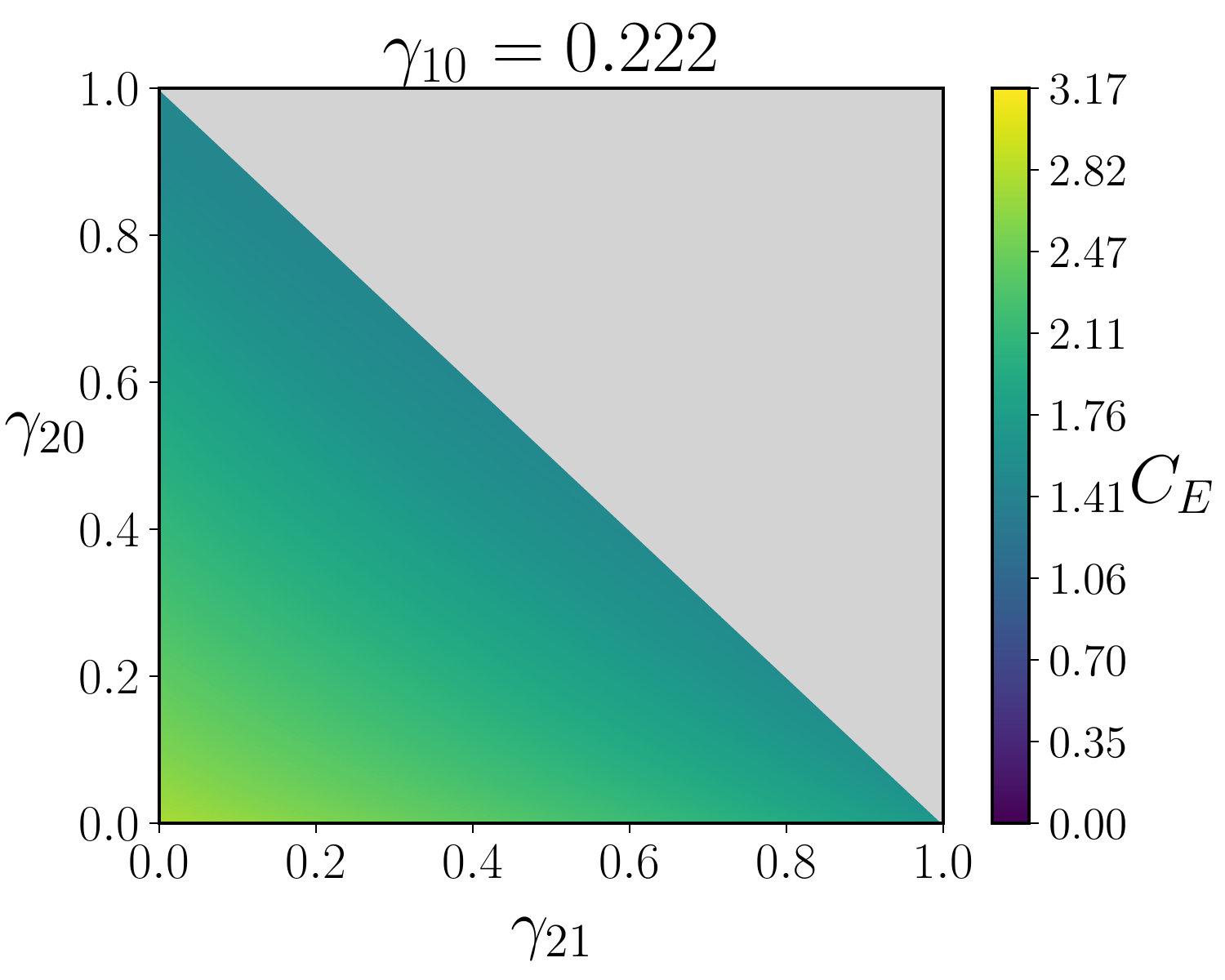}
     \end{subfigure}      
     \begin{subfigure}[t]{0.33\textwidth}
         \centering
         \hspace*{0.7cm}
         \includegraphics[width=1.15\textwidth]{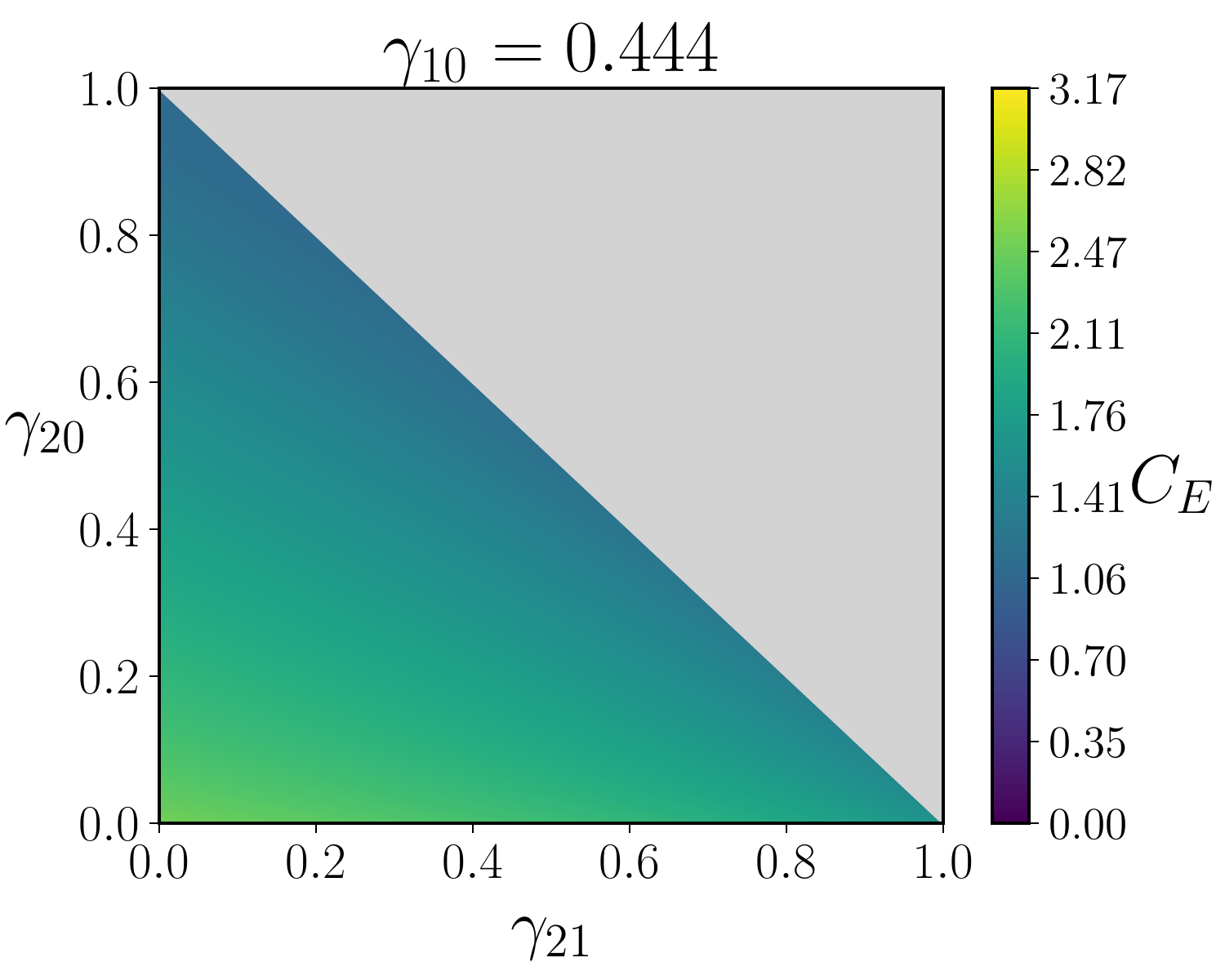}         
     \end{subfigure} 
       
     \hspace*{-1.5cm} 
     \begin{subfigure}[t]{0.33\textwidth}
         \centering
         \includegraphics[width=1.15\textwidth]{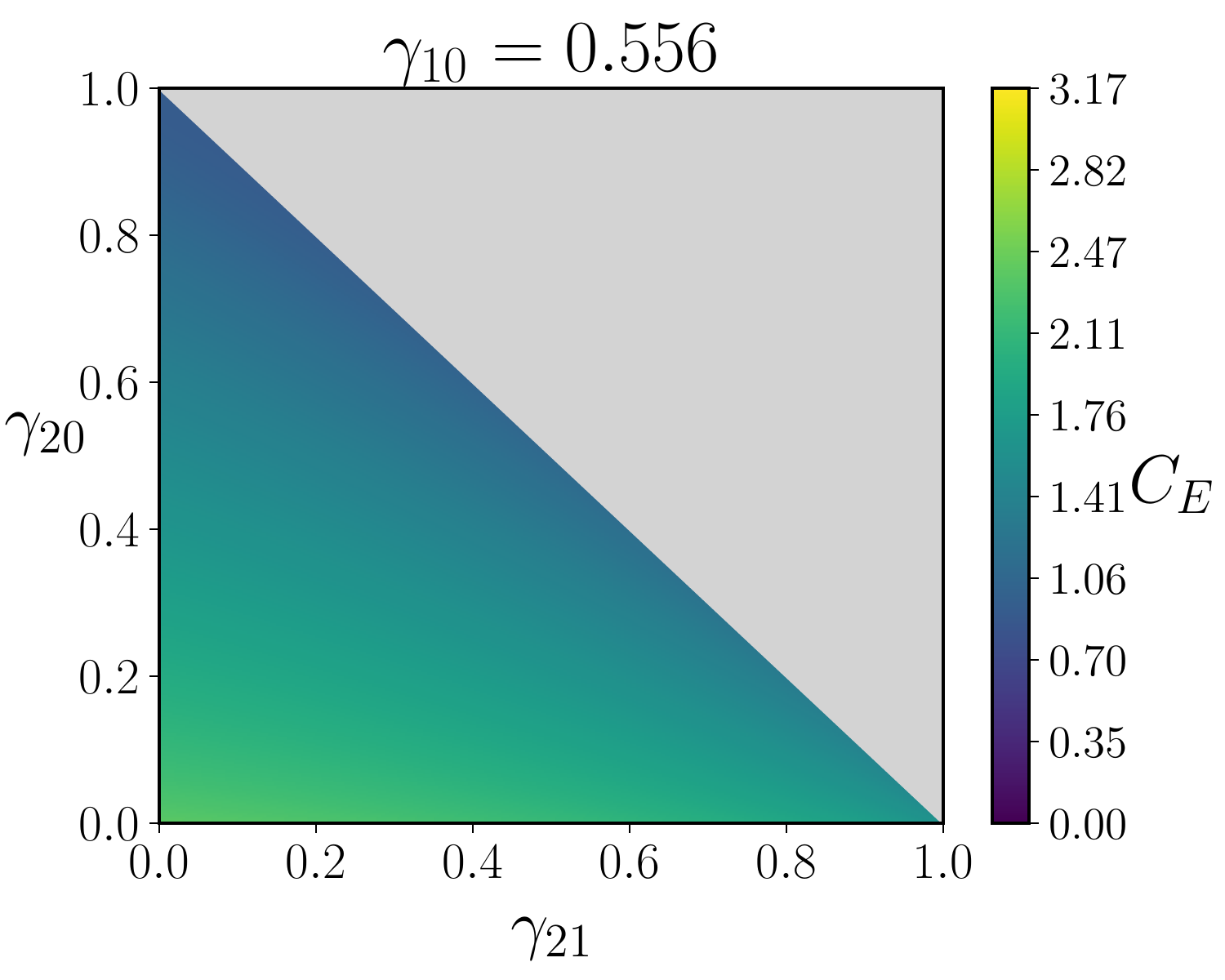}         
     \end{subfigure}     
     \hspace*{0.7cm}
     \begin{subfigure}[t]{0.33\textwidth}
         \centering
         \includegraphics[width=1.15\textwidth]{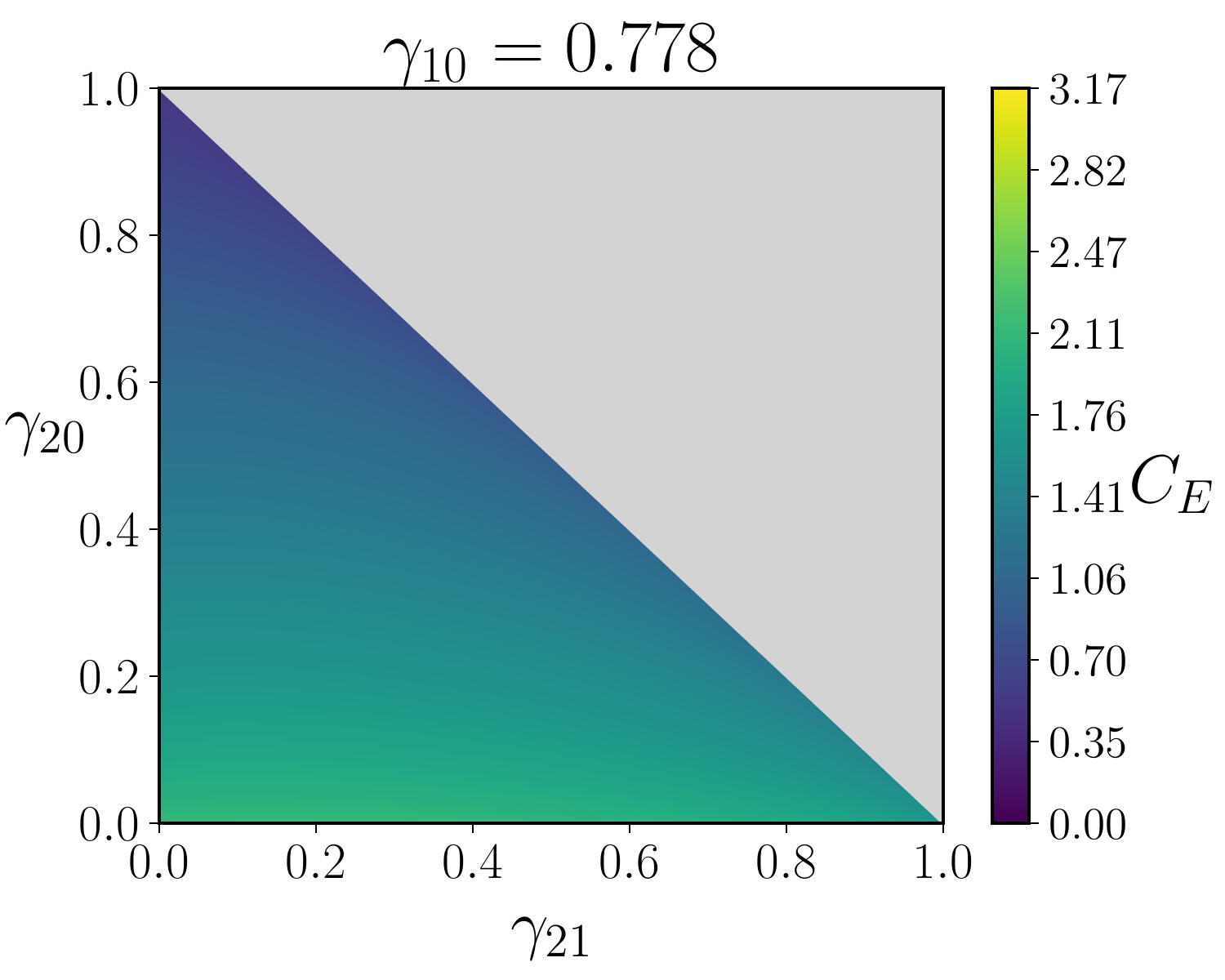}
     \end{subfigure}
     \begin{subfigure}[t]{0.33\textwidth}
         \centering
         \hspace*{0.7cm}
         \includegraphics[width=1.15\textwidth]{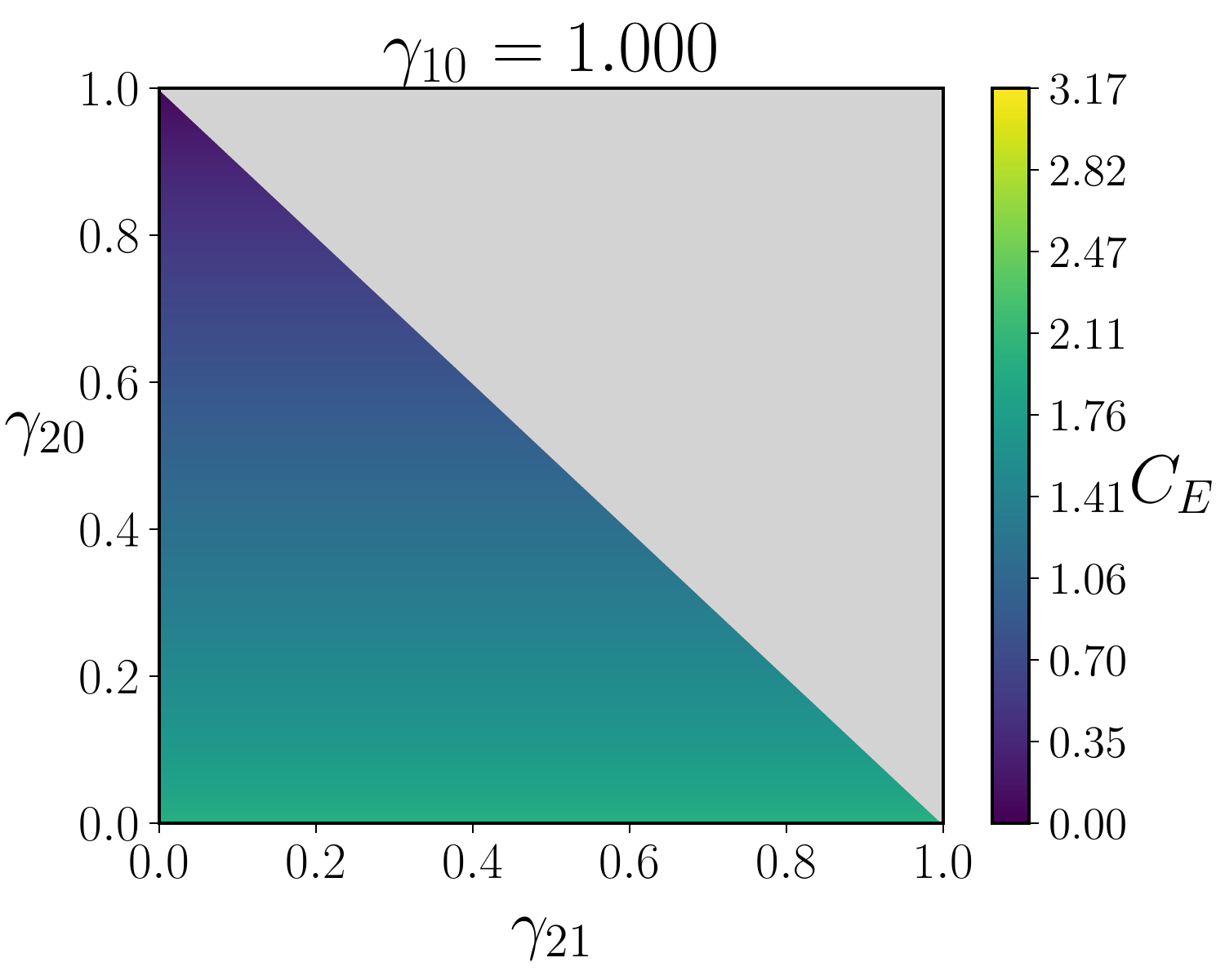}
     \end{subfigure}
        \caption{Entanglement assisted classical capacity $C_E(\Phi_{\Gamma})$ for different $\gamma_{10},$ $\gamma_{21}$ and $\gamma_{20}$. The grey area represents values of $\gamma_{20}$ and $\gamma_{21}$ s.t. $\gamma_{20}+\gamma_{21}>0$ for which the channel is not defined.}
        \label{fig: Ce capacity}
\end{figure}

To effectively compute these quantities we notice that the Shannon entropy is always concave w.r.t. the input state but, as we saw before, the coherent information can be proved concave only when the channel is degradable. Nevertheless the quantum mutual information is always concave in the input state \cite[chapter~13.4.2]{WILDEBOOK}. Therefore also $I(\Phi)$ can be maximized just over diagonal states if $\Phi$ is a covariant channel by following steps as in Eq.~(\ref{eq: coh info concave}).
 We report then in Fig.~\ref{fig: Ce capacity} the evaluation of $C_E(\Phi_{\Gamma})$ for a qutrit ReMAD channel at varying $\gamma_{20}$ and $\gamma_{21}$ for some instances of $\gamma_{10}$ values.

\section{$Q$ and $C_\text{p}$ in non-degradable regions}\label{app: Qcaps nondeg}

\subsection{$Q$ and $C_{\text{p}}$ for $\gamma_{10}=0$}\label{app: Q gamma_10}

In case $\gamma_{10}=0$ the actions of a generic qutrit ReMAD channel $\Phi_{\Gamma}^{(\gamma_{10}=0)}$ and of its complementary channel $\tilde{\Phi}_{\Gamma}^{(\gamma_{10}=0)}$ on a generic density matrix $\hat{\rho}$ reduce to
\begin{widetext}
\begin{eqnarray} \label{defgamma23} 
&\Phi_{\Gamma}^{(\gamma_{10}=0)}(\hat{\rho})=\begin{pmatrix}
\rho_{00}+\gamma_{20} \rho_{22} & \rho_{01} & \sqrt{1-\gamma_{21}-\gamma_{20}}\rho_{02}\\
\rho_{01}^* & \rho_{11}+\gamma_{21} \rho_{22} & \sqrt{1-\gamma_{21}-\gamma_{20}}\rho_{12}\\
\sqrt{1-\gamma_{21}-\gamma_{20}}\rho_{02}^* & \sqrt{1-\gamma_{21}-\gamma_{20}}\rho_{12}^* & (1-\gamma_{21}-\gamma_{20})\rho_{22}
\end{pmatrix},\\ \label{deftildegamma23} 
& \tilde{\Phi}_{\Gamma}^{(\gamma_{10}=0)}(\hat{\rho})=\left(
\begin{array}{ccc}
 1-(\gamma_{21}+\gamma_{20}) \rho_{22} & \sqrt{\gamma_{21}} \rho_{12} & \sqrt{\gamma_{20}} \rho_{02} \\
  \sqrt{\gamma_{21}} \rho_{12}^* & \gamma_{21} \rho_{22} & 0 \\
 \sqrt{\gamma_{20}} \rho_{02}^* & 0 & \gamma_{20} \rho_{22} \\
\end{array}
\right),
\end{eqnarray}
\end{widetext}

It's immediate to notice that $\Phi_{\Gamma}^{(\gamma_{10}=0)}$ has a noiseless subspace given by $\{\ket{0},\ket{1}\}$ and consequently
can establish the following lower bound: 
\begin{eqnarray} 
 C_\text{p}(\Phi_{\Gamma}^{(\gamma_{10}=0)}) \geq Q(\Phi_{\Gamma}^{(\gamma_{10}=0)})\geq \log_2(2)=1\;. \label{CONST23} 
\end{eqnarray} 
This implies that $\Phi_{\Gamma}^{(\gamma_{10}=0)}$ can't be antidegradable (we can reach the same conclusion by noticing that  $\tilde{\Phi}_{\Gamma}^{(\gamma_{10}=0)}$ has a kernel that is not included into the kernel of $\Phi_{\Gamma}^{(\gamma_{10}=0)}$ \cite{STR_DEG_CH}. For instance the former
  contains $\ket{0}\!\!\bra{1}$ while the latter doesn't). \\

\noindent To compute the capacities we follow the channel inversion method described in Appendix \ref{sec:channelinversion}. We find that $\Phi_{\Gamma}^{(\gamma_{10}=0)}$ can be inverted when $\gamma_{21}+\gamma_{20}=1$, 
and that $\tilde{\Phi}_{\Gamma}^{(\gamma_{10}=0)}\circ \Phi_{\Gamma}^{(\gamma_{10}=0)-1}$ is LCPTP when $\gamma_{21}+\gamma_{20}\leq \frac{1}{2}$,
 identifying then the region of degradability for the channel.   
 There, exploiting the channel covariance described in Sec. \ref{sec: Appendix averaged covariance}, we can compute $Q(\Phi_{\Gamma}^{(\gamma_{10}=0)})$ and as
\hspace*{-0.6cm}\vbox{\begin{align}\label{eq:IcohD23}
&Q(\Phi_{\Gamma}^{(\gamma_{10}=0)})\hspace*{-0.3cm}&&=C_\text{p}(\Phi_{\Gamma}^{(\gamma_{10}=0)})& \nonumber \\
& &&=\max_{p_1,p_2} \Big\{ -(p_1 +\gamma_{21}p_2)\log_2(p_1 +\gamma_{21}p_2) -[1-p_1-(1-\gamma_{20}) p_2]\log_2[1-p_1- (1-\gamma_{20}) p_2] \nonumber \\
& && \qquad\qquad -(1-\gamma_{21}-\gamma_{20})p_2\log_2((1-\gamma_{21}-\gamma_{20})p_2) +(1-(\gamma_{21}+\gamma_{20})p_2)\log_2(1-(\gamma_{21}+\gamma_{20})p_2)  \nonumber \\
& && \qquad\qquad +\gamma_{21} p_2\log_2(\gamma_{21} p_2)+\gamma_{20} p_2\log_2( \gamma_{20} p_2) \Big\} \; .
\end{align}}

We are therefore able to evaluate numerically the value of $Q$ and $C_\text{p}$ on the boundary of the degradability region $\gamma_{21}+\gamma_{20}= \frac{1}{2}$, where we are able to showing that there it's equal to the lower bound in Eq.~(\ref{CONST23}). 
 This, together with the closure under composition of channels $\Phi_{\Gamma}^{(\gamma_{10}=0)}$ and data processing inequalities, allows us to conclude that $Q(\Phi_{\Gamma}^{(\gamma_{10}=0)}), \; C_{\text{p}}(\Phi_{\Gamma}^{(\gamma_{10}=0)})=1$ in the entire parameter region $\gamma_{21}+\gamma_{20}\geq 1/2$.

\subsection{$Q$ and $C_{\text{p}}$ for $\gamma_{21}=0$}\label{app: Q gamma_21}

In the case of ReMAD channels with $\gamma_{21}=0$ we have that channels and complementary channels actions reduce to
\begin{widetext}
\begin{eqnarray}\label{eq:matrix form D13}
&\Phi_{\Gamma}^{(\gamma_{21}=0)}(\hat{\rho})=\begin{pmatrix}
\rho_{00}+\gamma_{10} \rho_{11}+\gamma_{20} \rho_{22}& \sqrt{1-\gamma_{10}} \rho_{01} & \sqrt{1-\gamma_{20}}\rho_{02}\\
\sqrt{1-\gamma_{10}}\rho_{01}^* & (1-\gamma_{10})\rho_{11} & \sqrt{1-\gamma_{10}}\sqrt{1-\gamma_{20}}\rho_{12}\\
\sqrt{1-\gamma_{20}}\rho_{02}^* & \sqrt{1-\gamma_{10}}\sqrt{1-\gamma_{20}}\rho_{12}^* & (1-\gamma_{20})\rho_{22}
\end{pmatrix},\\ \label{eq:matrix form D13 tilde}
& \tilde{\Phi}_{\Gamma}^{(\gamma_{10}=0)}(\hat{\rho})=\left(
\begin{array}{ccc}
 1-\gamma_{10} \rho_{11}-\gamma_{20} \rho_{22} & \sqrt{\gamma_{10}} \rho_{01} & \sqrt{\gamma_{20}} \rho_{02} \\
 \sqrt{\gamma_{10}} \rho_{01}^* & \gamma_{10} \rho_{11} & \sqrt{\gamma_{10}} \sqrt{\gamma_{20}} \rho_{12} \\
 \sqrt{\gamma_{20}} \rho_{02}^* & \sqrt{\gamma_{10}} \sqrt{\gamma_{20}} \rho_{12}^* & \gamma_{20} \rho_{22} \\
\end{array}
\right).
\end{eqnarray}
\end{widetext}

To compute the capacities we follow the channel inversion method described in Appendix \ref{sec:channelinversion}. We find that $\Phi_{\Gamma}^{(\gamma_{21}=0)}$ is invertible when $\gamma_{10},\gamma_{20}<1$, while $\tilde{\Phi}_{\Gamma}^{(\gamma_{21}=0)}\circ\Phi_{\Gamma}^{(\gamma_{21}=0)-1}$ is LCPTP when $\gamma_{10},\gamma_{20}\leq \frac{1}{2}$. From this follows that in this range of parameters the channels are degradable. Comparing  Eqs.~(\ref{eq:matrix form D13}) with (\ref{eq:matrix form D13 tilde}) we can also see that 
 \begin{eqnarray}
 \tilde{\Phi}_\Gamma^{({\gamma_{10},0,\gamma_{20}})}=\Phi_\Gamma^{(1-\gamma_{10},0,1-\gamma_{20})}\;. 
 \end{eqnarray}
 Therefore, by the same argument above, we have that $\Phi_{\Gamma}^{(\gamma_{21}=0)}$ is antidegradable for $\gamma_{10},\gamma_{20}\geq \frac{1}{2}$ 
 and that $Q(\Phi_{\Gamma}^{(\gamma_{21}=0)}), C_{\text{p}}(\Phi_{\Gamma}^{(\gamma_{21}=0)})=0$ for that range of values. 
To compute $Q(\Phi_{\Gamma}^{(\gamma_{21}=0)})$ and $C_{\text{p}}(\Phi_{\Gamma}^{(\gamma_{21}=0)})$ in the degradable region we exploit the channels covariance described in Sec.~\ref{sec: Appendix averaged covariance} to maximize only over diagonal inputs, getting
\begin{widetext}
\begin{align}\label{eq:cohID13}
& Q(\Phi_{\Gamma}^{(\gamma_{21}=0)})=C_{\text{p}}(\Phi_{\Gamma}^{(\gamma_{21}=0)}) \nonumber \\
&=\max_{\substack{p_1,p_2\in [0,1] \\ p_1+p_2\leq 1}} \Big\{-[1 - (1-\gamma_{10})p_1+(1-\gamma_{20})p_2]\log_2[1 - (1-\gamma_{10})p_1+(1-\gamma_{20})p_2]\nonumber\\ 
&\qquad\qquad\qquad -(1-\gamma_{10})p_1\log_2((1-\gamma_{10})p_1) -(1-\gamma_{20})p_2\log_2((1-\gamma_{20})p_2)\nonumber\\ 
&\qquad\qquad\qquad +(1-\gamma_{10}p_1-\gamma_{20}p_2)\log_2(1-\gamma_{10}p_1-\gamma_{20}p_2) +\gamma_{10} p_1\log_2(\gamma_{10} p_1)+\gamma_{20} p_2\log_2(\gamma_{20} p_2)
\Big\} \; .
\end{align}
\end{widetext}

 The capacities are also known on the boundaries of the parameters space, since when one of the remaining damping parameters is 0 $\Phi_{\Gamma}^{(\gamma_{21}=0)}$ reduces to a single-decay qutrit MAD, for which $Q$ and $C_\text{p}$ are known \cite{MAD}. 
  When one of the damping parameters is instead 1 we reduce to the MAD channel discussed in Appendix~\ref{sec:first}, for which $Q$ is already available. More precisely in Appendix.~\ref{sec:first} we compute $Q(\Phi^{(1,0,\gamma_{20})})$, verifying that it coincides with the capacity of a qubit ADC.
  
\noindent Since the value of $Q$ is known also on the boundaries of the degradable region, we can compare $Q({\Phi}_\Gamma^{(\gamma_{10},0,\gamma_{20})})$ at $\gamma_{20}=\frac{1}{2}$ and $\gamma_{20}=1$, for all $\gamma_{21}\geq 1/2$. We find that the two are the same, i.e. 
$Q({\Phi}_\Gamma^{(\gamma_{10},0,1)}) = Q({\Phi}_\Gamma^{(\gamma_{10},0,1/2})$, 
Accordingly, invoking the data-processing inequality and closure under composition, we can finally conclude that
\begin{eqnarray}  \label{GOODone} 
Q({\Phi}_\Gamma^{(\gamma_{10},0,1)}) = Q({\Phi}_\Gamma^{(\gamma_{10},0,\gamma_{20}})\; \qquad \forall \gamma_{20} \geq \frac{1}{2}\;,
\end{eqnarray} 
that allows us to evaluate the $Q$ and $C_{\text{p}}$ over the entire parameters region.

\subsection{$Q$ and $C_{\text{p}}$ for $\gamma_{10}=1$} \label{sec:first} 
We work here under the assumption of qutrit ReMAD channels with $\gamma_{10}=1$, for which the expression of channels and their complementary becomes
\begin{widetext} 
\begin{eqnarray}\label{eq:matr expr 1=123}
&&\Phi_\Gamma^{(1,\gamma_{21},\gamma_{20})}(\hat{\rho})=\begin{pmatrix}
1-(1-\gamma_{20}) \rho_{22}& 0 & \sqrt{1-\gamma_{21}-\gamma_{20}}\rho_{02}\\
0 & \gamma_{21} \rho_{22} & 0\\
\sqrt{1-\gamma_{21}-\gamma_{20}}\rho_{02}^* & 0 & (1-\gamma_{21}-\gamma_{20})\rho_{22}
\end{pmatrix},\\ \nonumber \\ 
& &\tilde{\Phi}_{{\Gamma}}^{(1,\gamma_{21},\gamma_{20})}(\hat{\rho})=\left(
\begin{array}{ccc}
 \rho_{00}+ (1-\gamma_{21}-\gamma_{20})\rho_{22} & \rho_{01} &\sqrt{\gamma_{20}} \rho_{02} \\
  \rho_{01}^* &  \rho_{11}+\gamma_{21} \rho_{22} &
    \sqrt{ \gamma_{20}} \rho_{12}\\
 \sqrt{\gamma_{20}} \rho_{02}^* & \sqrt{\gamma_{20}} \rho_{12}^* & \gamma_{20} \rho_{22} \\
\end{array}
\right) \; , \label{eq:matr expr 1=123ewe}
\end{eqnarray}
\end{widetext}
for $\gamma_{21},\gamma_{20}\in[0,1]$ such that $\gamma_{21}+ \gamma_{20}\leq 1$. 
The channel cannot be degradable since $\Phi_\Gamma^{(1,\gamma_{21},\gamma_{20})}$ has a kernel that is not included into the kernel of $\tilde{\Phi}_\Gamma^{(1,\gamma_{21},\gamma_{20})}$ \cite{STR_DEG_CH}. We are still able though to express exactly the value of its quantum capacity.  Specifically we are able to state that 
 \begin{eqnarray}
Q(\Phi_\Gamma^{(1,\gamma_{21},\gamma_{20})}) &=&C_{\text{p}}(\Phi_\Gamma^{(1,\gamma_{21},\gamma_{20})}) = Q^{(1)}(\Phi_\Gamma^{(1,\gamma_{21},\gamma_{20})}) ={\cal Q}(\gamma_{21},\gamma_{20}) \;, \label{exact}
\end{eqnarray} 
with ${\cal Q}(\gamma_{21},\gamma_{20})$ defined as 
\begin{eqnarray}\nonumber 
{\cal Q}(\gamma_{21},\gamma_{20})&\equiv&
\max_{\hat{\tau}_{\text{diag}}} \left\{ S(\Phi_\Gamma^{(1,\gamma_{21},\gamma_{20})}(\hat{\tau}_{\text{diag}})) -
 S(\tilde{\mathcal{D}}_{(1,\gamma_{21},\gamma_{20})}(\hat{\tau}_{\text{diag}}))\right\}  
 \\
 &=&\max_{p\in [0,1]} \Big\{  - (1-(1-\gamma_{20}) p) \log_2(1-(1-\gamma_{20}) p) - (1-\gamma_{21}-\gamma_{20})p \log_2 (1-\gamma_{21}-\gamma_{20})p)\nonumber \\
&& \qquad\quad\; + ( 1-(\gamma_{21}+\gamma_{20}) p) \log_2( 1-(\gamma_{21}+\gamma_{20}) p) +\gamma_{20} p \log_2 \gamma_{20} p)\Big\} \;, \label{eq:QCapacity12312}
\end{eqnarray}
where we are able to restrict the maximization to diagonal density matrices of form $\hat{\tau}_{\text{diag}}=(1-p) |0\rangle\!\langle 0|+ p |2\rangle\!\langle 2|$  of $\text{A}'$ associated with the subspace ${\cal H}_{\text{A}'}\equiv \mbox{Span}\{\ket{0},\ket{2}\}$.
We plot ${\cal Q}(\gamma_{21},\gamma_{20})$ in Fig.~\ref{fig: Q(21,20)}: 
we notice that when $\gamma_{20}\geq \frac{1-\gamma_{21}}{2}$ we have ${\cal Q}(\gamma_{21},\gamma_{20}) = 0$, coherently with the fact that in that parameter region the channel $\Phi_\Gamma^{(1,\gamma_{21},\gamma_{20})}$ has zero capacity, i.e. 
 \begin{eqnarray}
Q(\Phi_\Gamma^{(1,\gamma_{21},\gamma_{20})}) &=&C_{\text{p}}(\Phi_\Gamma^{(1,\gamma_{21},\gamma_{20})})  = 0 \quad   \forall \; 1-\gamma_{21}\geq  \gamma_{20}\geq \tfrac{1-\gamma_{21}}{2}\;.  \label{exact0}
\end{eqnarray}   
\begin{figure}[t!]
     \centering
     \includegraphics[width=0.5\textwidth]{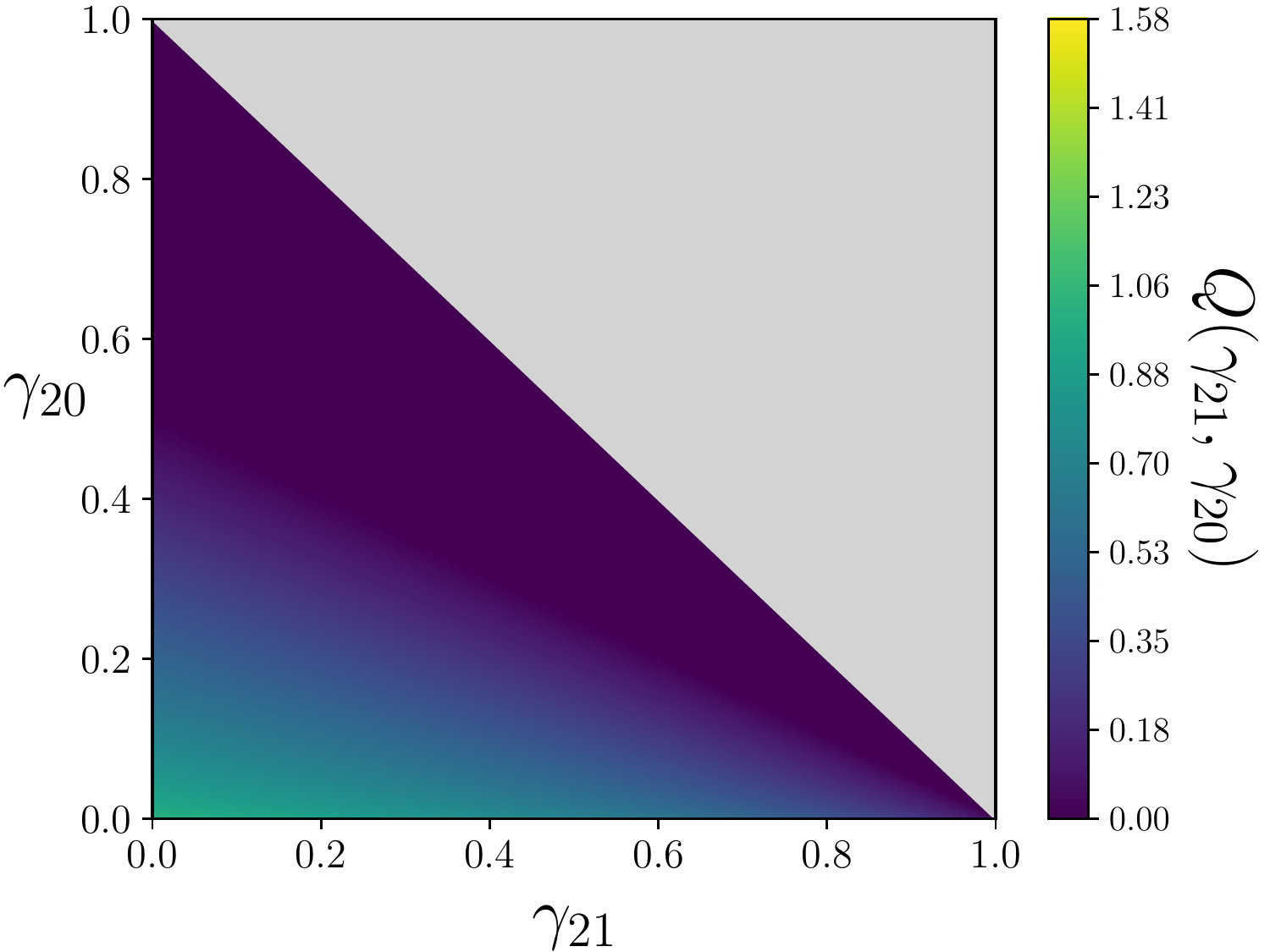}    
     \caption{ Numerical evaluation of $Q(\gamma_{21},\gamma_{20})$ obtained by performing the maximization in Eq.~(\ref{eq:QCapacity12312}). }
        \label{fig: Q(21,20)}
\end{figure}

To prove Eq.~(\ref{exact}) let us start by observing that ${\cal Q}(\gamma_{21},\gamma_{20})$ provides a natural lower bound for
$Q(\Phi_\Gamma^{(1,\gamma_{21},\gamma_{20})})$ and hence for $C_{\text{p}}(\Phi_\Gamma^{(1,\gamma_{21},\gamma_{20})})$:  
\begin{eqnarray} \label{lower1}
Q(\Phi_\Gamma^{(1,\gamma_{21},\gamma_{20})}) &\geq &  \max_{\hat{\rho}}  I_{\text{coh}}(\Phi_\Gamma^{(1,\gamma_{21},\gamma_{20})}, \hat{\rho}) \geq \max_{\hat{\tau}_{\text{diag}}}  I_{\text{coh}}(\Phi_\Gamma^{(1,\gamma_{21},\gamma_{20})}, \hat{\tau}_{\text{diag}}) = {\cal Q}(\gamma_{21},\gamma_{20})\;.
\nonumber 
\end{eqnarray} 
We need now to show that ${\cal Q}(\gamma_{21},\gamma_{20})$ can also upper bound  $Q(\Phi_\Gamma^{(1,\gamma_{21},\gamma_{20})})$.
To do this we construct a new channel $\Phi '^{(\gamma_{21},\gamma_{20})}$ 
with larger capacity than $\Phi_\Gamma^{(1,\gamma_{21},\gamma_{20})}$, i.e.
\begin{eqnarray} 
Q(\Phi_\Gamma^{(1,\gamma_{21},\gamma_{20})}) &\leq& Q(\Phi '^{(\gamma_{21},\gamma_{20})}) \;, \label{impo1} \\
C_{\text{p}}(\Phi_\Gamma^{(1,\gamma_{21},\gamma_{20})}) &\leq& C_{\text{p}}(\Phi '^{(\gamma_{21},\gamma_{20})}) \;, \label{impo1cp} 
\end{eqnarray} 
and for which we can show that 
\begin{eqnarray} Q(\Phi '^{(\gamma_{21},\gamma_{20})})=C_{\text{p}}(\Phi '^{(\gamma_{21},\gamma_{20})}) = {\cal Q}(\gamma_{21},\gamma_{20})\;.\label{impo2}  \end{eqnarray}  
For this purpose notice that, since the population of level $|1\rangle$ completely depleted, the output produced by $\Phi_\Gamma^{(1,\gamma_{21},\gamma_{20})}$ can be reproduced by the channel
$\Phi '^{(\gamma_{21},\gamma_{20})}: {\cal L}({\cal H}_{\text{A}'}) \rightarrow {\cal L}({\cal H}_{\text{A}}) $ operating on the two-levels Hilbert space ${\cal H}_{\text{A}'}\equiv \mbox{Span}\{\ket{0},\ket{2}\}$,
and producing qutrit states of A as outputs. In particular, calling $\hat{\tau}$ a generic density matrix on ${\cal H}_{\text{A}'}$ we have
\begin{equation}\label{eq:matrix form Phi1=123}
\Phi '^{(\gamma_{21},\gamma_{20})}(\hat{\tau})=\begin{pmatrix}
1-(1-\gamma_{20}) \tau_{22}& 0 & \sqrt{1-\gamma_{21}-\gamma_{20}}\tau_{02}\\
0 & \gamma_{21} \tau_{22} & 0\\
\sqrt{1-\gamma_{21}-\gamma_{20}}\tau_{02}^* & 0 & (1-\gamma_{21}-\gamma_{20})\tau_{22}
\end{pmatrix}
\end{equation} 
with associated complementary channel 
\begin{equation}  \label{eq:matrix form Phi1=123tilde}
\tilde{\Phi}'^{(\gamma_{21},\gamma_{20})}(\hat{\tau})=\left(
\begin{array}{ccc}
 1-(\gamma_{21}+\gamma_{20})\tau_{22} & 0 & \sqrt{\gamma_{20}}\tau_{02}\\
 0 & \gamma_{21} \tau_{22} & 0\\
 \sqrt{\gamma_{20}}\tau_{02}^* & 0 & \gamma_{20}\tau_{22}
\end{array}
\right),
\end{equation}
where for $i,j=0,2$ we set $\tau_{ij}\equiv \langle i| \hat{\tau} | j\rangle$.

\noindent $\Phi '^{(\gamma_{21},\gamma_{20})}$
 fulfills the inequality (\ref{impo1}) because $\Phi_\Gamma^{(1,\gamma_{21},\gamma_{20})}$, while producing the same output of $\Phi '^{(\gamma_{21},\gamma_{20})}$, is  also `wasting' resources in the useless level $|1\rangle$. Explicitly, notice that we can write
 \begin{eqnarray} 
\Phi_\Gamma^{(1,\gamma_{21},\gamma_{20})} = \Phi '^{(\gamma_{21},\gamma_{20})} \circ {\cal A} \;, 
\end{eqnarray} 
 being $\mathcal{A}: {\cal L}({\cal H}_{\text{A}}) \rightarrow
{\cal L}({\cal H}_{\text{A}'})$ is the LCPTP map bringing the input state of the qutrit A to the qubit $\text{A}'$ by transfer of the population of the level $\ket{1}$ to $\ket{0}$ followed by the erasure of $\ket{1}$ , i.e. 
\begin{eqnarray}\label{eq:matr expr 1=123adsf}
\mathcal{A}(\hat{\rho})&=&\begin{pmatrix}
\rho_{00} + \rho_{11} & \rho_{02} \\
\rho_{20} &\rho_{22} \\
\end{pmatrix},
\end{eqnarray}
 where $\rho_{ij} =\langle i | \hat{\rho} |j\rangle$ with $\hat{\rho} \in \mathfrak{S}({\cal H}_{\text{A}})$.  Equation~(\ref{impo1}) can then be derived as a consequence of the data processing inequality applied to $Q$ (and $C_{\text{p}}$). 
 The second part of the argument, i.e. Eq.~(\ref{impo2}), can instead be proved by noticing that, differently of the original channel $\Phi_\Gamma^{(1,\gamma_{21},\gamma_{20})}$ that is not degradable, $\Phi '^{(\gamma_{21},\gamma_{20})}$ is  degradable if 
\begin{eqnarray}0\leq  \gamma_{20}\leq ({1-\gamma_{21}})/{2}\;, \label{REDGED} \end{eqnarray}
and antidegradable otherwise, i.e. for  $(1-\gamma_{21})/2\leq \gamma_{20}\leq {1-\gamma_{21}}$.
This can be shown by observing that in the region identified by the inequality~(\ref{REDGED}) the quantity 
\begin{eqnarray} \bar{\gamma}_3\equiv
 \frac{1-\gamma_{21}-2\gamma_{20}}{1-\gamma_{21}-\gamma_{20}}\;, \end{eqnarray} 
 belongs to $[0,1]$ and can be used to build a single-decay qutrit MAD channel 
   $\Phi_\Gamma^{(0,0,\bar{\gamma}_3)}$ . Furthermore by direct calculation we get  
  \begin{eqnarray}\label{deg} 
 \Phi_\Gamma^{(0,0,\bar{\gamma}_3)} \circ \Phi '^{(\gamma_{21},\gamma_{20})} = \tilde{\Phi} '^{(\gamma_{21},\gamma_{20})}\;,
 \end{eqnarray}  
 which shows that $\Phi_\Gamma^{(0,0,\bar{\gamma}_3)}$ acts as the 
 degrading channel of $\Phi '^{(\gamma_{21},\gamma_{20})}$. 
   From Eqs.~(\ref{eq:matrix form Phi1=123}) and (\ref{eq:matrix form Phi1=123tilde}) 
 it is also evident that $\Phi '^{(\gamma_{21},\gamma_{20})}$ can be recovered from $\tilde{\Phi}'^{(\gamma_{21},\gamma_{20})}$ by the substitution $\gamma_{20}\rightarrow 1-\gamma_{21}-\gamma_{20}$. Consequently, using the same construction of Eq.~(\ref{deg}), we can conclude that $\Phi '^{(\gamma_{21},\gamma_{20})}$  is antidegradable when $(1-\gamma_{21})/2\leq \gamma_{20}\leq {1-\gamma_{21}}$.

Finally, to derive Eq.~(\ref{impo2}) we observe that as the original channel $\Phi_\Gamma^{(1,\gamma_{21},\gamma_{20})}$ also $\Phi '^{(\gamma_{21},\gamma_{20})}$ is covariant and accordingly we can express its capacity as 
\begin{align} \label{CAPQprime} 
\hspace*{-0.5cm}Q(\Phi '^{(\gamma_{21},\gamma_{20})})=C_{\text{p}}(\Phi '^{(\gamma_{21},\gamma_{20})})=\max_{\hat{\tau}_{\text{diag}}} \Big\{ S(\Phi '^{(\gamma_{21},\gamma_{20})}(\hat{\tau}_{\text{diag}})) - S(\tilde{\Phi}'^{(\gamma_{21},\gamma_{20})}(\hat{\tau}_{\text{diag}}))\Big\} = {\cal Q}(\gamma_{21},\gamma_{20}),
\end{align}
 where the last identity follows from the fact that $\Phi '^{(\gamma_{21},\gamma_{20})}(\hat{\tau}_{\text{diag}}))$ coincides with 
   $\Phi_\Gamma^{(1,\gamma_{21},\gamma_{20})}(\hat{\tau}_{\text{diag}})$  and by the fact that the positive component of the spectrum of 
 $\tilde{\Phi}'^{(\gamma_{21},\gamma_{20})}(\hat{\tau}_{\text{diag}})$ coincides with the one of $\tilde{\Phi}^{(1,\gamma_{21},\gamma_{20})}(\hat{\tau}_{\text{diag}})$ (strictly speaking the above derivation  holds true only in the
 degradable region~(\ref{REDGED}) of 
 $\Phi '^{(\gamma_{21},\gamma_{20})}(\hat{\tau}_{\text{diag}}))$: still since ${\cal Q}(\gamma_{21},\gamma_{20})$ nullifies for 
 $1-\gamma_{21}\geq \gamma_{20}\geq ({1-\gamma_{21}})/{2}$, we can apply~(\ref{CAPQprime}) also in the antidegradability region of the channel where
 $Q(\Phi '^{(\gamma_{21},\gamma_{20})})=0$). 
 
\noindent Finally, we notice that for $\gamma_{21}=0$ the effective channel in Eq.~(\ref{eq:matrix form Phi1=123})  can be replaced by the channel 
 \begin{equation}\label{eq:matrix form Phi1=123gamma2=0}
\Phi '^{\gamma_{20}}(\hat{\tau})=\begin{pmatrix}
1-(1-\gamma_{20}) \tau_{22}&  \sqrt{1-\gamma_{20}}\tau_{02}\\
\sqrt{1-\gamma_{20}}\tau_{02}^* & (1-\gamma_{20})\tau_{22}
\end{pmatrix}\;, 
\end{equation}
mapping the two-level system $\text{A}'$ into itself via a qubit ADC with damping parameter $\gamma_{20}$. Accordingly, following the same analysis done above it follows that $Q(\Phi_\Gamma^{(1,0,\gamma_{20})})$ and $C_{\text{p}}(\Phi_\Gamma^{(1,0,\gamma_{20})})$ coincide with the respective capacities of a qubit ADC, computed in Ref.~\cite{QUBIT_ADC}.

\clearpage

\bibliographystyle{apsrev}
\bibliography{ReMAD_bib}

\end{document}